\documentclass[suppldata]{interact}
\usepackage{comment}
\DeclareUnicodeCharacter{2061}{}

\usepackage{dcolumn}
\usepackage{bm}
\usepackage{multirow}
\usepackage{makecell}
\usepackage{longtable}
\usepackage{threeparttable}
\usepackage{flushend}
\usepackage{tablefootnote}
\usepackage{bbding}
\usepackage{pifont}
\usepackage{wasysym}
\usepackage{amssymb}
\usepackage{color, url}
\usepackage{braket}
\usepackage{graphicx}
\usepackage{braket}
\usepackage{amsmath}
\usepackage{float}
\usepackage{bbding}
\usepackage{hyperref}
\usepackage{makecell}
\usepackage{caption}
\usepackage{amsfonts}
\usepackage{multicol}
\usepackage{multirow}
\usepackage{hyphenat}
\usepackage{comment}
\hypersetup{
    colorlinks=true,
    linkcolor=blue,
    filecolor=magenta,      
    urlcolor=cyan,
}

\usepackage{ifpdf}
\usepackage{footnote}
\usepackage{epstopdf}
\usepackage[caption=false]{subfig}

\usepackage[numbers,sort&compress]{natbib}
\bibpunct[, ]{[}{]}{,}{n}{,}{,}
\renewcommand\bibfont{\fontsize{10}{12}\selectfont}
\makeatletter
\def\NAT@def@citea{\def\@citea{\NAT@separator}}
\makeatother

\theoremstyle{plain}

\theoremstyle{definition}

\theoremstyle{remark}

\begin{document}


\title{The Unified Effect of Data Encoding, Ansatz Expressibility and Entanglement on the Trainability of HQNNs}

\author{
\name{Muhammad Kashif\thanks{CONTACT Email: mkashif@hbku.edu.qa} and Saif Al-Kuwari}
\affil{Division of Information Computing, College of Science and Engineering, Hamad Bin Khalifa University, Qatar Foundation, Doha Qatar}
}

\maketitle

\begin{abstract}
In this paper, we propose a framework to study the combined effect of several factors that contribute to the barren plateau problem in quantum neural networks (QNNs), which is a critical challenge in quantum machine learning (QML). These factors include data encoding, qubit entanglement, and ansatz expressibility. To investigate this joint effect in a real-world context, we focus on hybrid quantum neural networks (HQNNs) for multi-class classification. Our proposed framework aims to analyze the impact of these factors on the training landscape of HQNNs. Our findings show that the barren plateau problem in HQNNs is dependent on the expressibility of the underlying ansatz and the type of data encoding. Furthermore, we observe that entanglement also plays a role in the barren plateau problem. By evaluating the performance of HQNNs with various evaluation metrics for classification tasks, we provide recommendations for different constraint scenarios, highlighting the significance of our framework for the practical success of QNNs.  
\end{abstract}

\begin{keywords}
quantum machine learning; entanglement; data encoding; quantum neural networks; trainability
\end{keywords}

\section{Introduction}

The quest to build practical quantum computers has intensified over the last few years. 
Several quantum devices, with around one hundred qubits, have already been developed.
These devices are known as noisy intermediate-scale quantum (NISQ) devices \cite{preskil:2018}. 
Although these devices are limited and susceptible to errors, they demonstrate clear advantage for specific applications as compared to the best existing classical computers \cite{Arute:2019, zhong:2020, Wu:2021, Madsen:2022}. 
As the NISQ devices require quantum routines of shallow depth and robustness against noise, a hybrid design space integrating classical and quantum processing has become a leading approach to realize the potential of quantum computing for a wide range of applications \cite{Bergholm:2018}.


In a hybrid design space context, variational quantum algorithms (VQAs) are the most popular class of algorithms. 
These algorithms utilize NISQ devices for evaluating the objective function through parameterized quantum circuits (PQCs) and classical devices for function optimization with respect to the target application. 
The VQAs have been studied for a wide range of applications, including quantum chemistry \cite{Peruzzo:2014}, state diagonalization \cite{LaRose:2019}, factorization \cite{anschuetz:2018}, quantum optimization \cite{Farhi:2014}, and quantum field theory simulation \cite{Klco:2018,Klco:2019}. 
Furthermore, these algorithms have also been studied in the context of noise resilience \cite{Sharma:2020}, trainability \cite{McClean:2018, Grant:2019,Cerezo:2021} and computational complexity \cite{McClean:2016,Biamonte:2021}. 
In other words, VQAs closely resemble machine learning (ML) algorithms as they also train a computer to learn patterns \cite{goodfellow:2016}.  
Therefore, VQAs have been proposed as a quantum analog of various ML algorithms \cite{Kubler:2019,Suzuki:2020, Date:2021,Farhi:2014,Date:2021a,Arthur:2020}. 
Consequently, the new field of quantum machine learning (QML) has emerged by merging quantum computation and ML.  


In recent years, a number of PQCs have been proposed for QML applications \cite{Cerezo:2021a}. 
Amongst them are the quantum neural networks (QNNs), which have extensively been explored  \cite{Farhi:2018,Havlicek:2019,beer:2020,Cong:2019,Du:2021a,Mitarai:2018,kashif:2021,Kashif:2022} as quantum extensions of classical deep neural networks (DNNs).
The basic building block in QNNs is quantum perceptron, which has been proposed in different ways \cite{Schuld:2020,Zhang:2020,Tacchino:2020,sharma:2020a,Killoran:2019,Torrontegui:2019,Wan:2017}. 
To this end, PQCs became a promising quantum analog of artificial neurons \cite{benedetti:2019,Hubregtsen:2020,Abbas:2021}.
Given the NISQ era limitations, hybrid quantum neural networks (HQNNs) are commonly used to analyze the potential quantum advantage in QNNs. %
HQNNs replicate the QNN's architecture by enclosing a typical QNN in some classical input pre- and output post-processing. The input preprocessing typically aims to downsize the input to cope with the limitation of NISQ devices (mainly in number of qubits), and is done via a classical neuron layer with fewer neurons or some dimensionality reduction algorithm. 
The output postprocessing is performed to interpret the output of enclosed QNN output in a meaningful way. The postprocessing is usually done via a classical neuron layer at the end, which also allows to apply the familiar non-linear activation functions to get the final output/prediction.
Although QNNs are being extensively explored for various applications, the literature still lacks solid and concrete statements about their quantum advantage \cite{Rebentrost:2014,Biamonte:2017,Chia:2022}. 
%

\subsection{Barren Plateaus} \label{BP-causes}
One of the most challenging problems in HQNNs is the phenomenon of barren plateaus (BP) \cite{McClean:2018,Grant:2019,Cerezo:2021,Wang:2021}. 
In BP, the cost function landscapes during HQNN’s training become exponentially flat with an increase in the 
system size.
In other words, the gradients of parameters subject to optimization vanish exponentially as a function of the number of qubits. 
This implies that the existence of BP in HQNN's training landscapes affects their \emph{trainability}\footnote{The trainability essentially ensures that the objective function is optimized after every training iteration until the model convergence.}, resulting in a significant performance degradation and consequently limiting HQNN's applicability in practice.

It has been demonstrated in \cite{McClean:2018} that a sufficiently random ansatz (a quantum subroutine consisting of a sequence of gates applied to specific wires) will experience the BP if its uniform unitaries distribution matches up to the second moment, i.e., it forms a unitary 2-design. 
Therefore, the choice of ansatz is central to the success of hybrid quantum-classical algorithms. 
%
%
Some frequently explored ansatzes in this regard are the Hamiltonian variational ansatz \cite{Wecker:2015}, coupled cluster ansatz \cite{Bartlett:2007,lee:2018,Cao:2019}, quantum alternating operator ansatz \cite{Farhi:2014,Hadfield:2019} and hardware-efficient ansatz \cite{Kandala:2017}. 
Ideally, the ansatz is required to be both trainable and expressible to reach an optimal solution. 
The \emph{expressibility} of the ansatz is specifically desired so that it can provide an accurate approximation to the solution.
Simultaneously, the training landscapes also need to have accessible-enough features to find the solution. 
The \emph{expressibility}  of a quantum ansatz implies how uniformly the given ansatz can explore the unitary space \cite{Holmes:2022}.
The authors in \cite{Holmes:2022} extend the original work reporting BP \cite{McClean:2018} to ansatz expressibility.
They propose the idea of problem-inspired ansatz rather than the hardware-efficient ansatz, which are of significant importance in the NISQ era. 
The expressibility of ansatz plays a significant role in overcoming BP to a certain extent. 
Ansatz with greater expressibility are more susceptible to BP and vice versa. Similarly, \emph{deep ansatz} are considered to be more expressible \cite{McClean:2018}.
Moreover, ansatz expressibility can directly be derived from the nature of the target problem (more expressible ansatz for complex problems and vice versa). 
%


In addition to ansatz expressibility, the problem of BP in HQNNs may also arise due to the type of entanglement used in PQCs \cite{Marrero:2020}, the noise levels in high-depth quantum circuits \cite{Wang:2021,Maciejewski:2021,preskil:2018,Alam:2019,Xue:2019} and how the data is being encoded into the PQCs \cite{Abbas:2021}. 
Entanglement is a fundamental property of quantum mechanics and is a key to constructing expressible quantum circuits. However, it can also be a potential source of BP, as discussed in \cite{Marrero:2020}. 
Data encoding is also a crucial step in HQNNs and is often considered to be the performance bottleneck, and it can also affect the trainability and expressive of HQNNs \cite{Schuld;2021}.


\subsection{Research Gap} 
As discussed in Section \ref{BP-causes}, data encoding, ansatz expressibility, and the entanglement are simultaneously used in HQNNs. 
While the BP dependence on all these concepts has separately been explored in different studies \cite{Holmes:2022,Marrero:2020,Schuld;2021,Abbas:2021}, to the best of our knowledge, their joint (holistic) effect (with respect to each other) on the trainability of QNNs from the aspect of BP has not been investigated yet.
Furthermore, existing work focuses on the standalone mathematical implementation (formulation or modeling) of QNNs, whereas the HQNNs are more relevant for practical applications on NISQ devices, which allows us to experiment  with real-world datasets. 
Therefore, a framework for HQNNs that allows to simultaneously analyze the effect of data encoding, ansatz expressibility, and the entanglement between qubits on the trainability of HQNNs is needed. 


\subsection{Proposed Framework}
\begin{figure}[htp]
    \centering
    \includegraphics[scale=0.5]{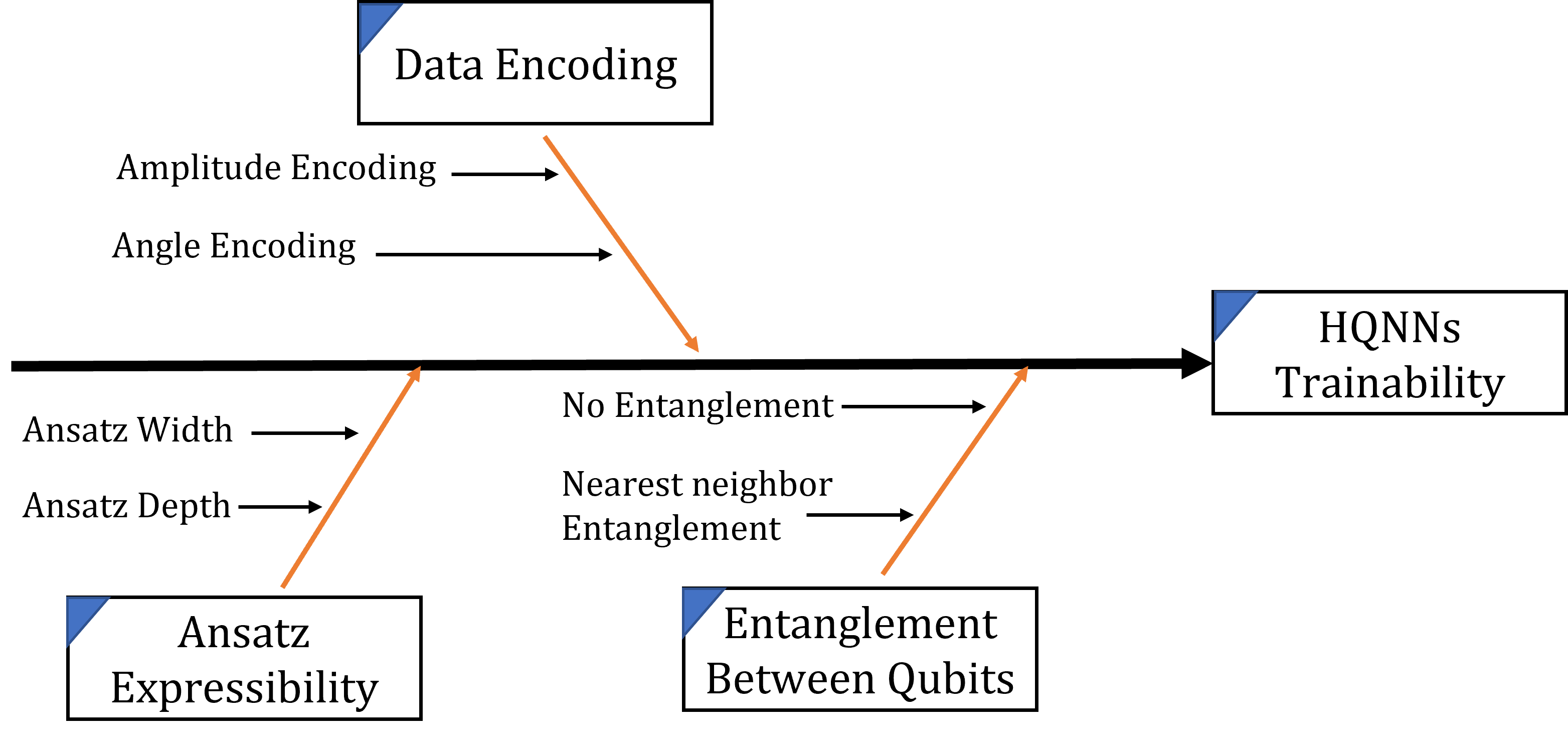}
    \caption{An Overview of Proposed Methodology}
    \label{fig:GA}
\end{figure}

In this paper, we propose a framework to perform an empirical analysis (based on data obtained from experiments) of the joint effect of data encoding, ansatz expressibility, and entanglement in HQNNs with respect to each another.  
We typically investigate the effects of aforementioned concepts in feed-forward HQNNs for a hardware-efficient periodic ansatz structure for a real-world application (multi-class classification).
An abstract illustration of our analysis is depicted in Figure \ref{fig:GA}. 

The classical to quantum feature mapping is achieved via the two frequently used data encoding strategies, called amplitude encoding and angle encoding. 
For the PQC, we consider two similar ansatz structures (to be used as hidden quantum layers in HQNN) differentiated only by the inclusion/removal of entanglement, which we name as entangled ansatz (shown as Nearest Neighbour Entanglement in Figure \ref{fig:GA}) and unentangled ansatz (shown as No Entanglement in Figure \ref{fig:GA}). 
Both the ansatz structures are separately experimented for each of the encoding scheme. 
We consider different width ($n$) of underlying quantum layers for the analysis of BP in HQNNs. 
Moreover, different depth ($n$) of quantum layers are considered for all the widths to analyze how the ansatz expressibility plays a role in trainability of HQNNs for both the ansatzes. 
%

We benchmark accuracy and loss convergence as evaluation metrics for the models used in this article. 
The HQNNs (with certain $n$ and $m$) achieving higher accuracy with faster and smoother convergence to the optimal solution are considered better than others. It typically implies that these models are not (yet) fully exposed to BP, and hence yields better performance.
In addition to the accuracy and loss convergence, we also evaluate the HQNNs for some additional performance parameters. 
These additional performance parameters are precision, recall score and  and F1-score. 

\subsection{Contribution}
We perform an extensive list of experiments to study the combined effect of data encoding, ansatz expressibility and entanglement between qubits on the trainability of HQNNs on a real world dataset. 
We typically use handwritten digit dataset from \emph{sklearn} \cite{scikit-learn}. 
The reason behind selecting this particular dataset over other famous similar datasets like MNIST \cite{lecun:1998}, is because of the smaller image size which is more suitable for NISQ devices.
Based on the achieved results, we observe that the problem of BP arises in HQNNs also, as the number of qubits increase, resulting in performance degradation. 
Furthermore, the occurrence of BP is dependent on the ansatz depth (expressibility) irrespective of the data encoding strategy. 
Consequently, we perform the trainability and expressibility analysis for both the data encodings first for entangled and then unentangled ansatz. 
This analysis provides an idea about the appropriate depth of quantum layers for the given width, when working with real-world applications. 
Moreover, the obtained results also provide an idea about which encoding strategy is slightly more advantageous (from BP and trainability aspect) than the other. 
In addition, we also observe that the entanglement between the qubits in quantum layers also plays a role in the trainability of HQNNs. 
However, it's impact (positive or negative) on the overall performance of underlying model is dependent on how the data is being encoded. 
Moreover, we also evaluate the HQNNs in terms of different evaluation metrics for classification applications (precision, recall and F1-score), which signifies the relevance of HQNNs in real-world applications. 
Lastly, we illustrate the significance of our proposed framework by   considering different constraint scenarios on the primary components of HQNNs (data encoding, ansatz expressibility and entanglement inclusion/removal),and provide recommendations for the optimal set of parameters.

It is important to note that the mathematical analysis of individual performance parameters (such as data encoding, ansatz expressibility, and the entanglement) exist in literature \cite{Schuld;2021,Abbas:2021,Holmes:2022,Marrero:2020}. 
Therefore, the mathematical formulation of various performance parameters in terms of trainability is out of the scope of this paper. 
However, a combined effect of all these non-trivial components (performance parameters) of HQNNs for a practical application has not yet been explored. 
One possible reason behind the lack of such a unified analysis is that it is challenging to theoretically analyze the effect of all these concepts (with respect to each other) simultaneously in a single framework. 
However, experimental investigation can provide the leverage of such unified analysis. 
Therefore, we attempt to experimentally/empirically analyze their joint (holistic or simultaneous) effect from practical viewpoint.

\subsection{Organization}
The rest of the paper is organized as follows: 
Section \ref{sec:HQNN} provides the necessary background of HQNNs along with its main components. 
The state-of-the-art on potential solutions and analysis of the BP problem in HQNNs is discussed Section \ref{sec:related_work}. 
The detailed methodology for the framework development of HQNNs analysis is presented in Section \ref{sec:methodology}. 
The experimentation details including the list of experiments performed is discussed in Section \ref{sec:exp_details}.
The analysis of proposed framework based on the obtained results is discussed in Section \ref{sec:results}. 
Section \ref{sec:significance} illustrates the significance of our proposed framework for various design constraints of HQNNs. Finally, Section \ref{conclusion} concludes the paper. 


\section{Hybrid Quantum Neural Networks} \label{sec:HQNN}
In this section, HQNNs are discussed in analogy with classical deep neural networks. 
Furthermore, the main components of HQNNs in correspondence with what is used in this paper, i.e., data encoding, observable measurement and cost function details are discussed in Section \ref{sec:data_enc} and Section \ref{sec:observable_costfucntion} respectively. Lastly, the ansatz trianability and expressibility are discussed in section \ref{sec:anstaz_train_express}.

\begin{figure}
\centering
\subfloat[DNN ]{\label{fig:DNN}%
\resizebox*{7cm}{!}{\includegraphics{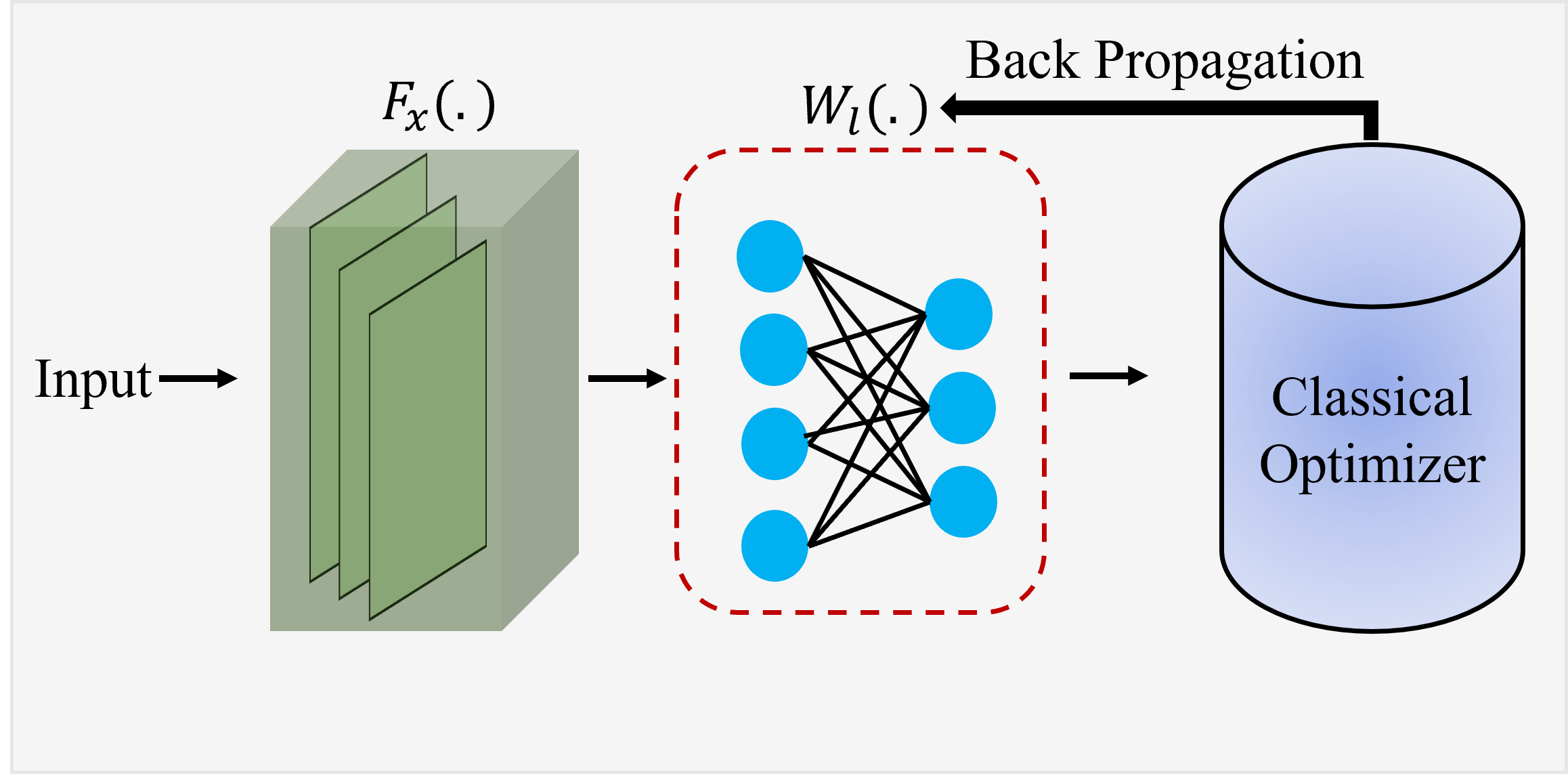}}}\hspace{5pt}
\subfloat[QNN \label{fig:QNN}]{%
\resizebox*{7cm}{!}{\includegraphics{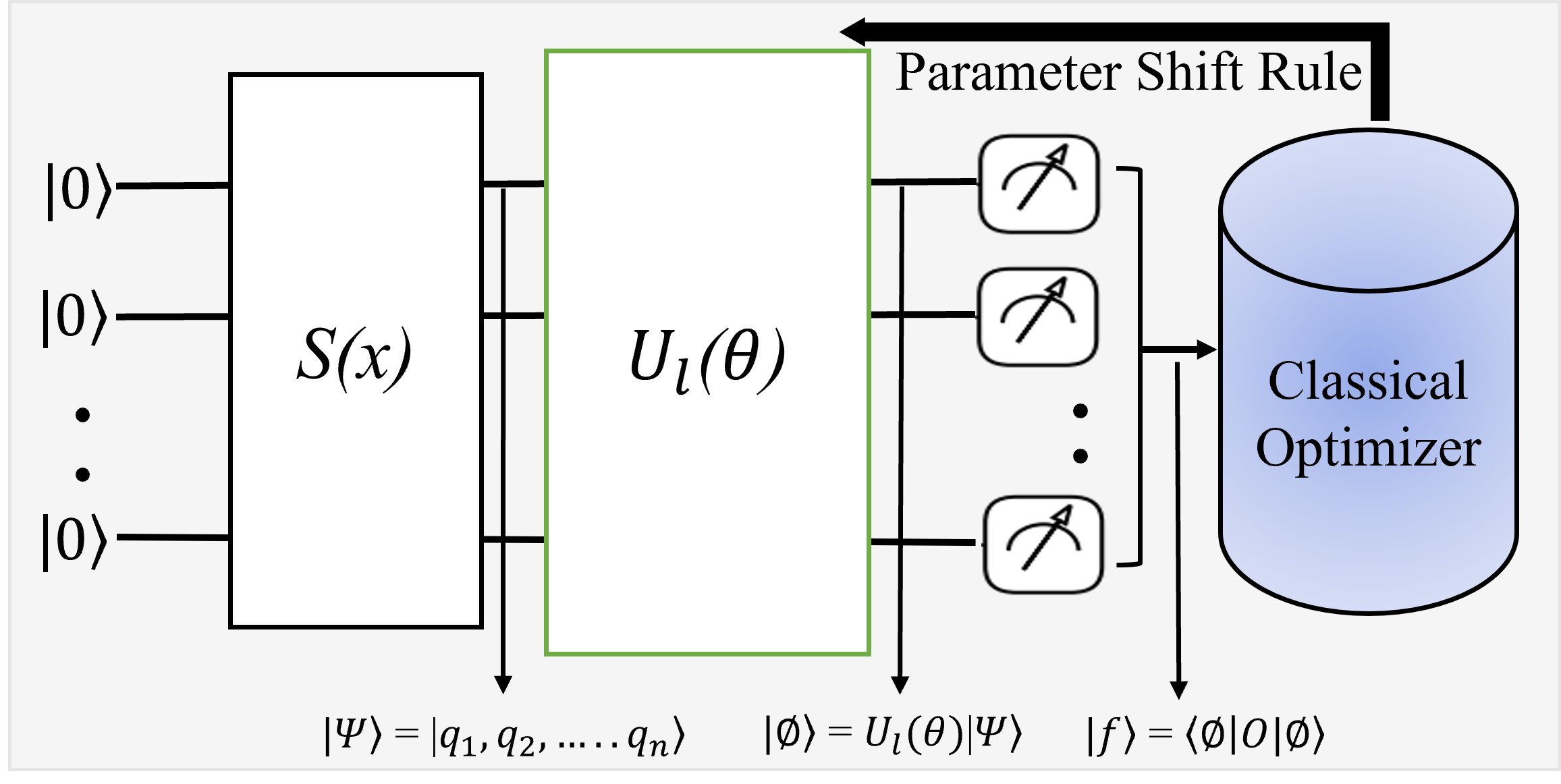}}}
\caption{Schematic Illustration of DNN and QNN} \label{fig:DNN_and_QNN}
\end{figure}

In a typical setting of classical DNNs (Figure \ref{fig:DNN}), the first step is to map the input data to the feature space via feature embedding layers $F_x(.)$. %
The embedded data is then trained through fully connected neuron layers $\prod_{l} W_l(.)$ to learn the inherent relationships between the input and output of a particular dataset. 
The number of layers and neurons in each layer can be customized according to the complexity of the target application. 


The quantum counterpart of DNNs (i.e. QNNs) has recently attracted attention due to the tremendous success of DNNs, and improved computation power QNNs may offer \cite{Du:2021}. 
QNNs have a structure similar to DNNs, as shown in Figure \ref{fig:QNN}. 
Analogous to other QML algorithms, QNNs also exploit PQCs, which are parameters (classically optimizable) dependent quantum circuits. 
Combining both architectures results in HQNN.  

The HQNNs work in five steps \cite{Arthur:2022}: (1) Input Downscaling: The first step in HQNNs working is input downscaling to cope with the limitations NISQ devices mainly in number of qubits. 
(2) Feature mapping:  Classical data points $x$ are mapped to $n-$qubit quantum state $\ket{\psi}$ represented by Equation \ref{eq1}, where $S(x)$ is the mapping function. 
(3) Training: The prepared quantum state is processed and trained using an ansatz $U$ via a series of single and multi-qubit unitaries as shown in Equation \ref{eq2}. Rotation angle in the ansatz is parameterized by vector $\theta$. 
(4) Measurement: The quantum state is then measured and the corresponding eigenvalue of the measurement observable $O$ is obtained as shown in Equation \ref{eq3}. 
(5) Classical postprocessing of the observable: the quantum state measurement results in a classical value and hence can be processed by a classical device. The postprocessing of measurement results is commonly done via a classical neuron layer. It also allows to apply a familiar non-linear activation functions like SoftMax, and an optimization routine to minimize the cost function. The steps $2-4$ mentioned above forms a typical architecture of QNN as shown in Figure \ref{fig:QNN}. Hence in a typical HQNN architecture, the QNN is completely replicated, making the analysis of quantum part (step $2-5$ above) in HQNNs valid for QNN also. 

\begin{equation} \label{eq1}
    \ket{\psi(x)} = S(x)\ket{0}^{\otimes n}
\end{equation}
    
\begin{equation} \label{eq2}
    \ket{\phi(x, \theta)} = U(\theta)\ket{\psi(x)}
\end{equation}

\begin{equation} \label{eq3}
    f(x,\theta) = \bra{\phi(x,\theta)} O \ket{\phi(x,\theta)}
\end{equation}
\label{Bibliography}

\noindent Generally, in QML, the underlying PQCs are iteratively executed for an input $x$ and a parameter vector $\theta$ to approximate its expectation value because of the probabilistic nature of quantum computation. 
In QNNs, this expectation value is usually considered as the output of the network \cite{Arthur:2022}.
In the following subsections, we provide a brief overview of different parts of HQNNs in correspondence with the approaches we use in this work.   

\subsection{Data Encoding} \label{sec:data_enc}
In QNNs, inputting the data in a way that quantum circuit can process has been a pressing challenge, and is often termed data encoding. The data encoding can be considered as a feature map that maps an input feature $x$ to the quantum system’s Hilbert space, thus creating a quantum state $\ket{\psi_x}$ \cite{Schuld:2019}. The encoding process is a crucial step while designing quantum algorithms and can significantly effects their performance\cite{Havlicek:2019,schuld:2018, Schuld:2019, Lloyd:2020}. In practice, the transformation ($x \longrightarrow \ket{\psi_x}$) is typically achieved through a unitary transformation ($S_x$), implemented using a variational circuit, whose parameters are dependent on the input data being encoded \cite{Havlicek:2019,Schuld:2019, LaRose:2020}.
The circuit ($S_x$) then acts on the initial state $\ket{\psi}$, which is usually a ground state, i.e., $\ket{\psi}=0^{\otimes n}$. The encoding is then realized as in Equation \ref{gen_enc_eq}.
 \begin{equation}
    x \mapsto E(x) = S_x \ket{\phi}\bra{\phi}S_x^\dagger = \ket{x}\bra{x} =: \rho_x 
        \label{gen_enc_eq}
\end{equation}
However, the transformation circuit $S_x$ is required to be hardware-efficient to accommodate the limitations imposed by the NISQ regime. Several encoding strategies have recently been proposed \cite{Lloyd:2020,Schuld:2019,LaRose:2020}. However, in HQNNs, the frequently used ones are amplitude and angle encoding.

\subsubsection{Amplitude Encoding}
In amplitude encoding, data is encoded into quantum state amplitudes. For $x \in \mathbb{R}^n$, the amplitude encoding maps $x \longrightarrow E(x)$ into the amplitudes of an $n-$qubit quantum state as shown in the Equation below:

\begin{equation} \label{eq5}
    \ket{\psi_x} = \sum_{i=1}^{N} x_i\ket{i}
\end{equation}

where $N=2^n$, $x_i$ is the $i^{th}$ element of $x$ and $\ket{i}$ is the $i^{th}$ computational basis state. For a classical dataset $\mathcal{D}$ with $M$ examples and $N$ features as shown in Equation below:
 \begin{equation}\label{eq6}
     \mathcal{D} =  \{  x^{(1)}, \dots x^{(m)}, \dots x^{(M)}\} 
 \end{equation}
where $x^{(m)}$ is an $N-$dimensional feature vector for $m=1, \dots M$. The amplitude can then easily be understood by concatenating $x^{(m)}$ in a single vector as follows:

 \begin{equation}\label{eq7}
     \alpha = \mathcal{C}_{norm} = \{x_1^{(1)}, \dots, x_N^{(1)}, x_1^{(2)}, \dots, x_N^{(2)}, x_1^{(M)}, \dots, x_N^{(M)}      \}
 \end{equation}
 
The factor $\mathcal{C}_{norm}$ is the normalization factor and must be normalized such that $|\alpha|^2=1$. Then the input can be represented in the computational basis using the following Equation:

\begin{equation}
    \ket{\mathcal{D}} = \sum_{i=1}^{2^n} \alpha_i\ket{i}
\end{equation}

where, $\alpha_i$ represents amplitude vector elements ($\alpha$) and $\ket{i}$ are the computational basis states. The total number of amplitudes being encoded are $N \times M$. A system of $n-$qubits can encode $2^n$ features. Therefore, amplitude encoding requires $n \geq log_2(NM)$⁡. The constants can be padded if the total number of features being encoded are greater than $2^n$ \cite{schuld:2018}.

\subsubsection{Angle Encoding}
Angle encoding, also called qubit encoding, encodes the input data features into rotation angle of qubits and has been used in various QML algorithms \cite{grant:2018, schuld:2018, Cao:2020}. For a feature vector $x = [x_1, x_2,…x_N]^T  \in \mathcal{X}^N$, the following Equation typically represents the angle encoding.

\begin{equation}
    \ket{x} = \bigotimes_{i=1}^{N} cos(x_i)\ket{0}+sin(x_i)\ket{1}
    \label{eq9}
\end{equation}
The angle encoding encodes $N$ features into an $n-$qubit system. The state preparation unitary for a single feature per qubit in qubit encoding is represented by the following Equation:

\begin{equation}
    S_{{x}_{j}} = \bigotimes_{i=1}^{N} U_i \hspace{0.2cm} where \hspace{0.2cm}
      U_i :=
    \begin{bmatrix}
 \cos(x_j^{(i)})  &  -\sin(x_j^{(i)})\\ 
 \sin(x_j^{(i)})   &\cos(x_j^{(i)}) 
\end{bmatrix}
\label{eq10}
\end{equation}
The angle encoding approach can also be generalized to encode two features in a single qubit; this is called dense qubit encoding \cite{LaRose:2020}. However, in this work, we adopt simple qubit encoding. 



\subsection{Observable Measurement and Cost Function} \label{sec:observable_costfucntion}
The qubits can be measured in different measurement bases. 
In this work, we use the eigenbasis of $\sigma^z$ for the expectation value of our PQC. 
For an $n-$qubit system, the observable measurement in $\sigma^z$ basis can be described as the tensor product of $n$ Pauli-Z matrices i.e., $O = \sigma^{\otimes n}$.  
The $\sigma^z$ observable returns $-1$ for odd parity quantum state and $1$ for even parity, keeping the overall expectation value of PQC in the range $[-1,1]$.

The training of quantum layers HQNNs is subjected to finding the parameter vector $\theta$ that minimizes the loss after every training iteration. 
Since we consider a multi-class classification problem, we use sparse categorical cross entropy as a cost function, as shown in the Equation below:

\begin{equation} \label{eq11}
    Cost = -\frac{1}{N}\sum_{i=1}^{N} [y_i log (\hat{y_i}) + (1-y_i)log(1-\hat{y_i})  ]
\end{equation}

\noindent where, $y_i$ are true labels and $\hat{y_i}$ are predictions. Classical optimization techniques, like gradient descent, can be used for the optimization of the cost function, which simply takes the partial derivatives of each parameter and decide the next minimum direction. However, the output of PQC, i.e. expectation value of measurement observable, needs to be differentiated for every parameter in the variational ansatz. The expectation value of a PQC can be differentiated with respect to each parameter using the parameter shift rule (Equation \ref{eq12}), first introduced for QML algorithms in \cite{Mitarai:2018} and extended in \cite{Schuld:2019a}. 

\begin{equation} \label{eq12}
    \frac{df}{d\theta_i} = \frac{f(\theta_i + s) - f(\theta_i - s)}{2}
\end{equation}
where, $s$ is the macroscopic shift and is determined by the corresponding gate’s eigenvalue, which is parameterized by $\theta_i$.

\subsection{Ansatz Trainability and Expressiblity} \label{sec:anstaz_train_express}

The ansatz can be thought of as a PQC consisting of single-qubit parameterized gates. 
It may or may not contain multi-qubit (entangling gates) depending on the target problem. 
These parameterized gates are dependent on adjustable parameters. 
In the context of HQNNs (or QML in general), these parameters are trained in data-driven tasks, which is analogous to the case of classical NNs. 
The network is said to be trainable as long as the optimization algorithm is able to minimize the loss (as per the defined cost function) in every training iteration. 
The trainability of the network is then compromised when the gradients of parameters are not accessible for further optimization, and the optimization algorithm can not reach the optimal solution. 

PQC can be considered expressive, if it can be exploited to uniformly explore the unitary group $\mathcal{U}(d)$. Thus, the expressiblity of PQC can be defined in terms of the following super-operator \cite{Holmes:2022}.
\begin{equation}
    \mathcal{A_\mathbb{U}}^{(t)}(.) := \int_{\mathcal{U}(d)} d\mu(V)V^{\otimes t}(.)(V^\dagger)^{\otimes t} - \int_{\mathbb{U}} dU U^{\otimes t}(.) (U^\dagger)^{\otimes t}
\end{equation}

\noindent where $d\mu(V)$ is the volume element of Haar measure and $dU$ is the volume element corresponding to to uniform distribution over $\mathbb{U}$. 
If $\mathcal{A_\mathbb{U}}^{(t)}(X)=0$ for all operators $X$ then averaging over elements of $\mathbb{U}$ agrees with averaging over Haar distribution up to the $t$-th moment. 
In this case, $\mathbb{U}$ forms a \textit{t-design}.

\section{Related Work} \label{sec:related_work}

In this section, we provide details of recent state-of-the-art solutions (mainly inspired by the BP problem in classical NNs) to potentially overcome the issue of BP in QNNs aiming to enhance their trainability. 
Additionally, some potential sources of BP in QNNs training landscapes are also discussed, where the BP problem is analyzed from the aspect of underlying ansatz.


The BP phenomenon in QNNs was first studied in \cite{McClean:2018}, where the random PQCs were initialized, and then the variance of partial derivatives was calculated i.e., $var[\partial C]=(\partial C^2 )-(\partial C)^2$. 
The authors then show that for deep circuits of order $poly(n)$, the variance exponentially vanishes with the number of qubits i.e., $var[\partial C]=\frac{1}{2^n}$, due to the fact that the circuit forms 2-design (sample all the unitaries in Hilbert space). 

Unlike the classical case, where a gradient-based backpropagation algorithm solves this trainability problem, its quantum analogous solution is challenging to implement \cite{Skolik:2021}. 
In HQNNs, the PQC is run on a quantum device, whereas its optimization is performed classically. The composition of two fundamentally different computation approaches makes it challenging to implement the backpropagation algorithms in QNNs.
Recently, the problem of BP in QNNs has caught significant attention. While some proposed solutions to potentially overcome the BP problem, others analyze the fundamental structure of QNNs to determine the parameters which give rise BP problem. In the following sections, we present some recent state-of-the-art focusing on tackling and analyzing the issue of BP in QNNs. 
%

\subsection{Potential Solutions of BP} \label{BP_Solutions}
Several solutions were recently proposed to potentially overcome the BP problem in QNNs. In \cite{grant:2018}, a small portion of the quantum circuit is initialized randomly while the remaining parameters are carefully chosen to implement the identity operation as a whole. This approach avoids the initialization on a plateau only for the first training step, and the learnability is still affected during the subsequent training iterations. 
Similarly, in \cite{Volkoff:2021}, a strategy to avoid BP was introduced by enforcing the assignment of multiple parameters in the circuit to reduce the total number of parameters subject to training. However, this approach limits the optimization process to a particular set of parameters and eventually increases the circuit depth for convergence. 

Inspired by the layer-wise training of classical NNs, which were shown to potentially prevent the BP caused by random initialization in classical NNs \cite{Bengio:2006}, a similar approach has also recently been used in QNNs \cite{Skolik:2021}. The layer-wise training approach in QNNs focuses on training a small subset of parameters in each training iteration by incrementally increasing the circuit depth, which results in larger gradients magnitude because of a smaller number of parameters as compared to training the complete circuit. 

A novel approach for mitigating BP in QNN is recently proposed in \cite{kashif:2023resqnets}, where the residual approach from classical NNs is exploited in QNNs. The study demonstrates that incorporating the residual approach in QNNs leads to a significant enhancement in their training performance.

Since BP is fundamentally the problem of vanishing gradients, it may seem that gradient-free optimization approaches can help overcome this issue. However, it has recently been proved that even gradient-free optimization cannot escape the BP issue in QNNs \cite{Arrasmith:2021}. In fact, it was shown that the cost function differences (deciding factor to make optimzation decisions in gradient-free approaches) are exponentially suppressed in BP.


\subsection{Analysis of BP} \label{BP_analysis_literature}
BP is a problem corresponding to the cost function landscapes, where the partial derivatives of parameters become exponentially flat with system size. As mentioned earlier, the initial study discusses the occurrence of BP for \emph{deep ansatz}, which usually is believed to be more expressible.

A recent study investigating the existence of BP for shallow circuits in comparison with deep circuits is in \cite{Cerezo:2021aa}. The authors show that by making BP dependent on the cost function, it can be extended to shallow circuits. To this end they study two cost functions namely: local cost function (measuring single qubit in a multi-qubit systems) and global cost function (measuring all qubits in a multi-qubit systems). The authors then conclude that for the global cost function, the QNNs will experience BP irrespective of underlying PQC's depth (($\mathcal{O}(1), \mathcal{O}(log(n)), \mathcal{O}(poly(log(n))), \mathcal{O}(poly(n))$)). On the other hand, in the case of local cost function, the gradients vanish at worst polynomially and are therefore trainable up to a depth of order ($\mathcal{O}(log(n))$). The BP starts appearing for depth of order ($\mathcal{O}(poly(n))$) for the local cost function and in between these regions, there is a transition region where gradients decay from polynomial to exponential. 
In a recent study conducted by \cite{Kashif:2023}, a follow-up investigation was conducted on the globality and locality of cost functions in real-world applications. The study posits that, in the context of multi-class classification, the use of a global cost function results in significantly improved performance as compared to local cost functions. However, for binary classification, both global and local cost functions demonstrate similar levels of effectiveness. These findings suggest that the choice of cost function must be carefully considered when designing classification models for multi-class scenarios, whereas in the case of binary classification, either approach may be equally viable.

Recent work \cite{Holmes:2022} analyzed the ansatz expressibility to gradient magnitudes and showed that the more expressible the ansatz is, the more likely it is to have a BP. On the other hand, ansatz with relatively lower expressibility would have a delayed BP up to a certain depth, but in principle, expressible ansatz is still favorable because they might provide solutions for multiple problems as compared to less expressible ansatz, which would be problem specific. Therefore, the greater expressibility in deep ansatzes makes them more susceptible to experiencing BP during training. 

Excess entanglement in the hidden layers of QNNs can also cause BP and hinder the learning process \cite{Marrero:2020}. Exploiting the volume law from quantum thermodynamics in \cite{Marrero:2020}, the authors show the existence of BP in the cost function landscape for both gradient-based and gradient-free optimization approaches. This observation was made for both feedforward QNNs and quantum Boltzmann machines. Moreover, the way of encoding data into the QNN, can also give rise to BP \cite{Abbas:2021}.

%

%
%
In this paper, we do not intend to propose a solution for BP problem. Instead, we focus on the analysis of various components of HQNNs inline with section \ref{BP_analysis_literature}.

\section{Methodology} \label{sec:methodology}
This section presents the detailed methodology to obtain the relevant results for the proposed analysis. We start with providing the details of dataset used for training the HQNNs and how is it preprocessed to cope up with the limitations of NISQ devices.
Afterwards, the details of QNNs construction right from qubit initialization to final measurement are presented.
Finally, the discussion on methodology is concluded by providing the details on the classical postprocessing of QNN results.

\begin{figure}[h]
\centering
\includegraphics[scale= 0.4]{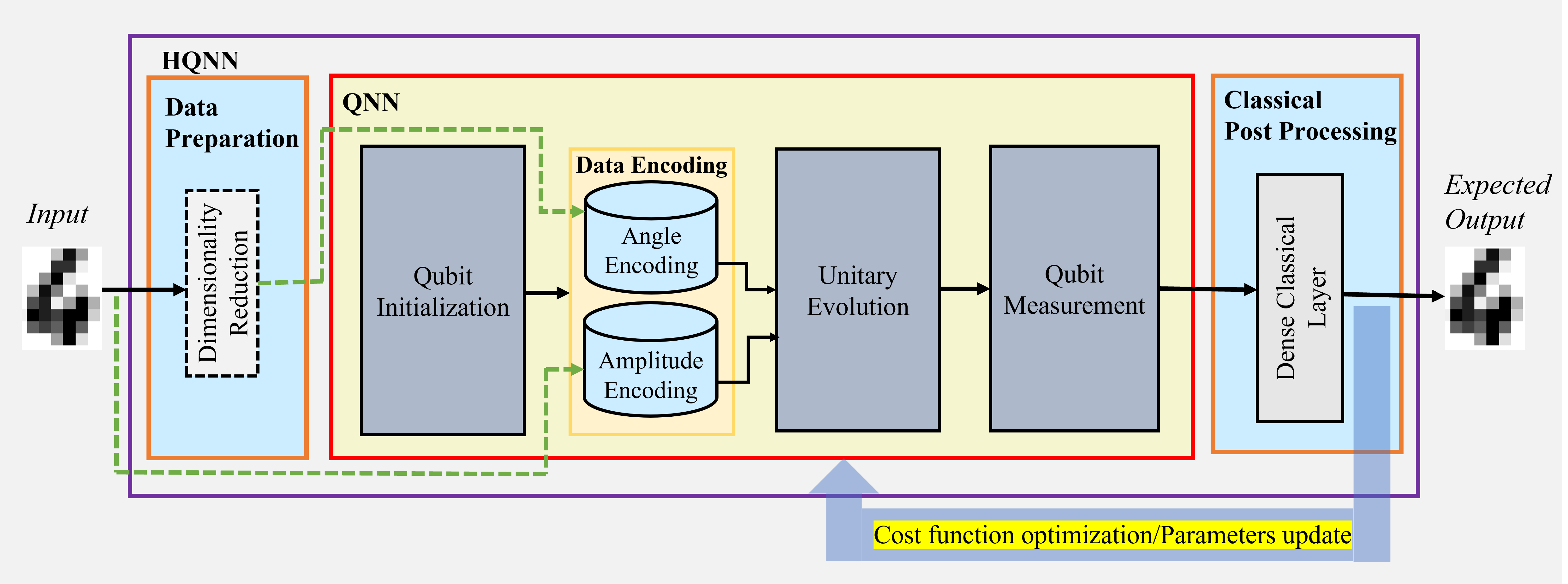}
	\caption{Proposed Methodology for the unified analysis of HQNN}
	\label{fig:HQNN_general_architecture}
\end{figure}

The typical architecture of HQNN for the unified analysis of joint effect of Data Encoding, Ansatz Expressibility, and Entanglement on the Trainability, used in this paper, is depicted in Figure \ref{fig:HQNN_general_architecture}. 
The HQNN architecture completely replicates QNN's architecture from Figure \ref{fig:QNN}. 
We call the architecture (used in this paper) hybrid because of two reasons; 1) the optimization is classically performed, which is also the case in general standalone QNNs (in NISQ regime). 
Secondly, we add a classical neuron layer at the end of the quantum layer(s). The advantage of the classical layer is that it allows the application of familiar non-linear activation functions from traditional ML. Furthermore, it enables to experiment on real-world datasets. 
The HQNN used in this paper have three primary ingredients namely; data preparation, QNN construction, and classical post-processing. Here, we present the typical workflow of HQNN used in this paper. In particular we present the details of every step of HQNN, from input to the output.

\subsection{Data Preparation} \label{sec:data_prep}
The first step in the workflow of HQNN used in this paper is to prepare the data such that it can be efficiently encoded into quantum states for further processing. 
The data preparation here is in context of the limitations of NISQ devices (mainly in number of qubits). 
Depending upon the size of input data and choice of encoding scheme, it is sometimes required to reduce the input features dimmension. 
In such a case, the input (image(s) in our case) is first passed to the dimensionality reduction algorithm before encoding.
We also use dimensionality reduction for one of the encoding schemes to cope up with the restrictions of NISQ era, details of which are presented in the following section, where we explain data encoding used in this paper as part of QNN construction.

\subsection{QNN Construction} \label{sec:QNN_construction}
Once the data is prepared, the next important step in HQNN workflow is the construction of QNN, which again has four ingredients namely; qubit initialization, data encoding, unitary evolution and qubit measurement .


\subsubsection{Qubit Initialization} 
It is the first step of QNN construction, which typically defines the total number of qubits or in other words, the width of quantum layers. 
A qubit can be initialized in any random state. However, we initialize all the qubits in the default ground state i.e., $\ket{\psi}^{\otimes n}=\ket{0}^{\otimes n}$, which is a relatively common practice. 


\subsubsection{Data Encoding}
Once the qubits are defined, the next step is the data encoding (classical to quantum feature mapping), such that the subsequent quantum layer(s) \footnote{we use the terms quantum layers, PQC, and ansatz interchangeably throughout this paper} can process the input. 
We use two frequently used data encoding strategies in QNNs, i.e., amplitude and angle encoding.
As discussed above the limited number of qubits in NISQ devices enforces to reduce the input feature dimension, which depends on the input feature size and the way data is being encoded in to quantum states. 
The input feature dimension for the dataset we have used is $64$ (details in section \ref{sec:exp_details}). 
In context of the data encoding used in this paper, we reduce the input feature dimensions every time the angle encoding is used because angle encoding needs $n$ qubits to encode $n$ features (see section \ref{sec:data_enc}). 
For amplitude encoding, the input features dimensionality reduction is not performed because the amplitude encoding can encode $2^n$ features in $n$ qubits (see section \ref{sec:data_enc}). 

\subsubsection{Unitary Evolution}
\begin{figure}[htp]
\centering
\includegraphics[scale= 0.5]{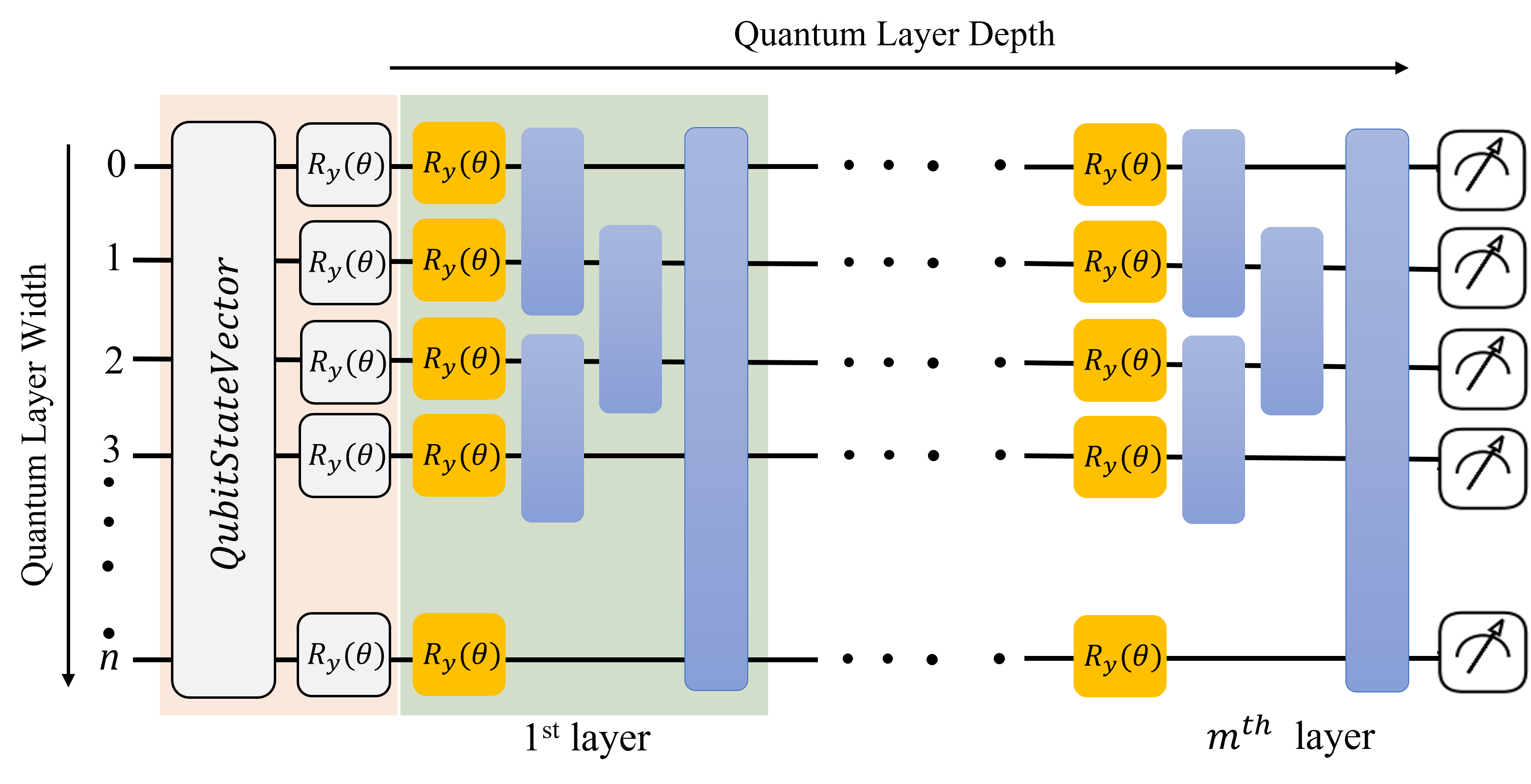}
	\caption{Quantum layer ansatz structure with entanglement. The light pink highlighted region is the data encoding part and either the QubitStateVector or $R_y(\theta)$ rotation gates are used depending on the encoding technique used. The former is used in case of amplitude encoding and the later in case of angle encoding. The green shaded area is the actual quantum ansatz used in training. The parameterized rotation unitaries in yellow boxes are trainable quantum parameters and the vertical blue bars represents two-qubit unitaries}
	\label{fig:ansatz1}
\end{figure}

The unitary evolution primarily consists of single-qubit parameterized (parameter dependent) gates and multi-qubit gates (for qubit entanglement). 
There is no hard-and-fast rule to insert a combination of gates for unitary evolution and different gate combinations are heuristically used for a given problem. 
We use two similar ansatz structures (entangled and unentangled ansatz) for unitary evolution differing only in entanglement inclusion/removal for unitary evolution. The complete details, from data encoding to unitary evolution to measurement for the quantum layers we have used in this paper, are presented below:
\paragraph{Entangled Ansatz.}
The entangled ansatz structure consists of a single-qubit parameterized unitaries ($R_y(\theta)$) and two-qubit unitaries (CNOT) for qubit entanglement. Taking into account the limitations of NISQ devices, in this paper we only consider the nearest-neighbor entanglement, where the last qubit in the underlying system is considered neighbor to the first qubit, as shown in Figure \ref{fig:ansatz1}. 

\begin{figure}[H]
\centering
\includegraphics[scale= 0.6]{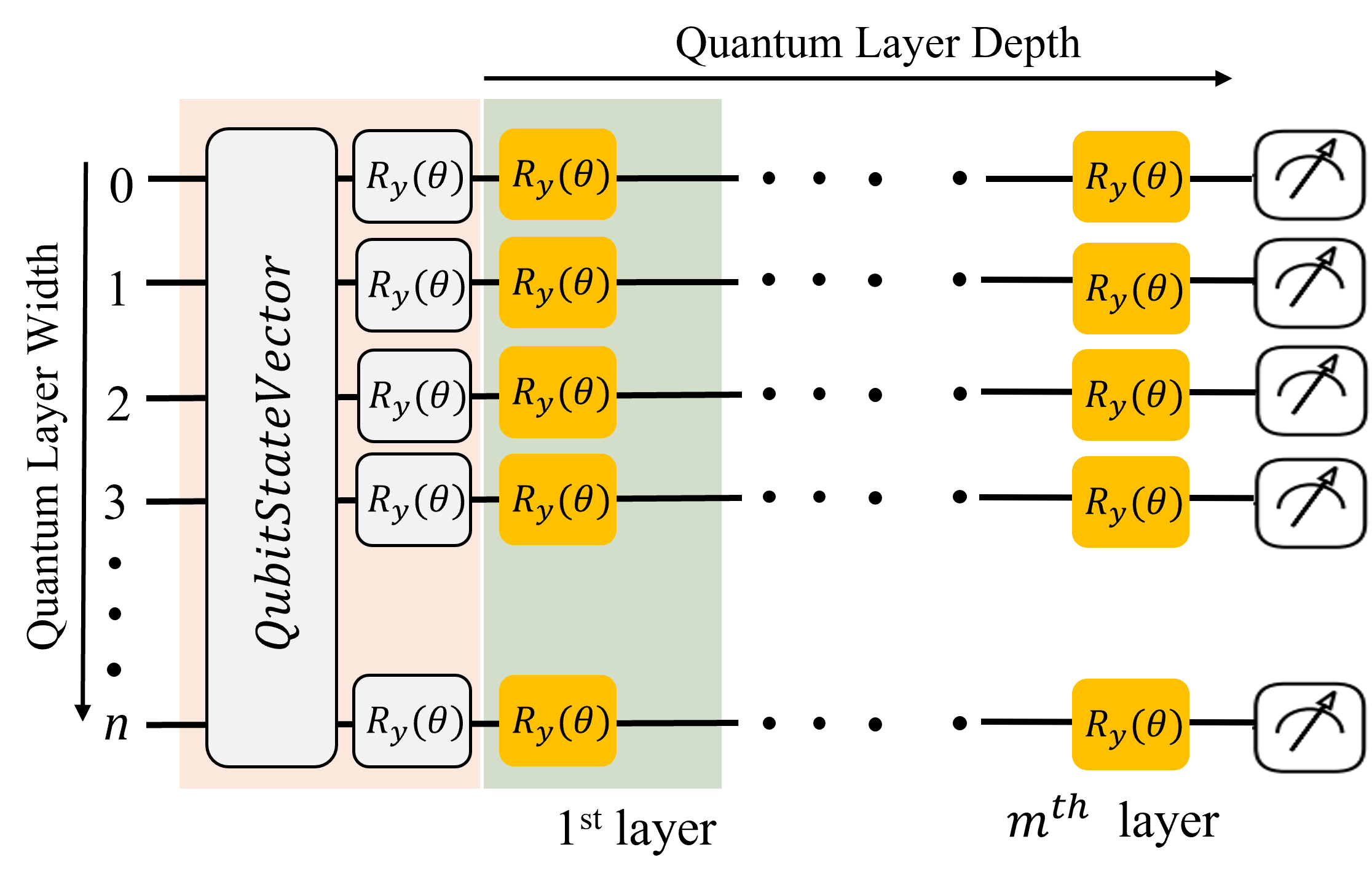}
	\caption{Quantum layer structure without entanglement. The light pink and gray highlighted regions represents the same as in Figure \ref{fig:ansatz1}}
	\label{fig:ansatz3}
\end{figure}

\paragraph{Unentangled Ansatz.}
Entanglement is an important property in quantum mechanics with an anticipated potential to enhance various applications including quantum machine learning. We now update our ansatz structure containing only the single-qubit parameterized unitaries with no entanglement as shown in Figure \ref{fig:ansatz3}. 
The motivation of entanglement exclusion comes from the fact that in QNNs trainable parameters are only the single-qubit unitaries, which are optimized during the training. 
 
We attempt to answer the question: does the unentangled ansatz can potentially delay the issue of BP to enhance the trainability of HQNNs because if it is also identified as a potential source of BP? 
We show that whether or not entanglement is significant in HQNNs and is dependent on the type of data encoding. For angle encoding, the entanglement does not aid towards better performance, in fact without any entanglement the overall performance improves. On the other hand, for amplitude encoding, the result shows the opposite case.

\subsubsection{Qubit Measurement}
For qubit measurements, different measurement bases can be used; however, the most frequently used are the default computational basis, i.e., the eigenbasis of $\sigma^z$. We also use the computational basis to get the output of QNN. 

\subsection{Classical Post Processing} \label{sec:classical_postprocess}
The last step in HQNN workflow is classical post-processing of QNN because the result of qubit measurement is a classical value rather than a quantum state. 
To post-process the QNN results, we use a classical dense neurons layer (every neuron is connected to every qubit's measurement output from PQC). 
The cost function is then defined based on which the optimizer tends to find the solution for the given problem. 
The optimizer updates the trainable parameters in every training iteration and retrains the PQC on updated parameters. The process is repeated until an optimal solution is found.

In this paper, the classical specifications (input, neuron layer at the end with non linear function, and classical optimization) of HQNN are fixed, and the primary focus is to experiment with enclosed QNNs (quantum layers).
Until here, all the ingredients to construct the HQNN (used in this paper) are presented. 
The following section present the experiment details including the classical components used in HQNN.


\section{Experimental Setup} \label{sec:exp_details}

The architectural details of HQNN used in this paper are presented in previous section (section \ref{sec:methodology}).
This section  provides the experimentation details about how that proposed HQNN architecture is trained and evaluated. 
In particular, the details about the dataset used for training, 
details of classical hyperparameters of HQNN 
and list of experiments 
performed to obtain the results for proposed analysis is also presented. 


\subsection{Data Preprocessing} \label{sec:data_Preprocess}
We consider a multi-class (image) classification problem. The sklearn digit dataset is used for training and evaluating the HQNN, which contains images of handwritten digits \cite{sklearn}. The reason of using this particular dataset over other similar yet popular datasets such as MNIST, is the smaller input feature size, which is more suitable in NISQ era.
This dataset has a total $10$ classes with $180$ samples per class, resulting in a total of approximately $\cong1797$ samples. Furthermore, each data point is an image of $8 \times 8$ resulting in feature dimension of $64$. 
Moreover, 75\% of the samples are used for training and the remaining 25\% for testing the HQNN. 
The data is encoded using both amplitude and angle encoding separately for each experiment. 
As discussed in section \ref{sec:methodology}, we reduce the input feature dimension before passing it to PQC, in case of angle encoding, where input features must be equal to that of total number of qubits. 
For input dimensionality reduction, we use principal component analysis (\emph{PCA}) from sklearn library (a popular ML library for the Python programming language), which reduces large datasets to smaller ones without losing the most important information in the dataset. 


\subsection{Hyperparameters Specifications } \label{sec:params_specs}
Given the nature of target problem (multi-class classification), we use \emph{categorical cross entropy} as the cost function.
For classical optimization of the cost function, we use \emph{Adam Optimizer} \cite{Kingma:2014}, with an initial learning rate of $0.01$. 
Since the dataset we used consists of a total $10$ classes, therefore the last classical neuron layer in HQNN has a total of 10 neurons. Furthermore, the non-linear activation function used is \emph{SoftMax}, which is general case in multi-class classification problems. .

The models used in this paper are set to train for the maximum of $100$ training iterations (epochs). However, to avoid overfitting, we schedule the learning rate as the training progresses using the \emph{early stopping} method from keras library (an open-source software library that provides a Python interface for artificial neural networks). The early stopping method monitors the validation loss for three consecutive iterations and if there is no improvement, the learning rate is reduced by a factor of $0.1$,  setting the new learning rate as: $new learning rate = previous learning rate \times 0.1$. If the validation loss is not improved for four consecutive training iterations, the early stopping method forcefully stops the training to avid overfitting. The input training data for all the experiments is passed in the batches of $16$. Finally, we use pennylane (a cross-platform Python library for differentiable programming of quantum computers) \cite{Bergholm:2018} for training the HQNNs.  


\subsection{List of Experiments} \label{sec:exp_list}
\begin{figure}[H]
\centering
\includegraphics[scale= 0.4]{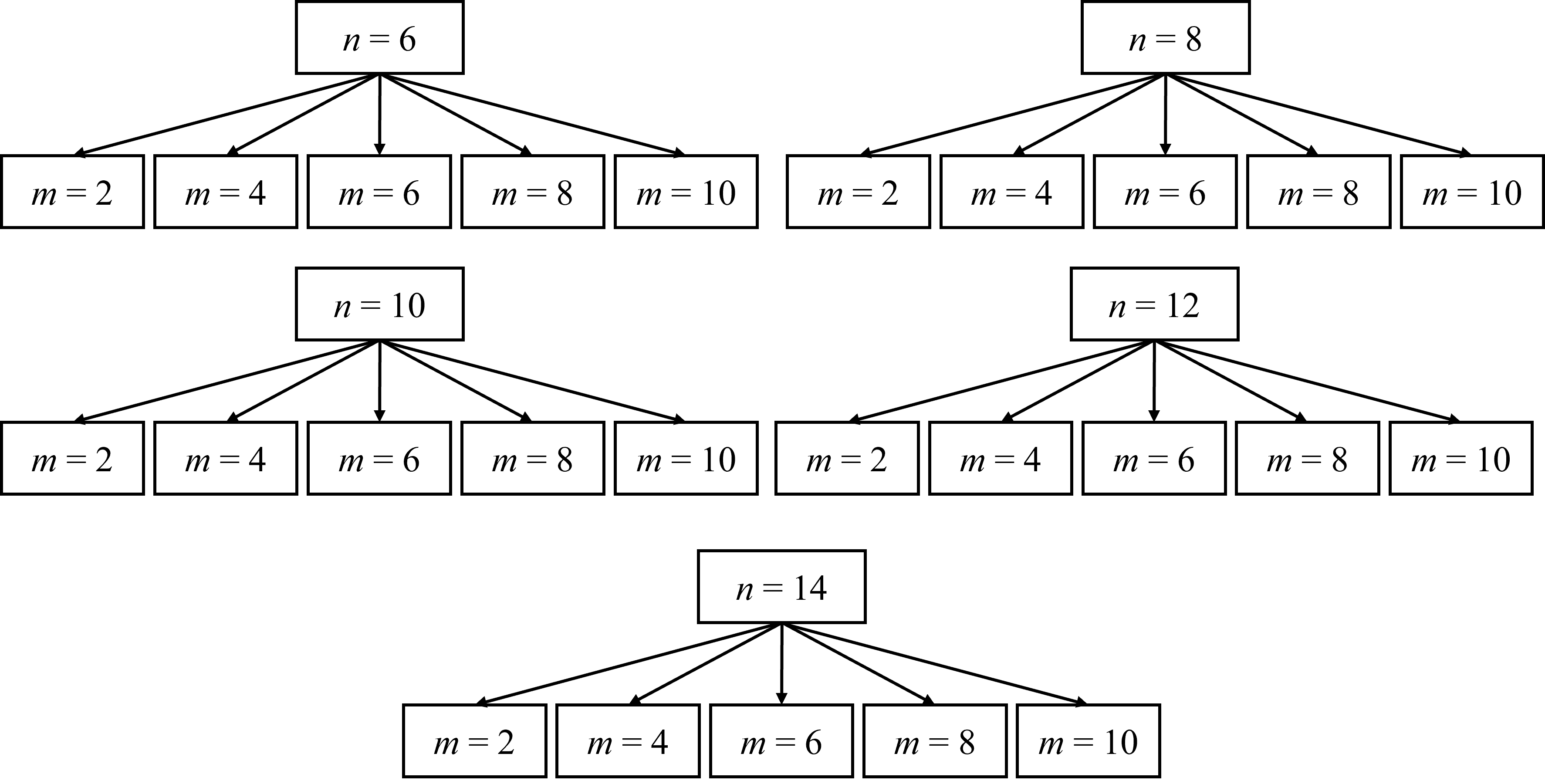}
	\caption{List of experiments performed in this work for different $n$ and $m$.  }
	\label{fig:experiments}
\end{figure}

The list of training experiments for our proposed analysis is shown in Figure \ref{fig:experiments}. Since the structure and size of the quantum layers are the primary focus of our analysis, we experiment with different widths ($n$) and depths ($m$) of quantum layers. $n$ denotes the number of qubits, and $m$ denotes the periodic repetition of quantum layers before the measurement.
We restricted to use the maximum width of $n=14$ and the maximum depth of $m=10$ because of the two reason: 1) the overall accuracy is starting to decline for bigger $n$ and $m$ which we speculate would further decline by increasing $n$ and $m$ because of the so-called phenomenon of BP, and 2) Since we use classical machines to simulate qubit systems, and even for a simple dataset (considered in this paper), it takes around $70-80$ hours to train for the maximum width ($n=14$) with maximum depth ($m-10$).

Stemming from the fact that BP exists in standalone QNNs for sufficiently random and expressible PQCs \cite{McClean:2018}, we first empirically analyze the existence of BP in HQNNs. We typically analyze whether or not, the addition of classical neurons layer can mitigate BP to any extent.
For that purpose, we train the HQNN for different widths ($n$) and depths ($m$) of the underlying PQC, 
Bigger values of $n$ and $m$ results in greater number of trainable parameters and generally considered as more expressible ansatz and vice versa. 
Changing the $n$ and $m$ of PQC helps to analyze the existence of BP in HQNNs. The issue of BP is analyzed for both the encodings separately with both ansatz structures (entangled and unentangled). This provides a general idea of which encoding strategy works well with which ansatz.
Upon observing the BP dependence on ansatz depth (expressibility) we then perform trainability vs. expressibility analysis, again for both encodings individually for each ansatz structure.   
This analysis provided an overall idea about how the ansatz expressibility is playing the role in the occurrence of BP and how deep an ansatz can be for a given width before the performance starts declining.
Further, it can also help to identify if there is any relationship between $n$ and $m$ to achieve a relatively better overall performance.
We then compare the performance of both the anstaz structures for both the encoding to understand the role of entanglement in HQNNs.
We conclude our analysis by providing a brief application mapping of HQNNs with other important evaluation metrics for classification tasks, i.e., precision, recall and F1-score.
  
\section{Results and Discussion} \label{sec:results}
In this section, we present our experimental analysis. 
We start by demonstrating the existence of BP in HQNNs
for both the ansatz structures individually with both the encodings used in this paper. 
Based on our analysis, we concur that BP depends on expressibility of quantum layers in HQNNs. We then perform the the trainability vs. expressibility analysis of HQNNs 
i.e., how the quantum layer(s) depth and width are related to BP and possibly effect the network's overall performance. 
The trainability vs. expressibility analysis reveals that the entanglement in underlying ansatz also plays a role in the overall performance of HQNNs. We then briefly compare both ansatz structures with both encodings used in this paper, to understand the role of entanglement.
Finally, to diversify our analysis, we evaluate the HQNNs in terms of other application-oriented evaluation metrics for classification tasks.

\subsection{Demonstration of BP Existence in HQNNs} \label{sec:BP_results}
In this section, we analyze the existence of BP in the training landscapes of HQNNs. 
We benchmark mean accuracy of HQNNs for the analysis of BP. The BP analysis is performed for both the encodings individually for both entangled and unentangled ansatz structures. The existence of BP hinders the learning process of HQNNs eventually resulting in lower accuracy. Consequently, for our analysis the higher accuracy would essentially entail that the model is not yet fully exposed to BP and vice versa.


\subsubsection{BP Demonstration for Entangled Ansatz with Amplitude Encoding} \label{sec:BP_results_entangled_amp}
\begin{figure}[H]
\centering
\subfloat[Mean accuracy as a function of $n$ \label{fig:AmpE_ansatz1a}]{%
\resizebox*{7cm}{!}{\includegraphics{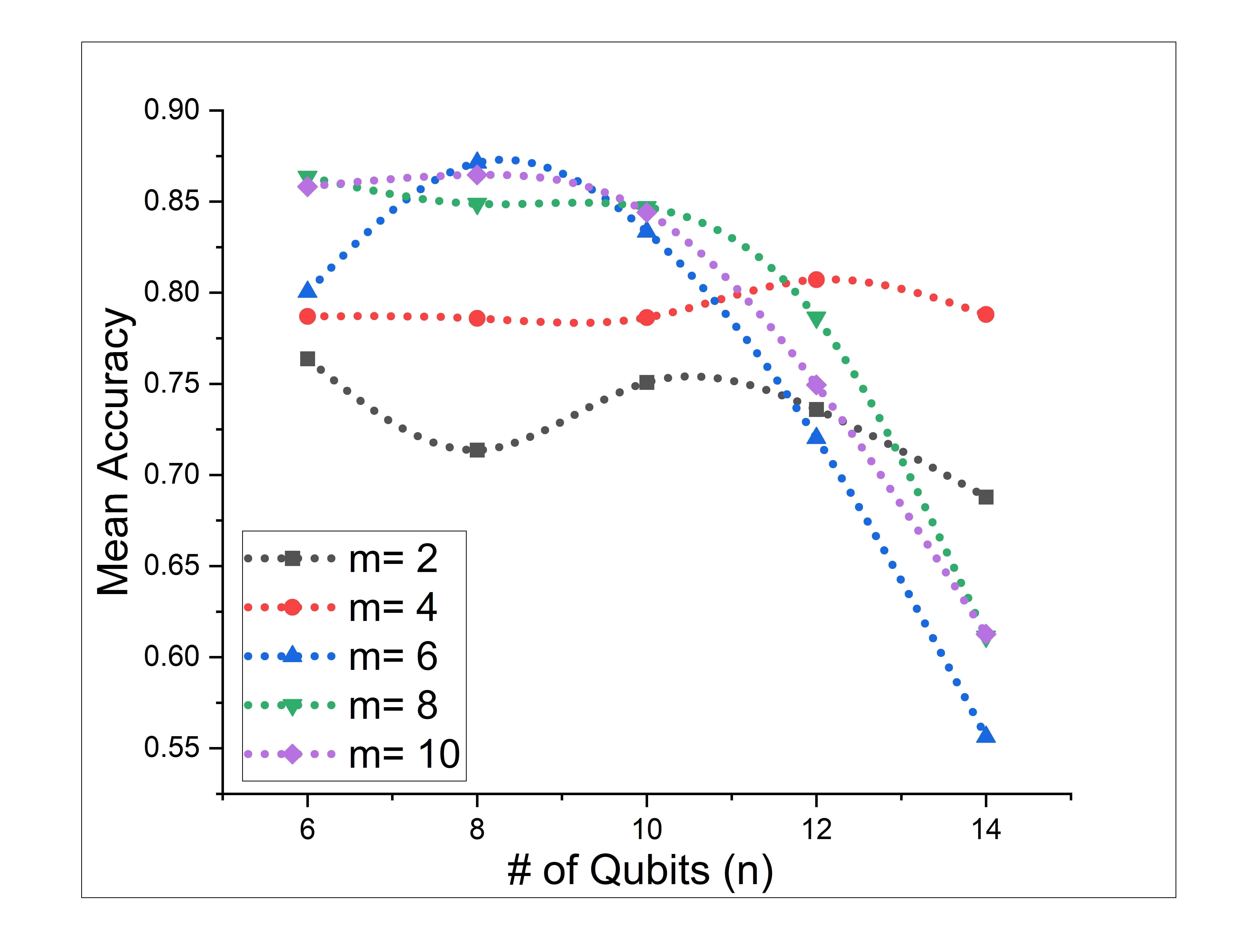}}}\hspace{5pt}
\subfloat[Mean accuracy as a function of $m$ \label{fig:AmpE_ansatz1b}]{%
\resizebox*{7cm}{!}{\includegraphics{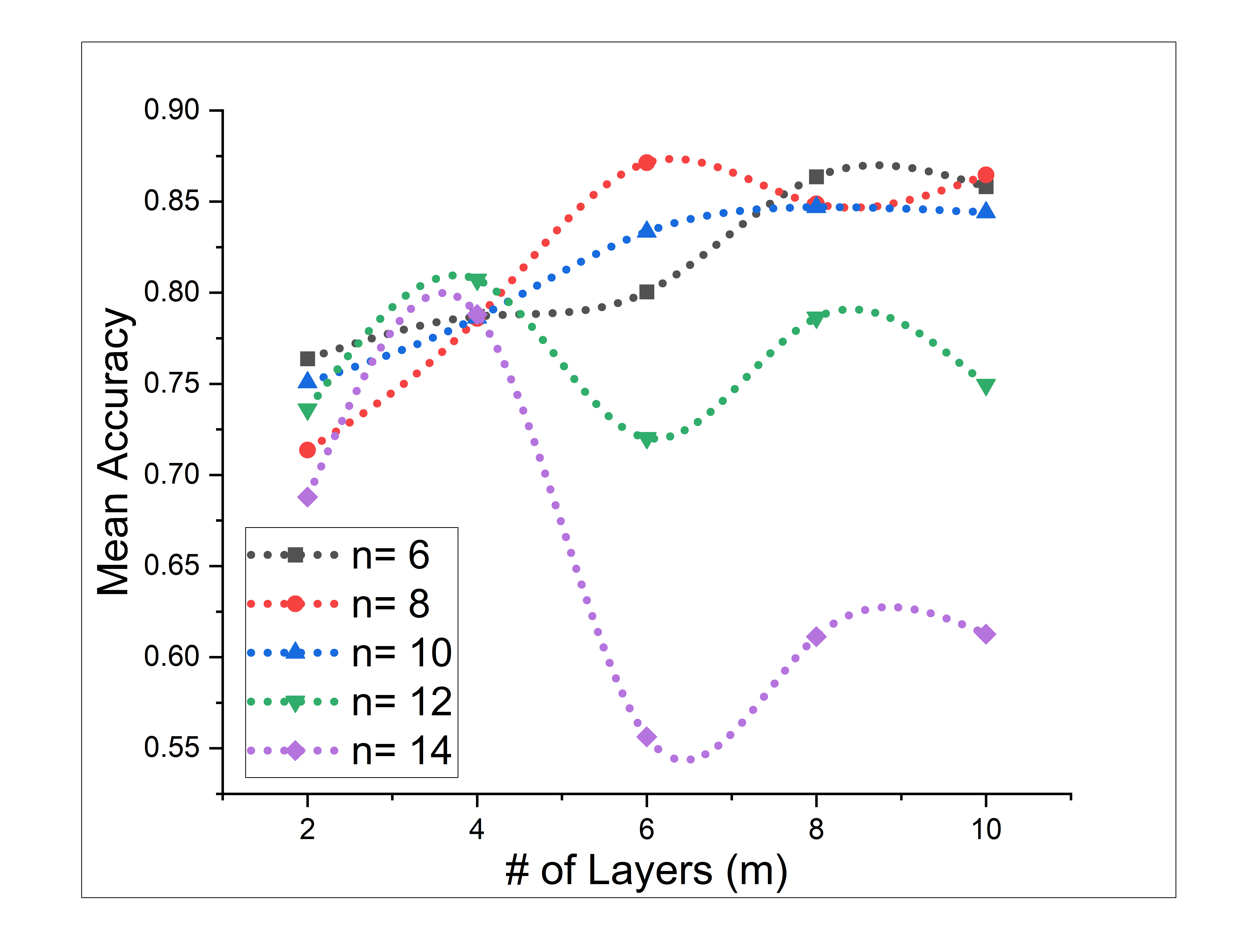}}}
\caption{Entangled Ansatz: Mean accuracy for different $m$ and $n$ with amplitude embedding.} \label{fig:mean_acc_AMPE_WE}
\end{figure}

The entangled ansatz contains single qubit unitaries and nearest neighbor qubit entanglement, as shown in Figure \ref{fig:ansatz1}. The original input feature size to be encoded into quantum states is $64$, as discussed in section \ref{sec:exp_details}. Hence, with amplitude encoding, the minimum number of qubits required to encode these features is $n=6$, which is a reasonable width for quantum layer(s) considering NISQ era. 
Therefore, while using amplitude encoding we do not apply PCA for feature reduction and all the input features are directly encoded. 
We then vary $n$ and $m$ according to Figure \ref{fig:experiments}, to analyze the occurrence of BP. The mean accuracy for all the experiments are shown in Figure \ref{fig:mean_acc_AMPE_WE}.

To analyze the effect of width ($n$) irrespective of depth ($m$), the mean accuracy of different training experiments is individually plotted for fixed $m$ and variable $n$ in Figure \ref{fig:AmpE_ansatz1a}.
The \emph{general} accuracy trend in Figure \ref{fig:AmpE_ansatz1a} is declining. This decline in performance, as the number of qubits ($n$) increase, indicates the possible presence of BP in training landscapes.
On the other hand, to observe the effect of depth ($m$) irrespective of the width ($n$), the mean accuracy is individually plotted for fixed $n$ and variable $m$ as shown in Figure \ref{fig:AmpE_ansatz1b}.
we observe that there is a \emph{trade-off} between quantum layer's depth and width to achieve a relatively better performance. The term \emph{trade-off} here essentially means that for smaller $n$ the \emph{allowable} $m$ is relatively greater and vice versa. 
One possible reason behind the trade-off is the phenomenon of overparameterization, which in HQNNs (used in this paper), is a result of an increase in $n$ and $m$. Overparameterization 
is vital in classical deep learning because of the complexity of modern day applications, and is often helpful in learning most intricate relationships in the input data. 
However, HQNNs are not much in favor of overparameterizing the network, and an optimal number of parameters (optimal $n$ and $m$) must be determined to achieve a relatively better performance.   

\subsubsection{BP Demonstration for Entangled Ansatz with Angle Encoding} \label{sec:BP_results_entangled_angle}
 
\begin{figure}[H]
\centering
\subfloat[Mean accuracy as a function of $n$ \label{fig:AngE_ansatz1a}]{%
\resizebox*{7cm}{!}{\includegraphics{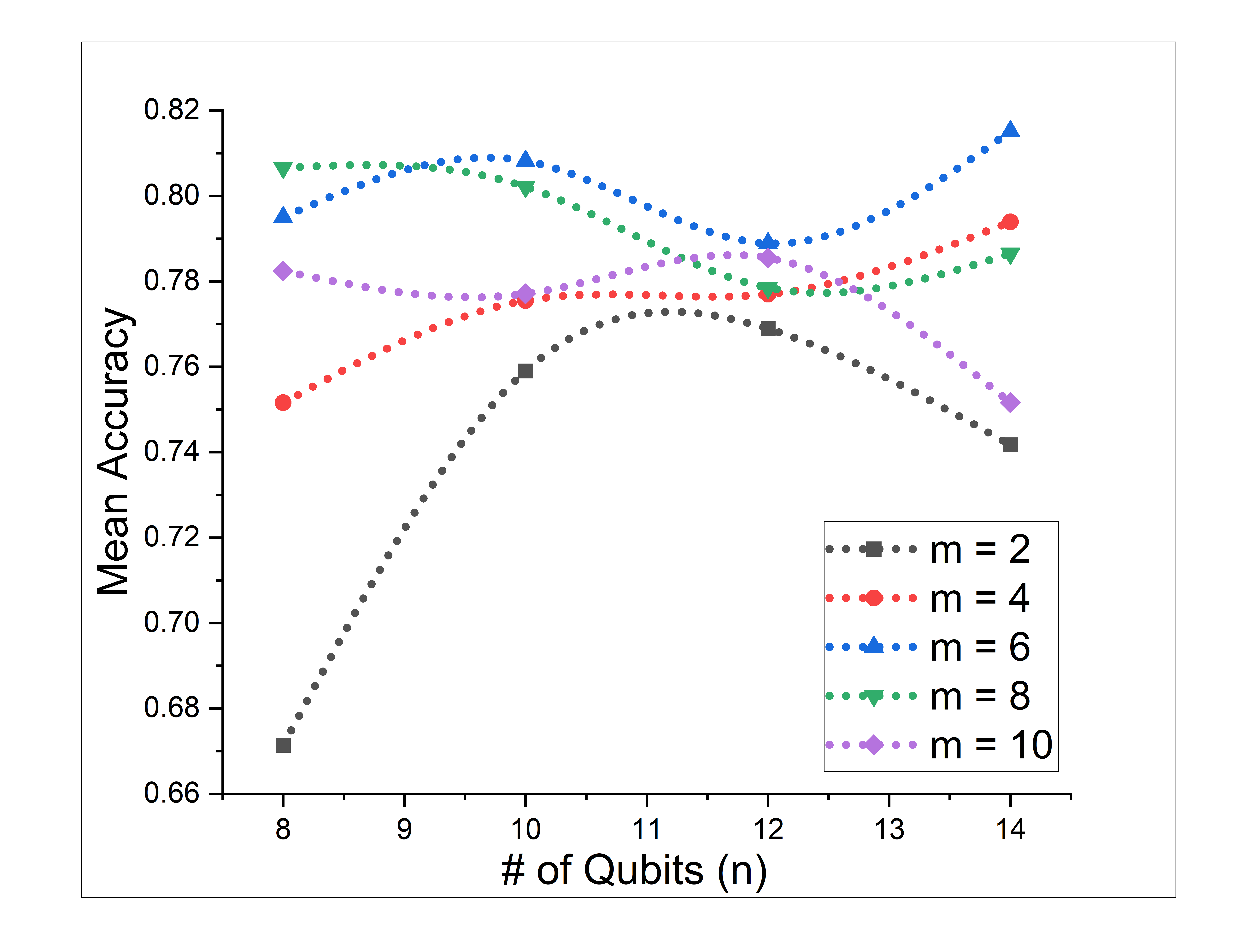}}}\hspace{5pt}
\subfloat[Mean accuracy as a function of $m$ 
 \label{fig:AngE_ansatz1b}]{%
\resizebox*{7cm}{!}{\includegraphics{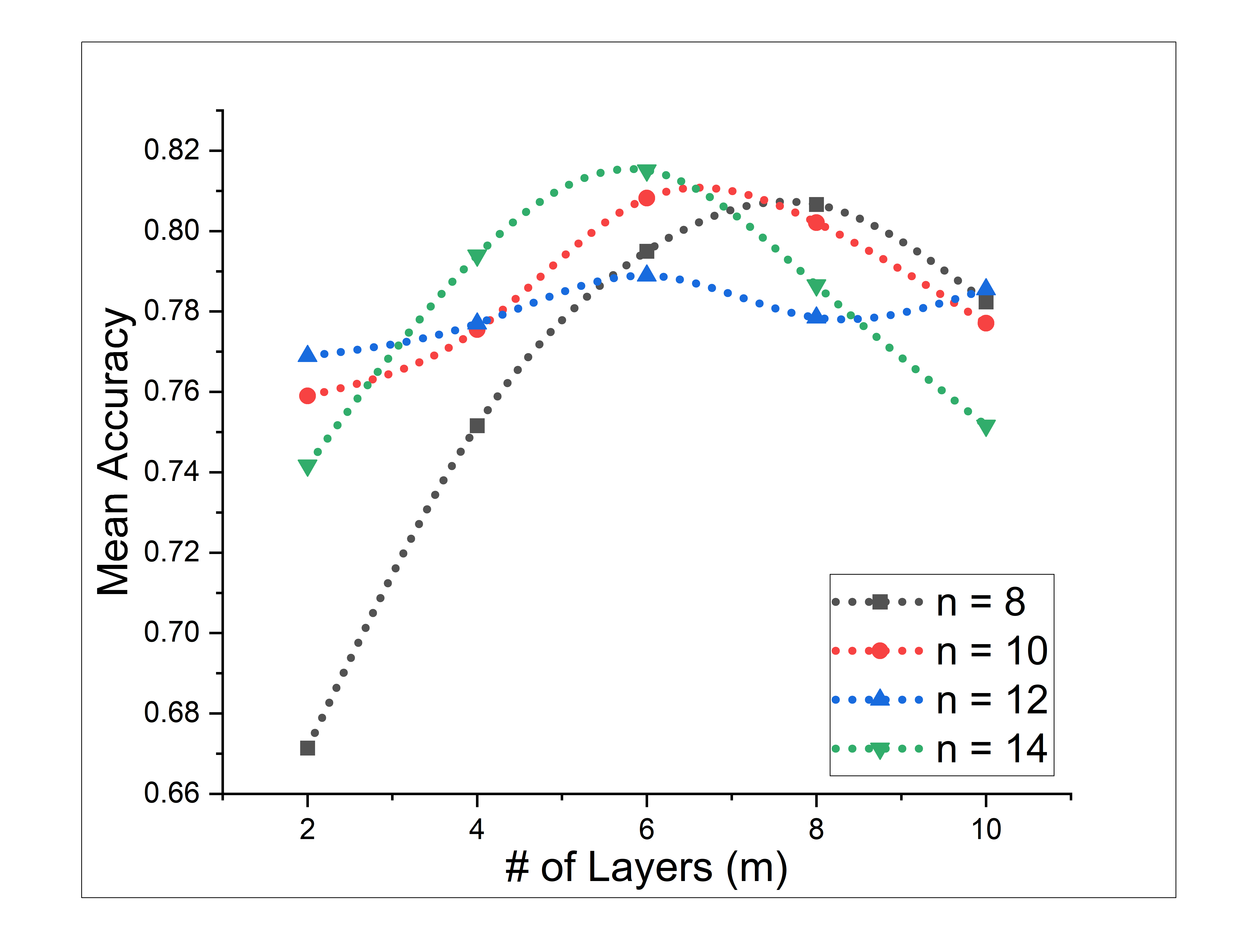}}}
\caption{Entangled Ansatz: Mean accuracy for different $m$ and $n$ with angle encoding. } \label{fig:mean_acc_AngE_WE}
\end{figure}

In angle encoding, the number of qubits or quantum layer's depth ($n$) must be equal to that of input feature dimension for successful  classical-to-quantum feature mapping, as discussed in section \ref{sec:data_enc}. 
Since the input feature size of our data is $64$, we need at least $64$ qubits to directly encode the data into the quantum states. 
Such a big number of qubits is not suitable for NISQ era. 
Therefore, we apply PCA to reduce the input feature dimension to experiment with angle encoding approach. The reduction conforms with the width of quantum layers ($n$) from Figure \ref{fig:experiments}. However, considering both the NISQ limitations and not losing much information while dimensionality reduction, the minimum width in angle encoding is $n=8$. 
Analogous to amplitude encoding, various experiments are performed for different $n$ and $m$. 
The mean accuracy for all the experiments is shown in Figure \ref{fig:mean_acc_AngE_WE}.

In classical ML, it is always conjectured that when we have more input features to train the model, the overall performance is improved, given a fairly complex underlying model. %
In angle encoding, the size of input features is directly proportional to the size of $n$. It typically means that every increase in $n$ results in greater input feature dimensions.
Hence, following traditional ML, the wider quantum layers should yield a better performance, since they are fed with relatively enhanced feature dimensionality compared to quantum layers of shallow width. 
However, this is not the case in HQNNs because an increase in $n$ reduces the accuracy after a certain $m$, as shown in Figure \ref{fig:mean_acc_AngE_WE}. To analyze the effect of quantum layer(s) width, the mean accuracy is plotted as a function of $n$, as shown in Figure \ref{fig:AngE_ansatz1a}. 
It can be observed that as $n$ increases the accuracy tends to reduce, particularly for bigger depths ($m=8$ and $10$). For relatively smaller depths ($m=4,6$), the accuracy improvement is negligible as $n$ increases exhibiting the presence of BP. 
To analyze the effect of quantum layer(s) depth, the mean accuracy is plotted as a function of $m$, as shown in Figure \ref{fig:AngE_ansatz1b}. 
It can be observed that in general, for a certain $n$ the accuracy tends to improve up to certain $m$ and then starts declining. Hence, in case of angle encoding also, there is a \emph{trade-off} between $n$ and $m$ to achieve a better overall performance. 

\subsubsection{BP Demonstration for Unentangled Ansatz with Amplitude Encoding} \label{sec:BP_results_unentangled_amp}

\begin{figure}[H]
\centering
\subfloat[Mean accuracy as a function of $n$]{\label{fig:AmpE_NE_1}%
\resizebox*{7cm}{!}{\includegraphics{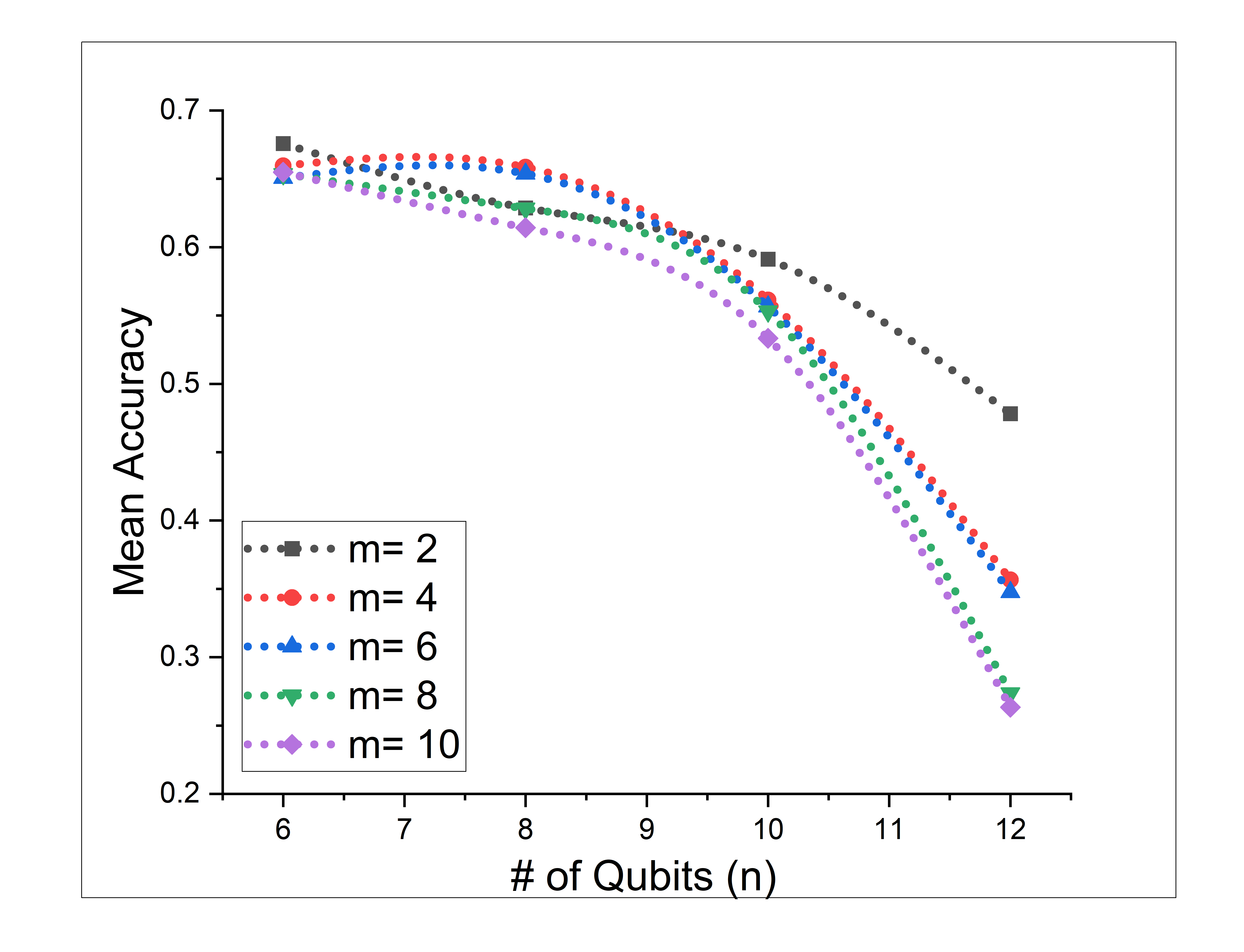}}}\hspace{5pt}
\subfloat[Mean accuracy as a function of $m$]{\label{fig:AmpE_NE_2}%
\resizebox*{7cm}{!}{\includegraphics{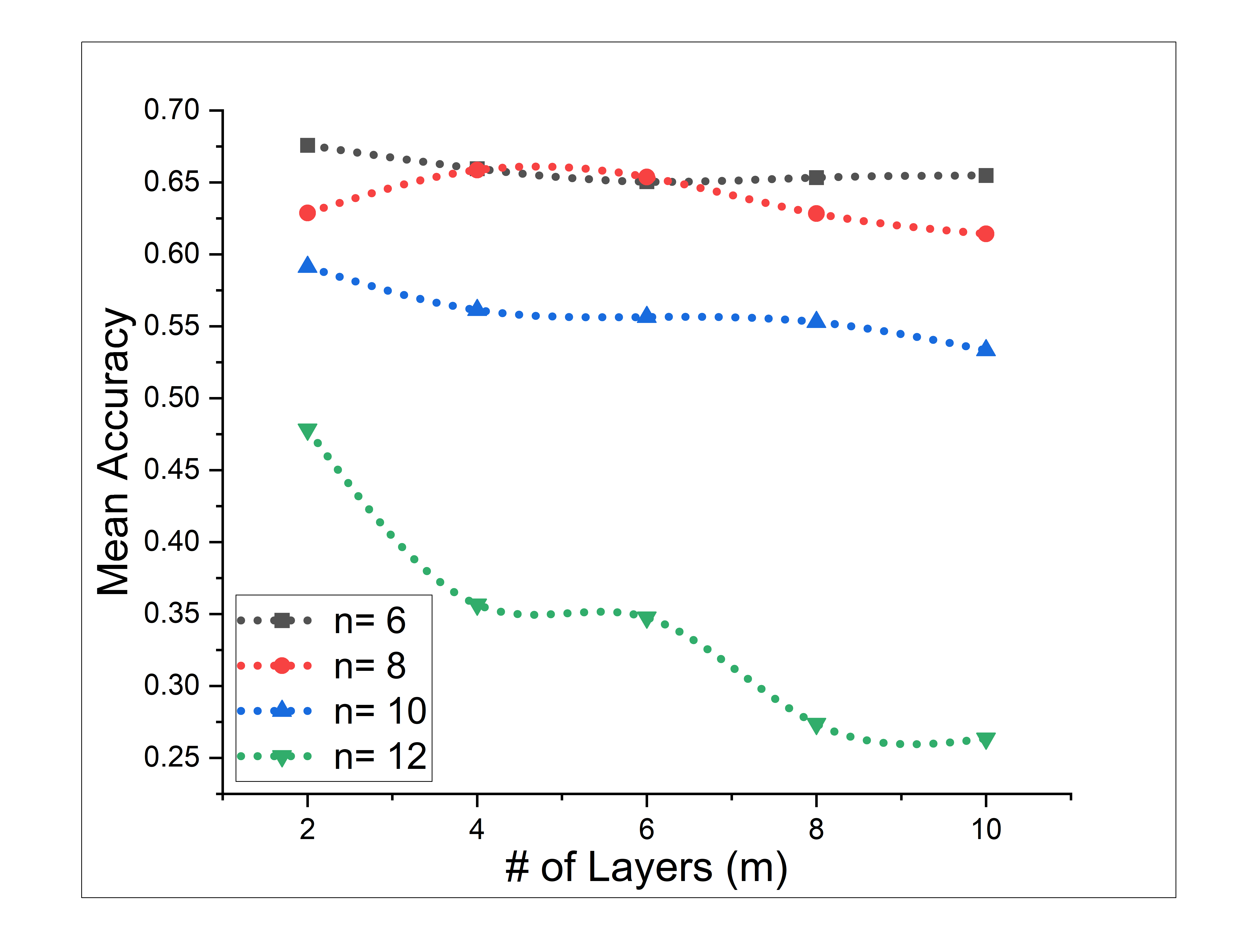}}}
\caption{Unentangled Ansatz: Mean accuracy fordifferent $n$ and $m$ with amplitude encoding.} \label{fig:ampE_mean_acc_NE}
\end{figure}

For amplitude encoding, the input feature size is unchanged i.e., $64$. We start with minimum required width of quantum layers ($n=6$) and perform all the experiments from Figure \ref{fig:experiments}. For the sake experimental simplicity, we skip experimenting for $n=14$ because we already observed the BP going from $n=6$ to $n=12$. The mean accuracy for all the training experiments are shown in Figure \ref{fig:ampE_mean_acc_NE}.

The mean accuracy of all the training experiments is plotted as a function of $n$ in \ref{fig:AmpE_NE_1} to analyze how the increase in number of qubits effect the performance of HQNN when no entanglement is included in quantum layers. 
It can be observed that analogous to the case of entangled ansatz, the accuracy starts declining as $n$ increases irrespective of $m$. The accuracy decline is a clear indication towards the occurrence of BP. 
Furthermore, in unentangled ansatz the accuracy decline is quite evident with increases in $n$ leading to a conclusion that no entanglement in quantum layers makes HQNNs more susceptible to BP.  
Similarly, the mean accuracy of all the training experiments is also plotted as a function of $m$ (Figure \ref{fig:AmpE_NE_2}) which helps in analyzing the role of $m$ in HQNNs for a fixed $n$. 
We observe that unlike the case of entangled ansatz with amplitude encoding, where the allowed circuit depth $m$ varies for different $n$, in case of unentangled ansatz, increase in $m$ results in further reduction (or a negligible improvement) in accuracy for all $n$.
The performance decline with an increase in $m$ is more evident for bigger $n$, leading to a conclusion that in case of unentangled ansatz smaller $n$ with smaller $m$ is more appropriate while constructing quantum layers. Bigger $n$ and $m$ makes unentangled quantum layers more prone to BP.  

\subsubsection{BP Demonstration for Unentangled Ansatz with Angle Encoding} \label{sec:BP_results_unentangled_ang}

\begin{figure}[H]
\centering
\subfloat[Mean accuracy as a function of $n$]{\label{fig:AE_NE_1}%
\resizebox*{7cm}{!}{\includegraphics{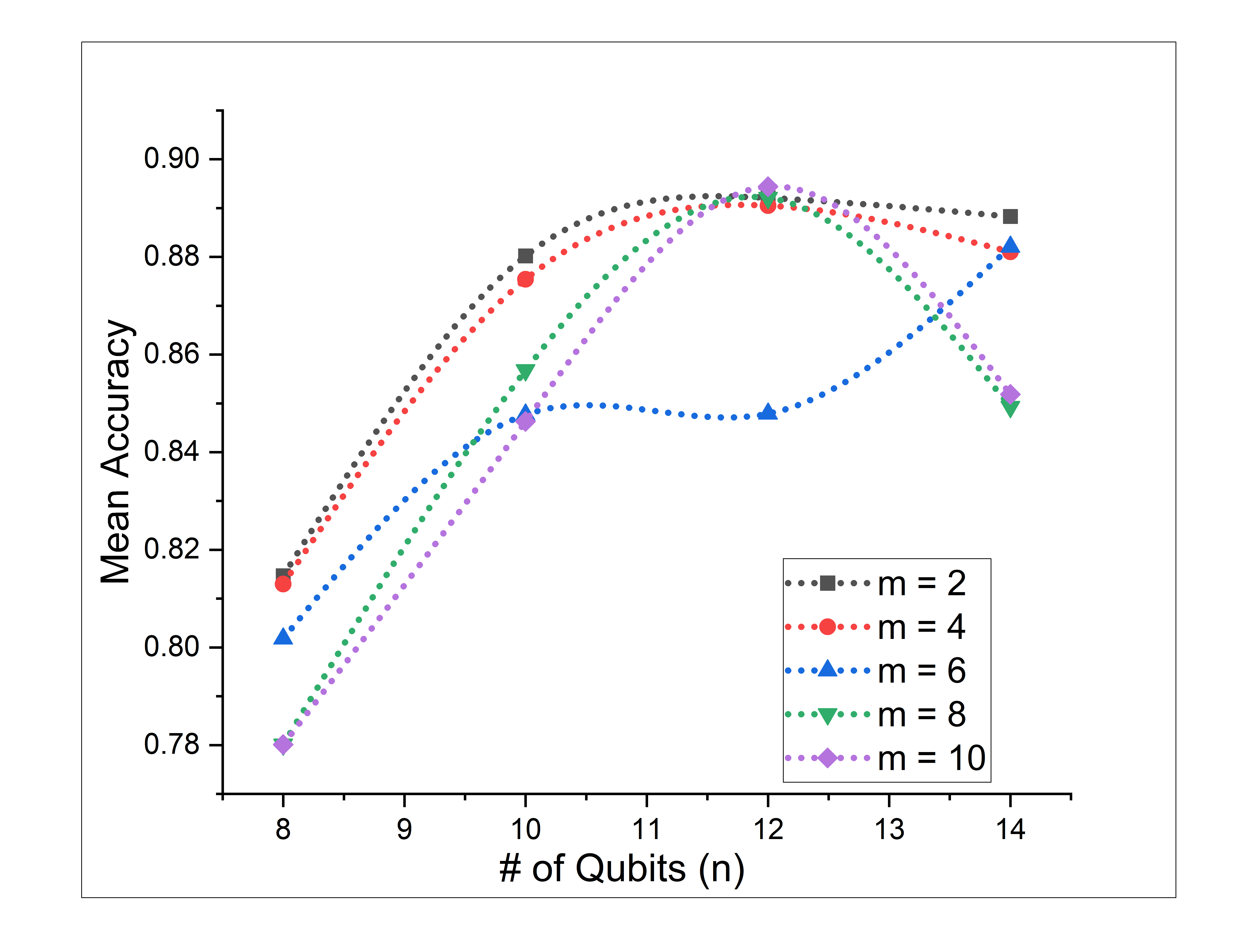}}}\hspace{5pt}
\subfloat[Mean accuracy as a function of $m$]{\label{fig:AE_NE_2}%
\resizebox*{7cm}{!}{\includegraphics{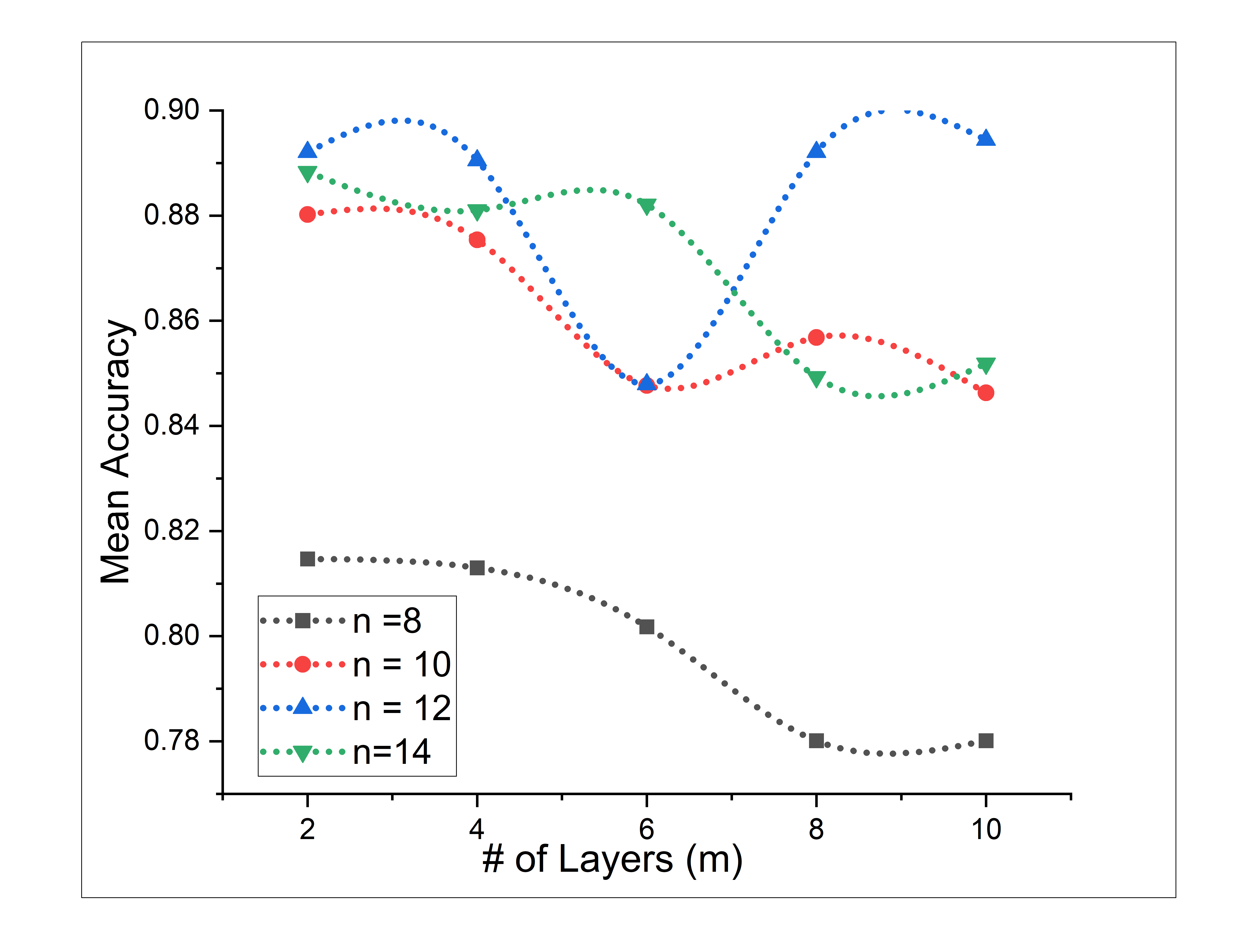}}}
\caption{Unentangled Ansatz: Mean accuracy for angle encoding.} \label{fig:AngE_mean_acc_NE}
\end{figure}

Analogous to the BP analysis in case of entangled ansatz with angle encoding, 
the input feature dimension is reduced to conform with the reasonable width of quantum layer(s). We start with the minimum of width of $n=8$. The mean accuracy for all the experiments is shown in Figure in \ref{fig:AngE_mean_acc_NE}.

Analyzing the effect of $n$ from BP aspect, the accuracy improves in general, as shown in Figure \ref{fig:AE_NE_1}. This is opposite to the results of entangled ansatz (Figure \ref{fig:AngE_ansatz1a}), where the accuracy starts decreasing after certain $m$, despite an increase in number of input features (with an increase in $n$). 
This leads to a conclusion that unentangled ansatz is less susceptible to BP than entangled ansatz when encoding the data in qubit rotation angles. 
The results in figure \ref{fig:AE_NE_2} helps analyzing the effect of $m$ on HQNN performance when the data is encoded via angle encoding. 
It can be observed that the accuracy deteriorates in general with an increase in $m$. 
The smallest depth ($m=2$) performs better than all other $m$, for all $n$ except $n=12$. however, even in that case ($n=12$) the accuracy improvement from $m=2$ to $m=10$ is not quite significant.

\subsection{Trainability Vs Expressibility of HQNNs} \label{sec:results_TvsE}
A more careful analysis of results in previous section shows that the decline in performance does not have a consistent relationship (quadratic, exponential etc.) but is dependent on the quantum layer(s) depth $m$. Consequently, a more insightful analysis of how deep quantum layer(s) can be, for a given width before experiencing the BP and eventually leading to the performance decline, would be important. 
In this section, we conduct an empirical analysis on trainability vs. expressibility of HQNNs, again for both ansatz structures individually with both encodings.  
We benchmark model's overall accuracy and loss convergence as evaluation metrics for trainability vs. expressibility analysis.  

\subsubsection{Trainability Vs. Expressibility of Entangled Ansatz with Amplitude Encoding} \label{sec:results_TvsE_entangled_amp}
\begin{figure}[H]
\centering
\includegraphics[scale= 0.6]{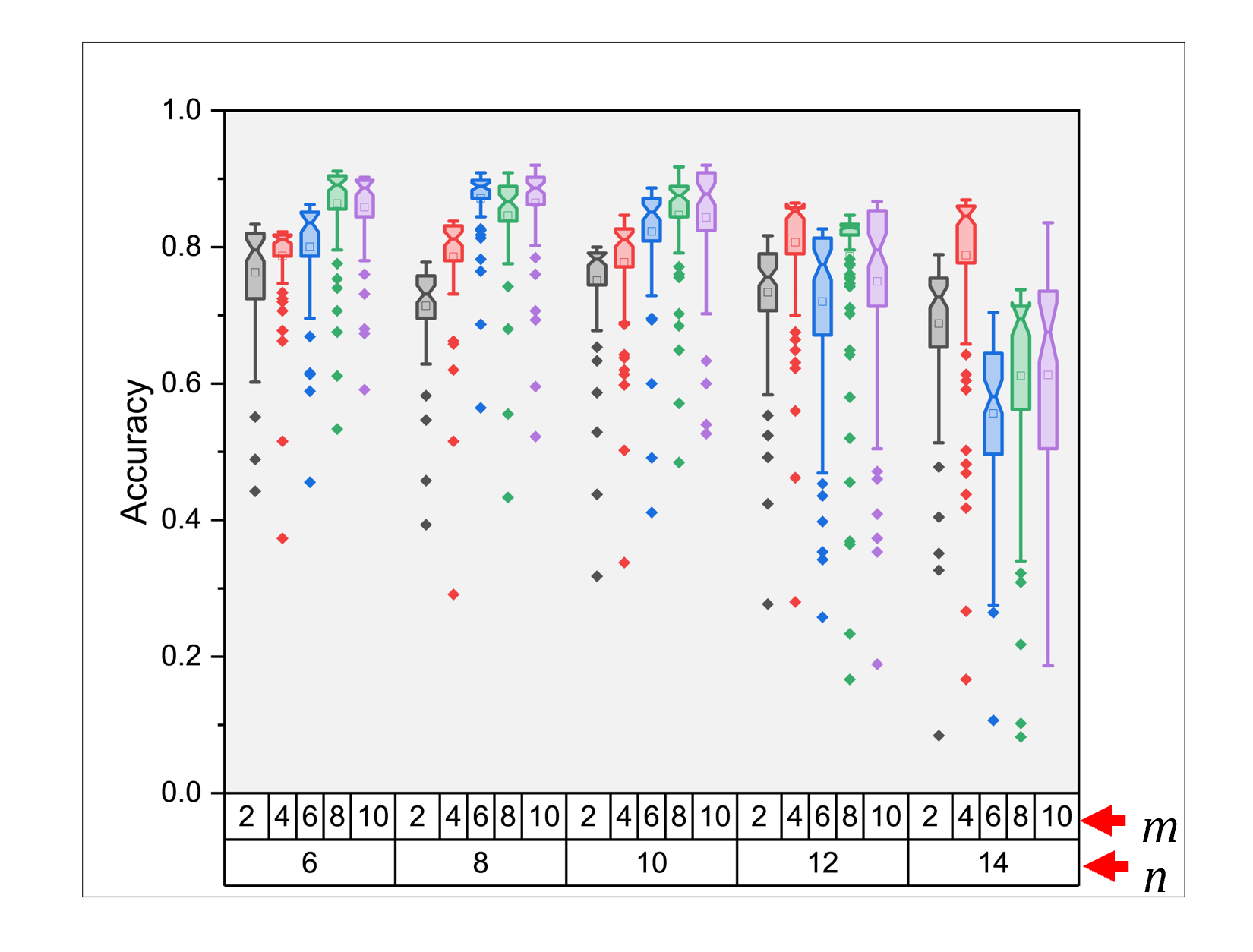}
	\caption{Entangled Ansatz: Accuracy trends for all $n$ and $m$ for amplitude encoding. }
	\label{fig:AmpE_WE_all_accuracies}
\end{figure}

The trainability vs. expressibility analysis of HQNNs gives an insight of how expressible a given ansatz can be (before the gradients start vanishing) for a particular width with better or at least equivalent performance. 
We performed different training experiments for different $n$ and $m$, as shown in Figure \ref{fig:experiments}. The model is evaluated based on the overall accuracy and convergence. The results are shown in Figure \ref{fig:AmpE_WE_all_accuracies}. 

\begin{figure}[htp]
\centering
\subfloat[$n=6$]{\label{fig:AmpE_1a}
\includegraphics[width=.3\textwidth]{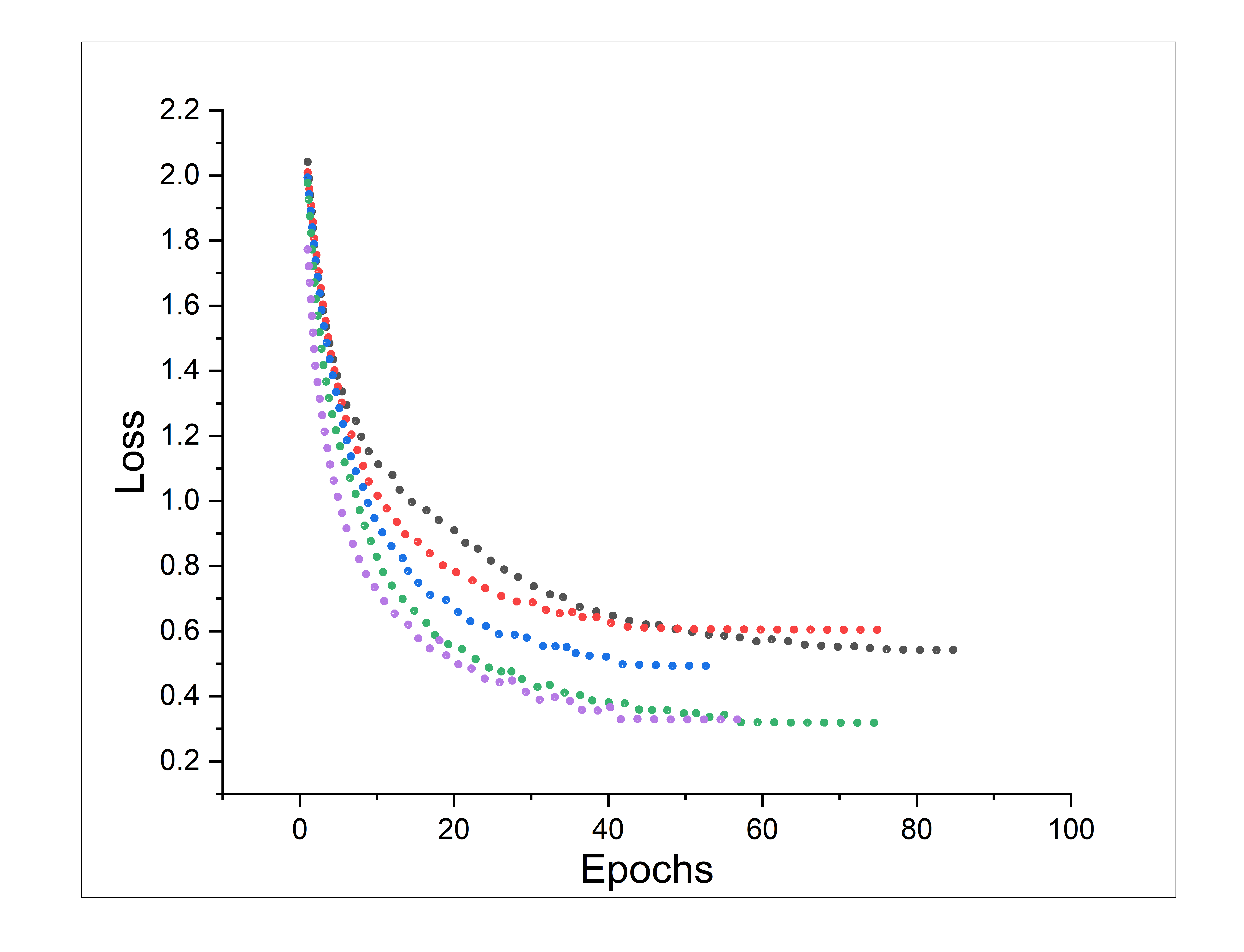}}\quad
\subfloat[$n=8$]{\label{fig:AmpE_2a}
\includegraphics[width=.3\textwidth]{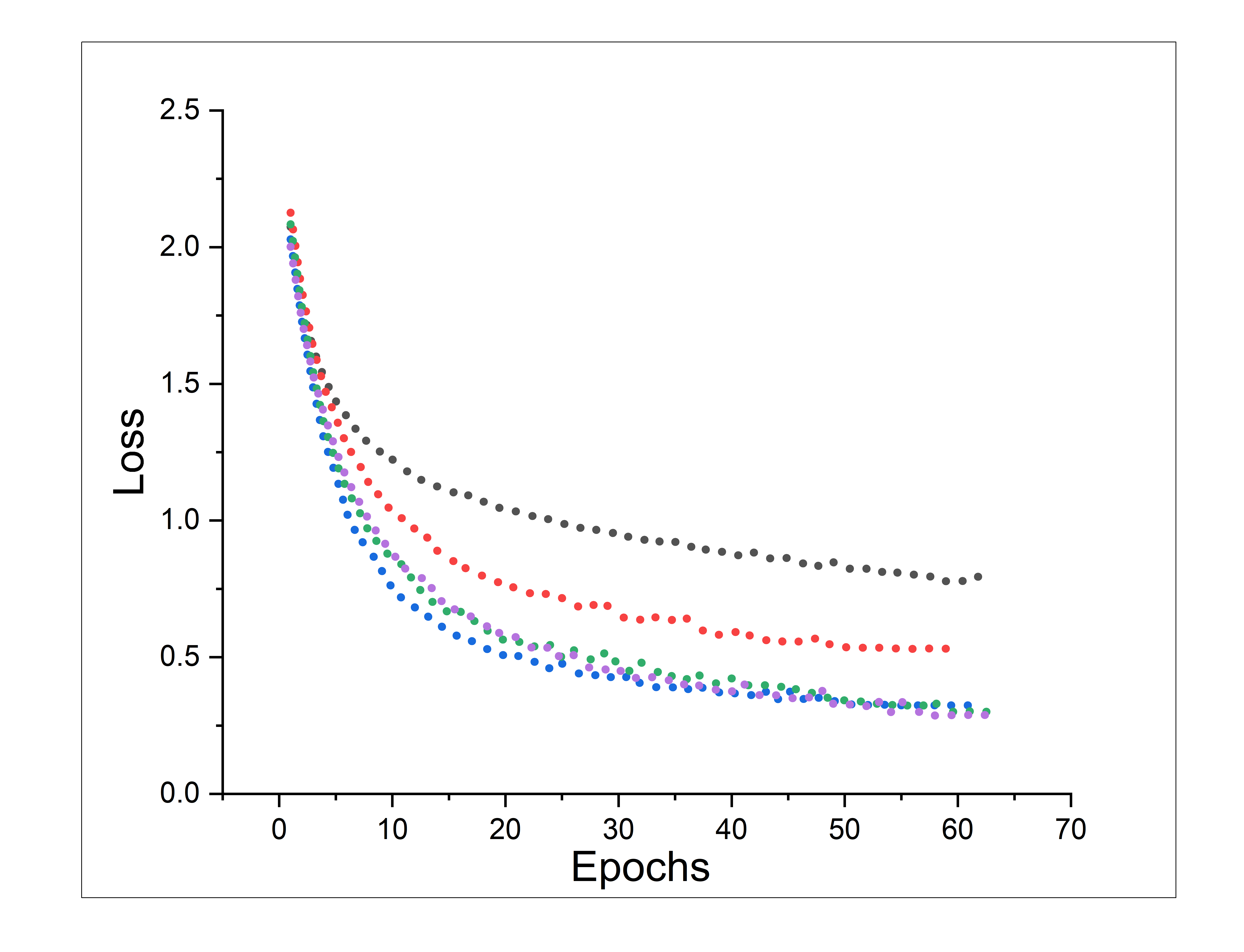}}\quad
\subfloat[$n=10$]{\label{fig:AmpE_3a}
\includegraphics[width=.3\textwidth]{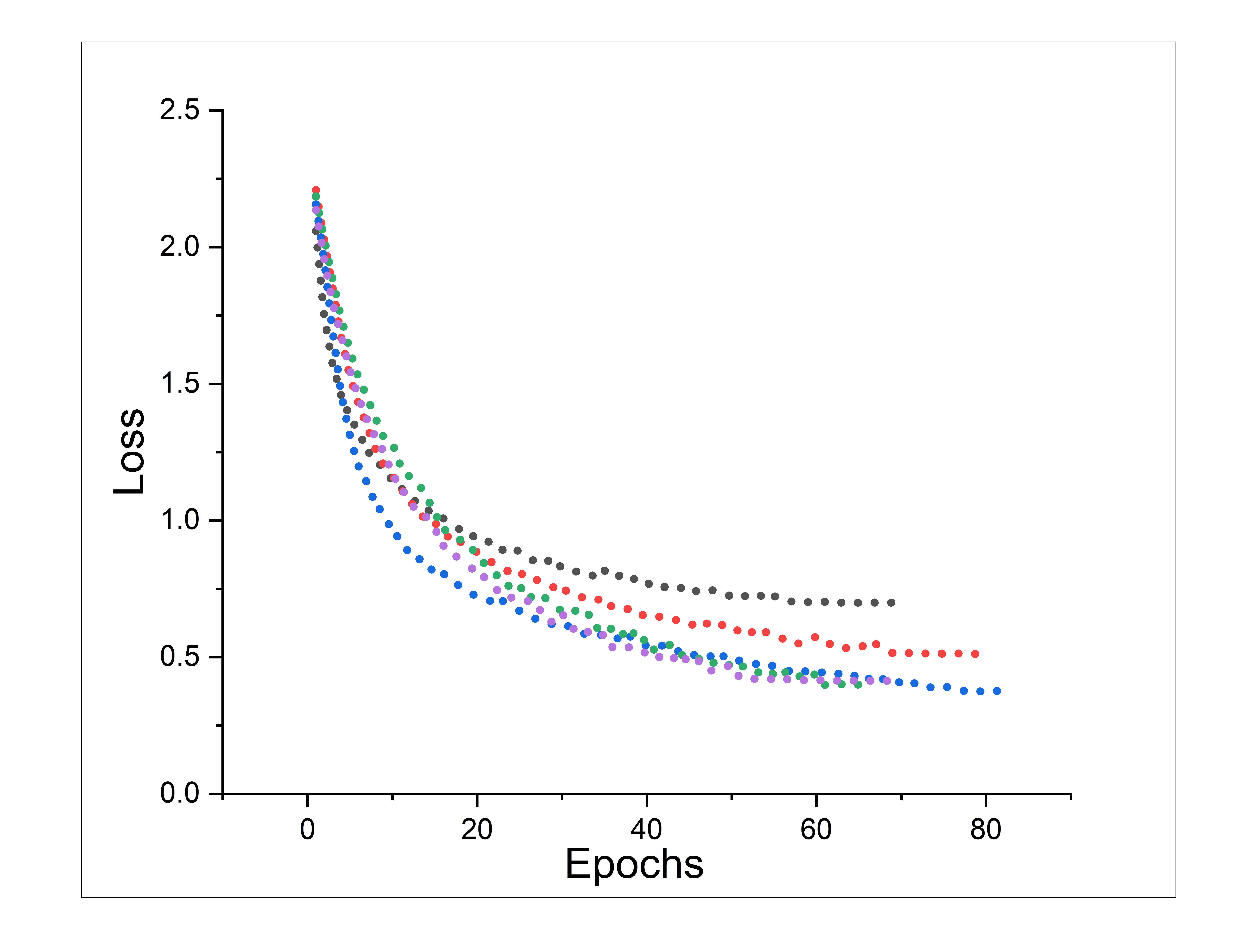}}

\medskip
\subfloat[$n=12$]{\label{fig:AmpE_4a}
\includegraphics[width=.3\textwidth]{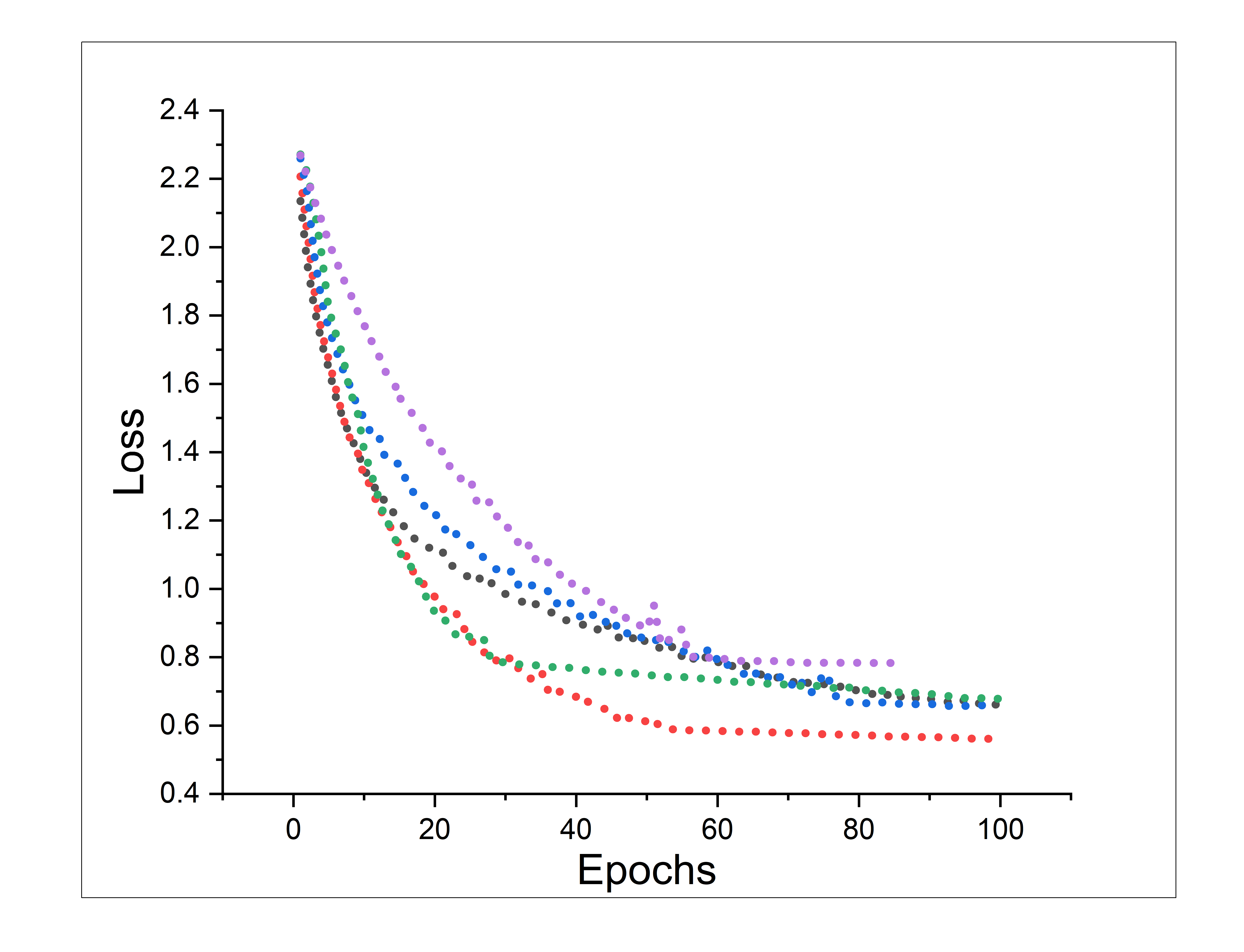}}\quad
\subfloat[$n=14$]{\label{fig:AmpE_5a}
\includegraphics[width=.3\textwidth]{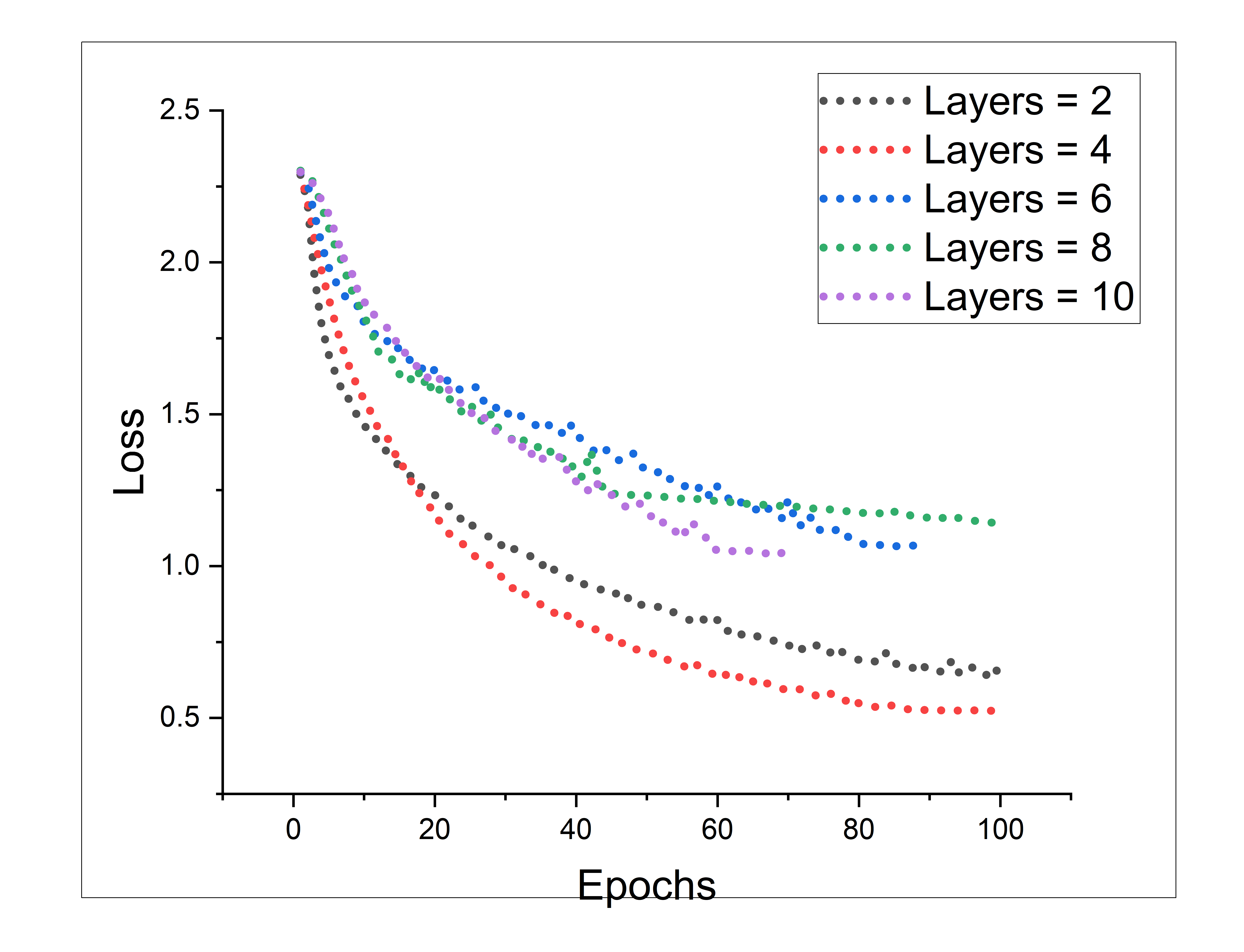}}

\caption{Entangled Ansatz: Loss trends for all $n$ and $m$ for amplitude encoding. }
\label{fig:Loss_all_AmpE_WE}
\end{figure}

In Figure \ref{fig:AmpE_WE_all_accuracies}, for $n=6$ the deeper quantum layers i.e., $m=8$ and $m=10$ has better results in terms of accuracy, however, convergence is better in case of $10$ layers as shown in Figure \ref{fig:AmpE_1a}. 
When we increase $n$ to $8$ and $10$ qubits, the circuit depth of $m=6, 8, 10$ have relatively better accuracy. However, in both cases ($n=8$ and $n=10$) the model converges relatively faster when $m=6$ as shown in Figure \ref{fig:AmpE_2a} and \ref{fig:AmpE_3a}, respectively, reducing the appropriate depth of quantum layers when more qubits are used. 
When $n$ is increased to $12$ qubits, the appropriate circuit depth is further reduced to $m=4$, both in terms of accuracy and convergence, as shown in Figures \ref{fig:AmpE_WE_all_accuracies} and \ref{fig:AmpE_4a}, respectively. However, the corresponding reduction in quantum layers depth becomes more evident when we further increase $n$ to $14$ qubits, as the model clearly converges faster for $m=4$ as shown in Figure \ref{fig:AmpE_5a}.

Considering the results shown in \ref{fig:AmpE_WE_all_accuracies} and \ref{fig:Loss_all_AmpE_WE} we concur that the BP phenomenon in HQNNs is not only the function of qubits (gradients vanish exponentially with number of qubits and the network becomes untrainable as we increase the qubits) but it is also dependent on how expressible the quantum layer(s) are.
If we analyze the accuracy in Figure \ref{fig:AmpE_WE_all_accuracies} as a whole, we observe that as we increase $n$ the performance starts deteriorating. Although, we observe a trade-off between $n$ and $m$ (bigger $n$ tends to reduce $m$ and vice versa), a careful analysis of individual accuracy reveals that smaller $n$ and relatively bigger $m$ are more appropriate to achieve better performance because the achieved accuracy is higher and are skewed on the higher side. For instance, for $n=6$ and $m=8$ almost 75\% of the accuracy are higher than the highest achieved accuracy for $n=14$ and $m=4$. 
Furthermore, the accuracy for bigger $n$ have more variance (more spread in accuracy) than smaller $n$, showing non-robust learning, in case of wider quantum layers.


\subsubsection{Trainability Vs. Expressibility of Entangled Ansatz with Angle Encoding} \label{sec:results_TvsE_entangled_ang}
\begin{figure}[H]
\centering
\includegraphics[scale= 0.6]{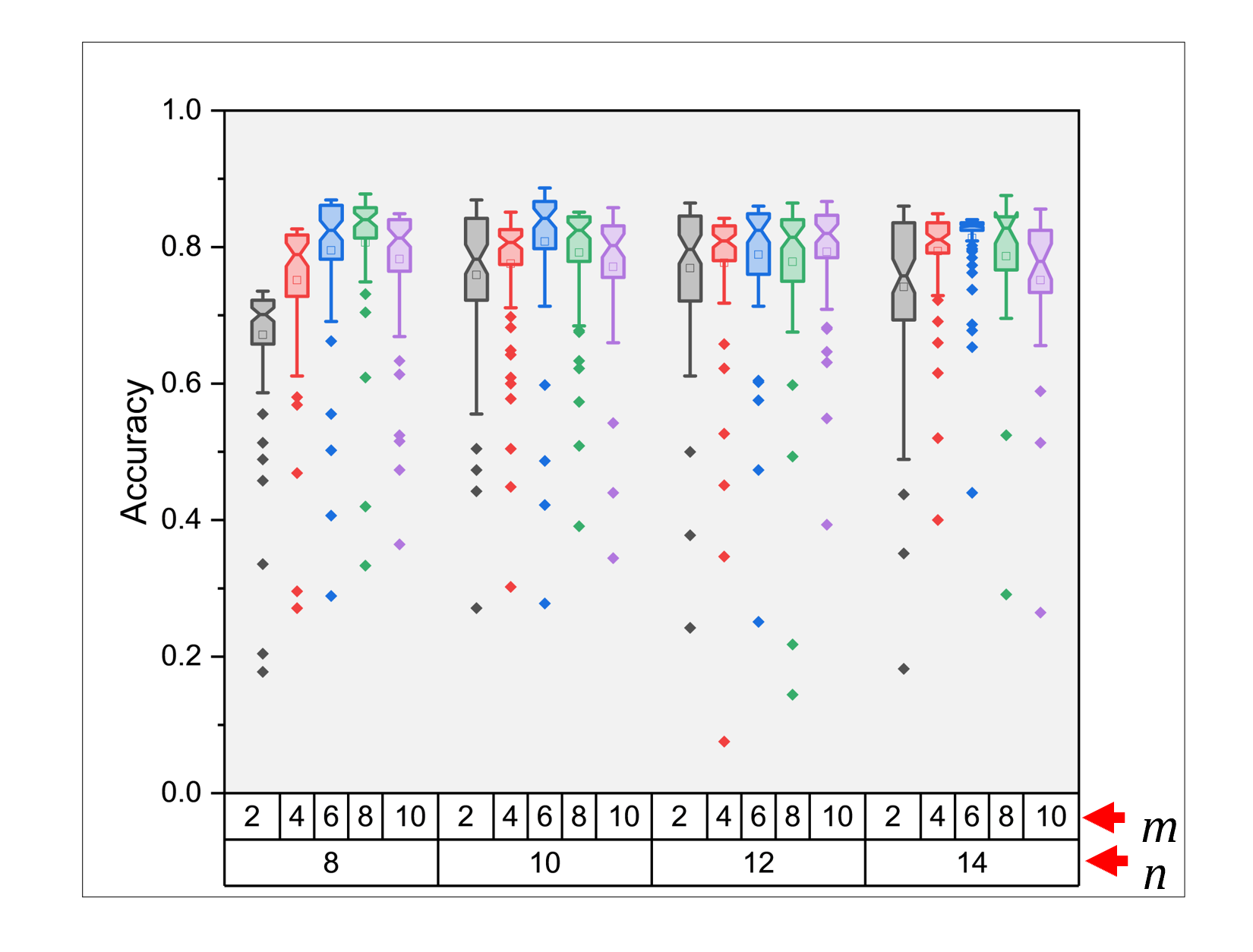}
	\caption{Entangled Ansatz: Accuracy trends for all $n$ and $m$ for angle encoding. }
	\label{fig:AngE_WE_all_accuracies}
\end{figure}

\begin{figure}[H]
\centering
\subfloat[$n=8$]{\label{fig:AE_aa}
\includegraphics[width=.45\textwidth]{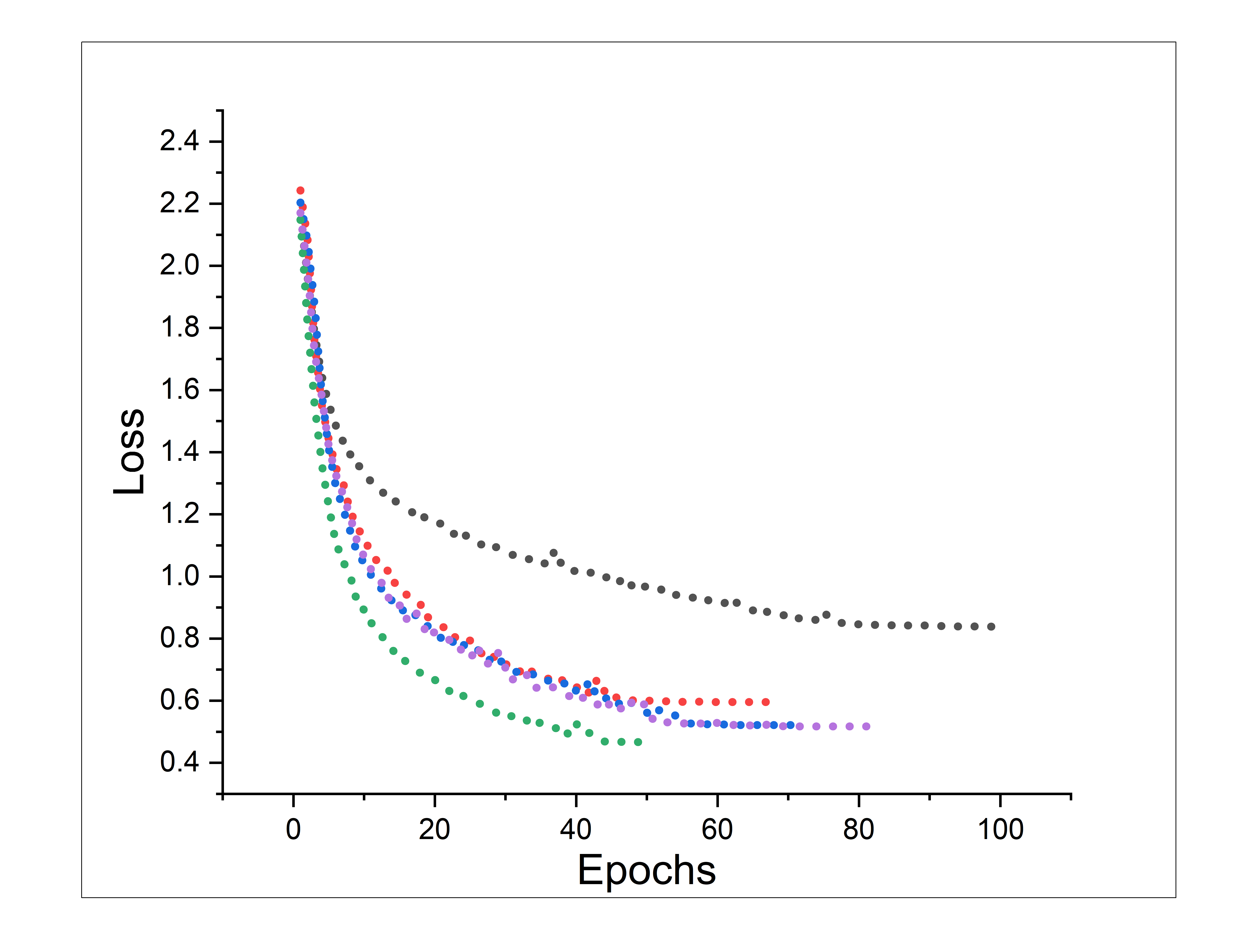}}\quad
\subfloat[$n=10$]{\label{fig:AE_bb}
\includegraphics[width=.45\textwidth]{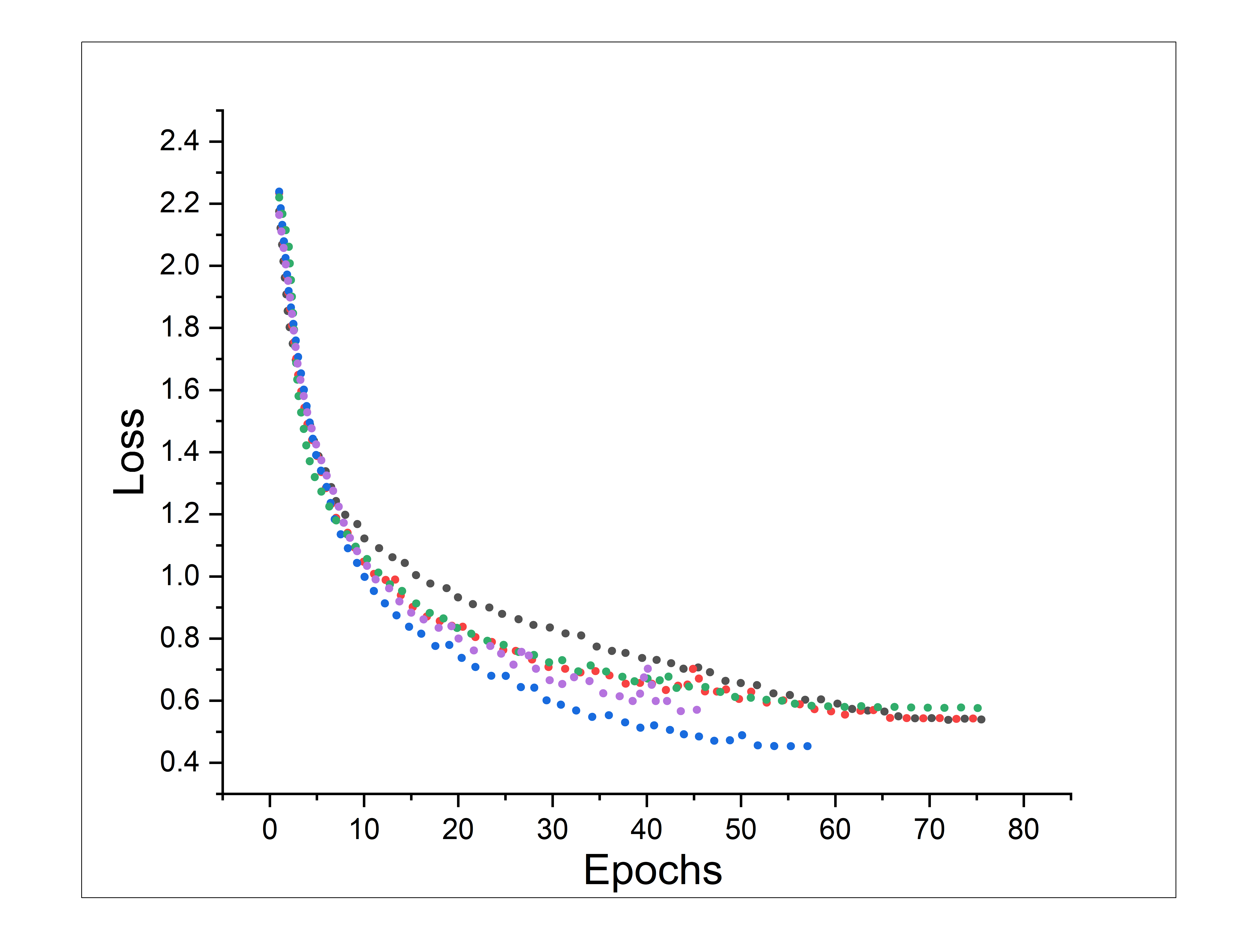}}

\medskip
\subfloat[$n=12$]{\label{fig:AE_cc}
\includegraphics[width=.45\textwidth]{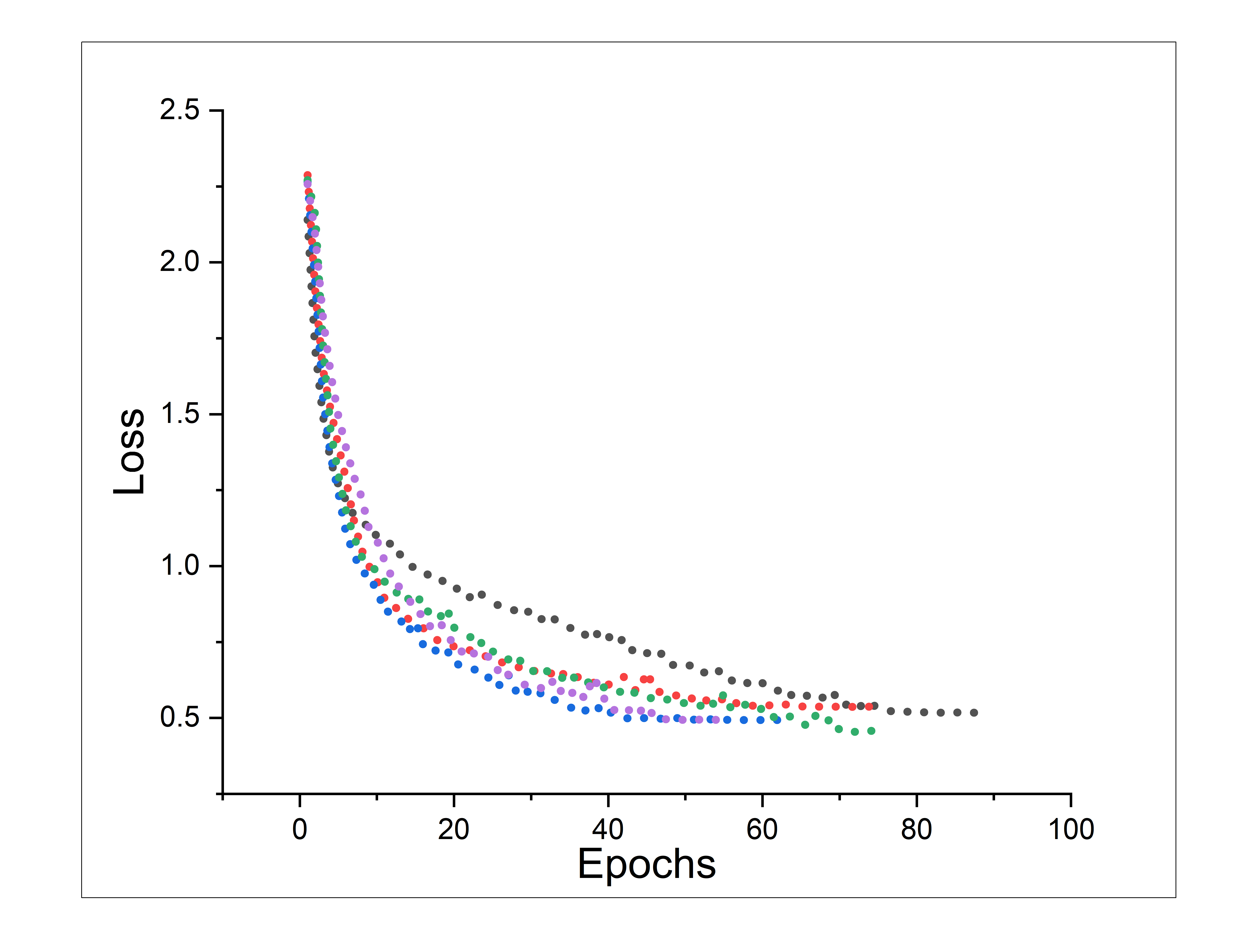}}\quad
\subfloat[$n=14$]{\label{fig:AE_dd}
\includegraphics[width=.45\textwidth]{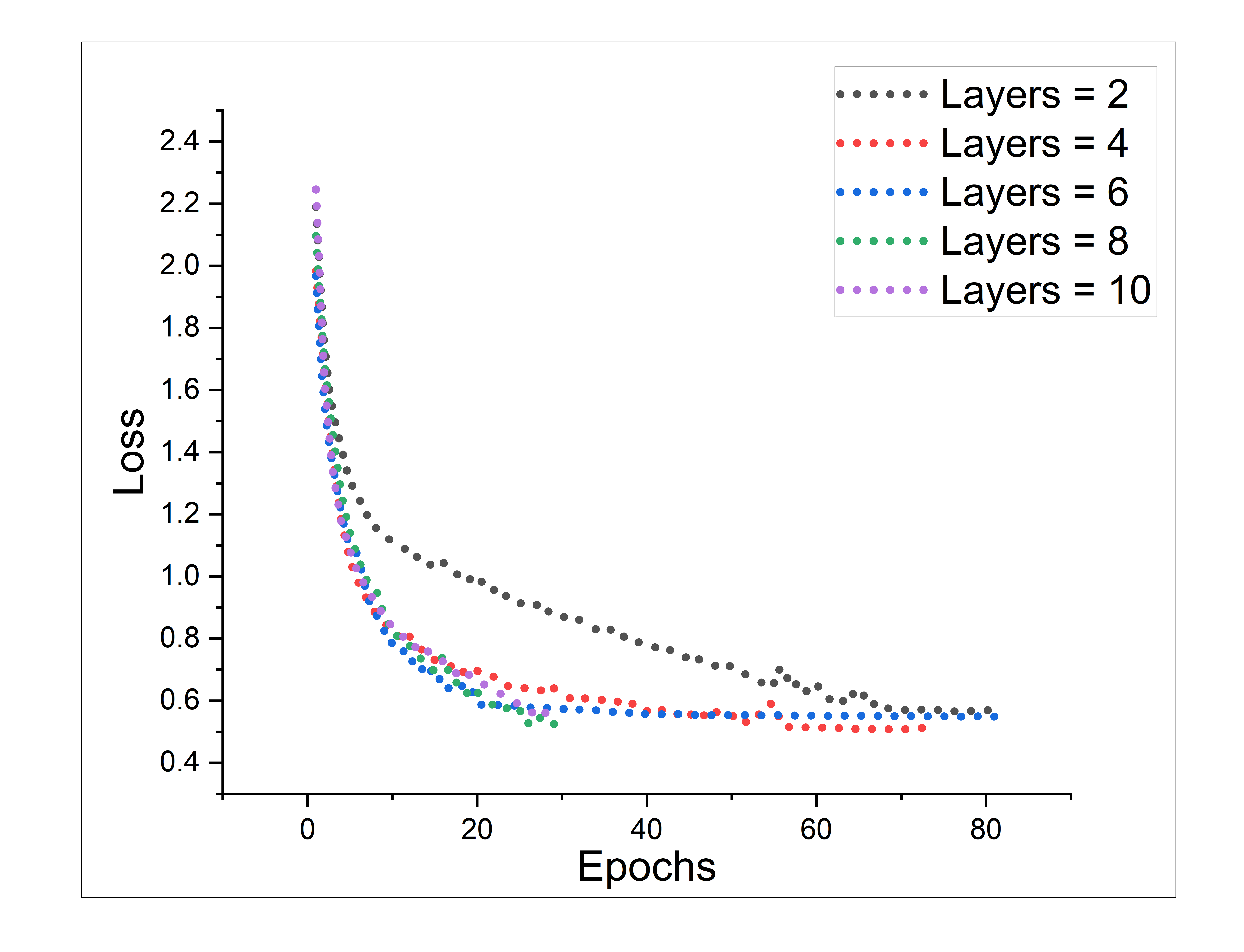}}

    \caption{Entangled Ansatz: Loss trends for all $n$ and $m$ for angle encoding.}
    \label{fig:Loss_all_AngE_WE}
\end{figure}

We now present the analysis of how a different data encoding strategy i.e., angle encoding, where the input data features are encoded in qubit rotation angles can affect the trainability in QNNs for certain $n$ and $m$. We encode the data in $RY$ rotations which is then passed to the quantum layer(s).  For the case of angle encoding, as explained in Section \ref{sec:data_enc} the number of features being encoded should be equal to that of the number of qubits. Our input features are images of size $8\times 8$ and hence $64$ features in total. We need 64 qubits to encode 64 features which is not suitable in NISQ era. Therefore, we apply PCA and reduce the input feature dimension to $8, 10, 12$ and $14$ features. 
The quantum layer(s) width ($n$) is equal to that of dimension of input features, and we experiment with different quantum layers having variable depths ($m=2,4,6,8,10$), similar to that of analysis with amplitude encoding and shown in Figure \ref{fig:experiments}, except for $n=6$, because We tend to keep $n$ bigger while applying PCA so that, not much information is lost. The accuracy and loss for all the experiments are shown in Figure \ref{fig:AngE_WE_all_accuracies} and \ref{fig:Loss_all_AngE_WE} respectively.

Based on the results obtained, we observe that for $n=8$, circuit of depth of $m=8$ is better in terms of both accuracy ( Figure \ref{fig:AngE_WE_all_accuracies} and convergence (Figure \ref{fig:AE_aa}) whereas the smallest depth quantum layer ($m=2$) is worst among all different values of $m$..
When the circuit width is increased to $n=10$ is, the circuit depth with better performance both in terms of better accuracy and convergence is $m=6$ reduced from $m=8$. 
Even though the input feature dimension is also increased (from $8$ to $10$), the circuit depth tends to reduce mainly because the gradients magnitude gets smaller.
For $n=12$, almost all the layers except $m=2$ have comparable performance in terms of accuracy, but $m=6$ has slightly better convergence rate than others. However, the individual accuracy dropped from $n=8$ and $n=10$.  
Similarly, for $n=14$, circuit depth of $m=14$ is betters in terms of both accuracy and convergence rate. Unlike amplitude encoding, the allowable circuit depth is more for bigger $n$, when the data is encoded in qubit rotation angles, and this is because of the greater number of input features to train as compared to smaller $n$.

\subsubsection{Trainability Vs. Expressibility of Unentangled Ansatz with Amplitude Encoding} \label{sec:results_TvsE_unentangled_amp}

\begin{figure}[H]
\centering
\includegraphics[scale= 0.6]{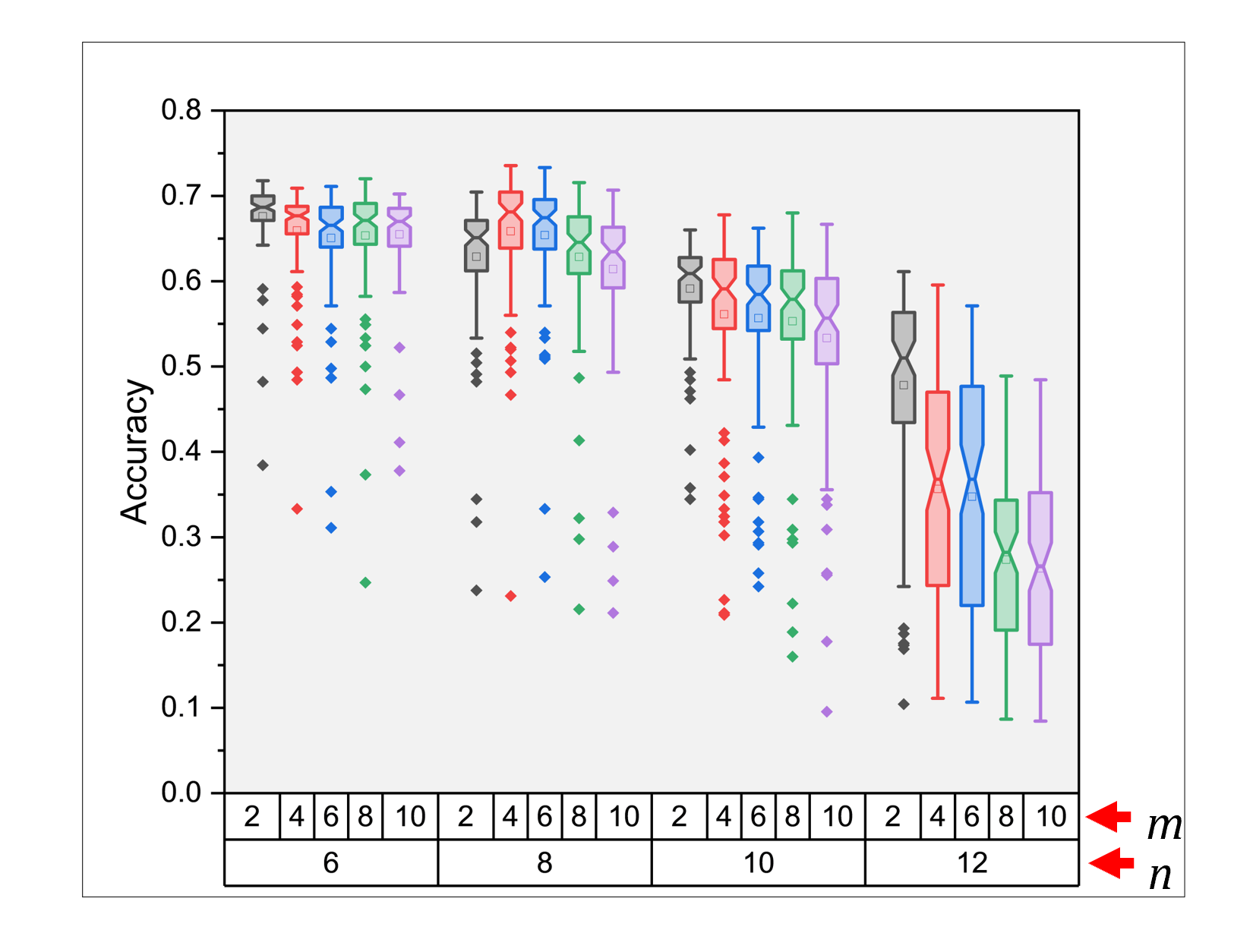}
	\caption{Unentangled Ansatz: Accuracy trends for all $n$ and $m$ for amplitude encoding. }
	\label{fig:AmpE_NE_all_acc}
\end{figure}

\begin{figure}[H]
\centering

\subfloat[$n=6$]{\label{fig:loss_ampE_NE_n=6}
\includegraphics[width=.45\textwidth]{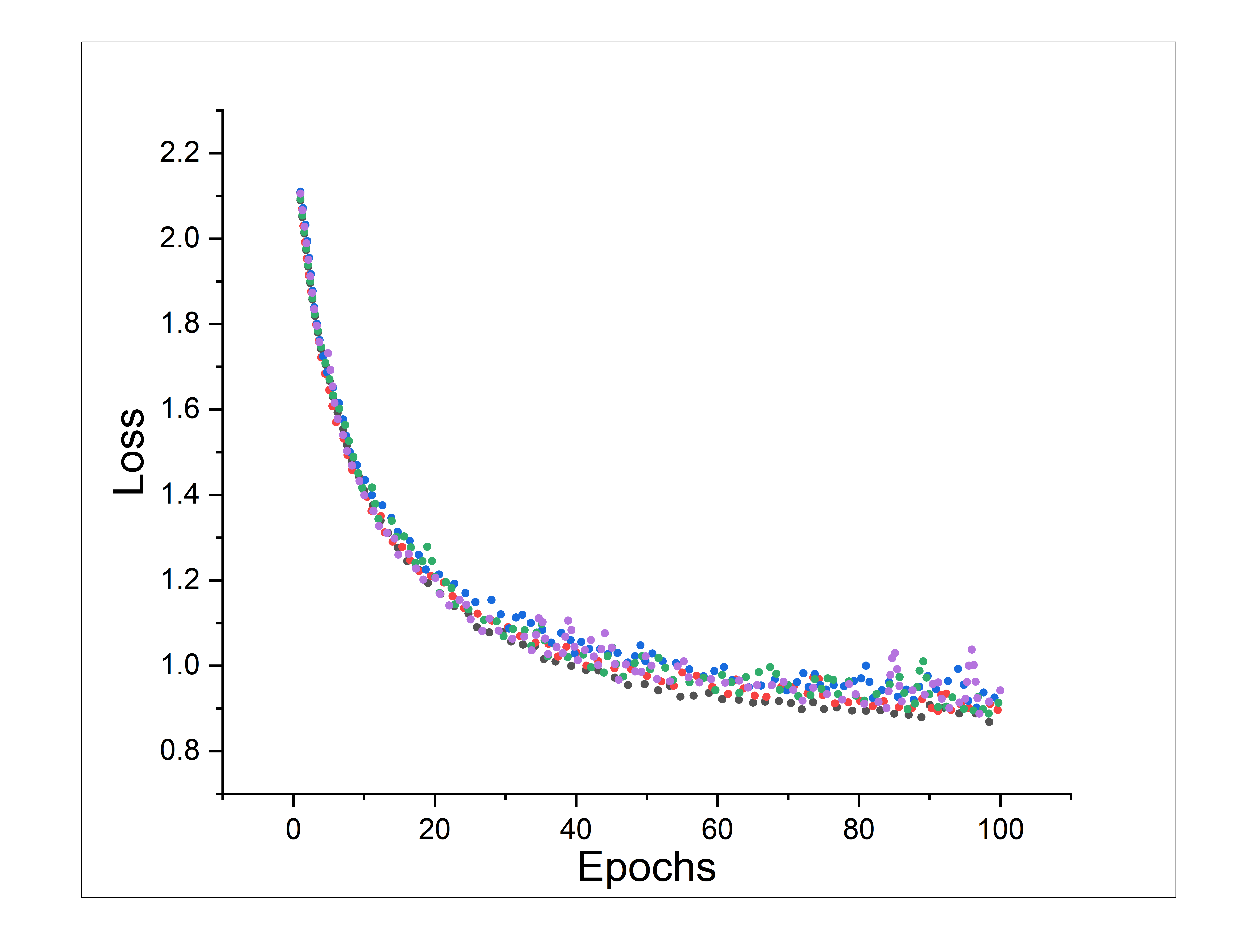}}\quad
\subfloat[$n=8$]{\label{fig:loss_ampE_NE_n=8}
\includegraphics[width=.45\textwidth]{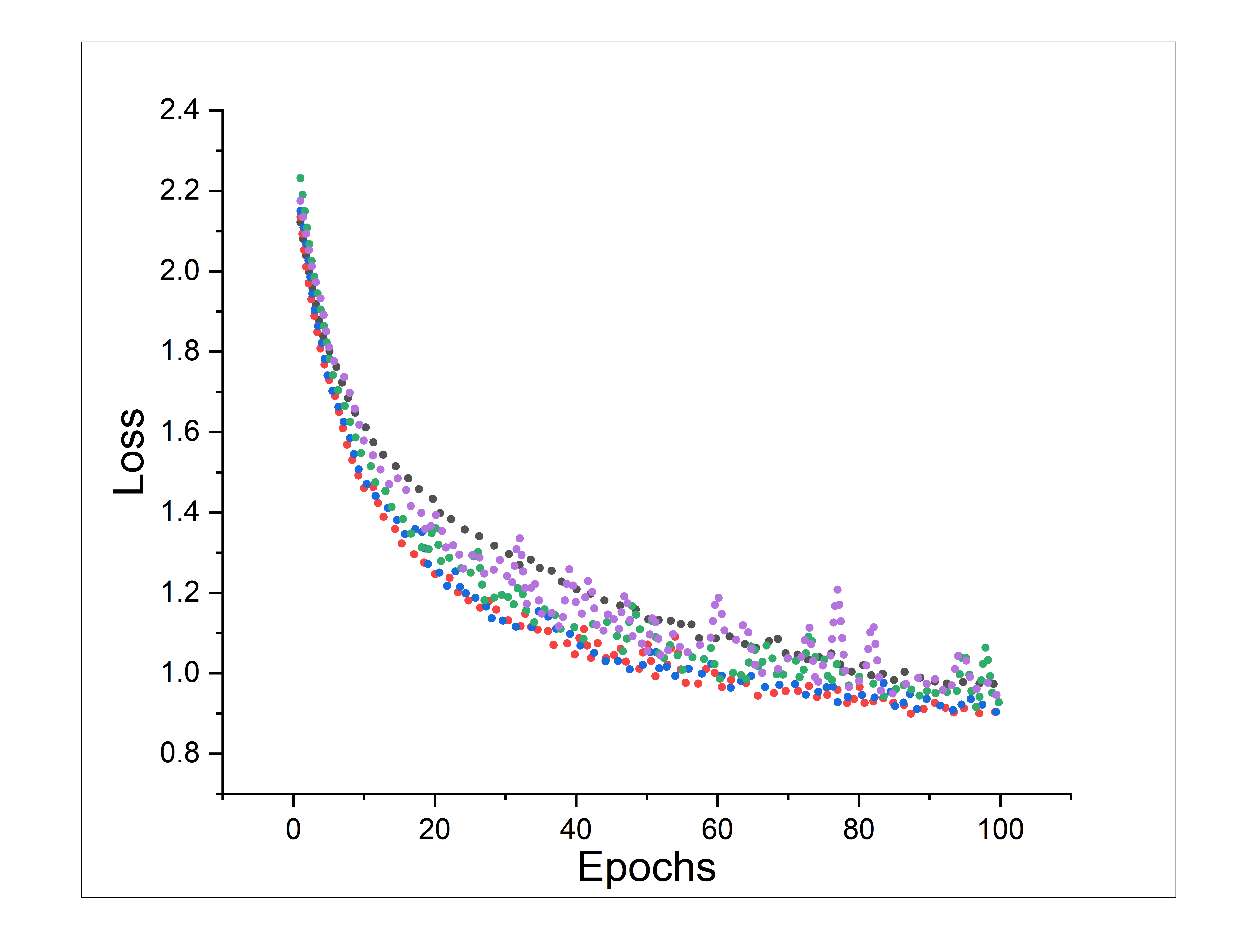}}

\medskip
\subfloat[$n=10$]{\label{fig:loss_ampE_NE_n=10}
\includegraphics[width=.45\textwidth]{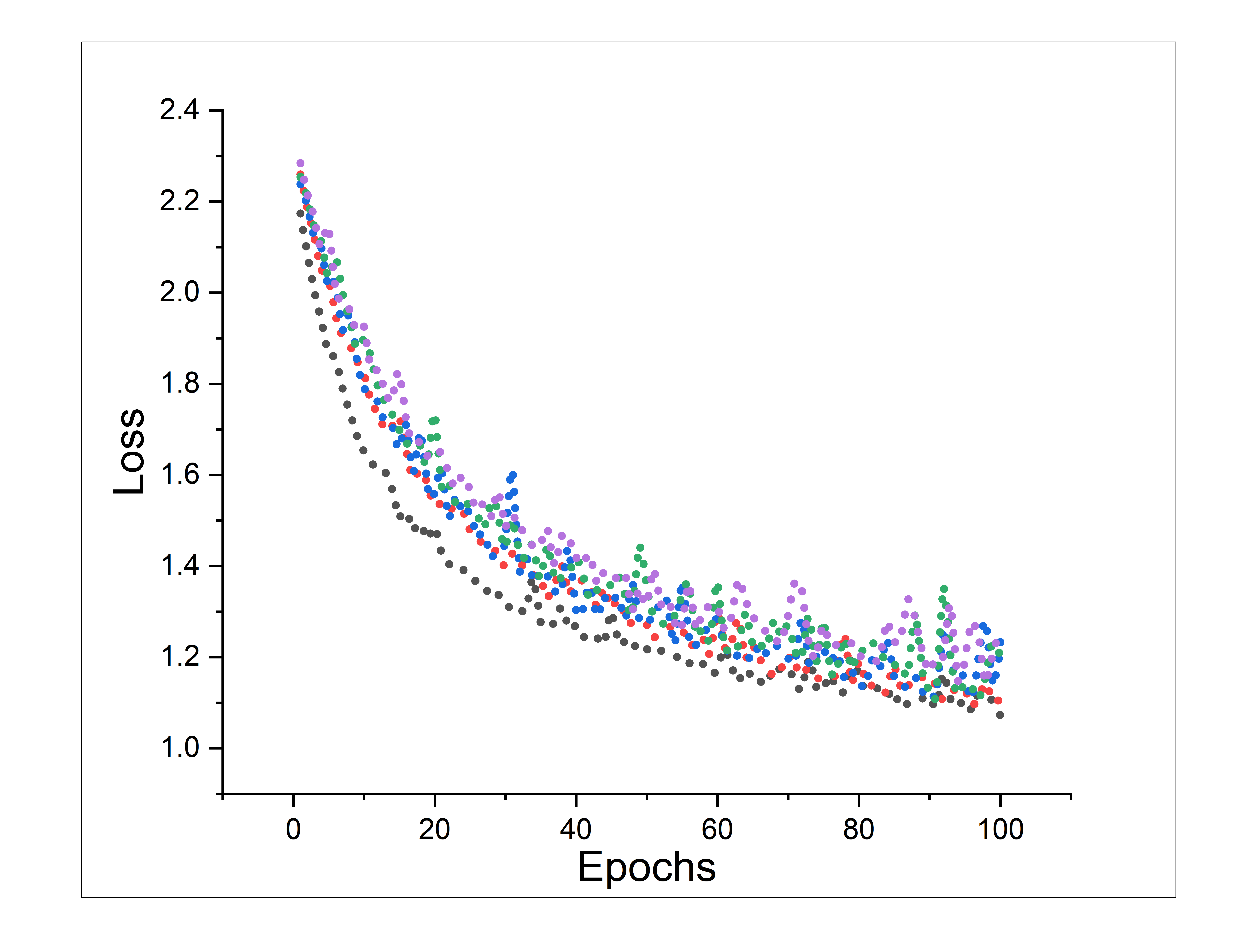}}\quad
\subfloat[$n=12$]{\label{fig:loss_ampE_NE_n=12}
\includegraphics[width=.45\textwidth]{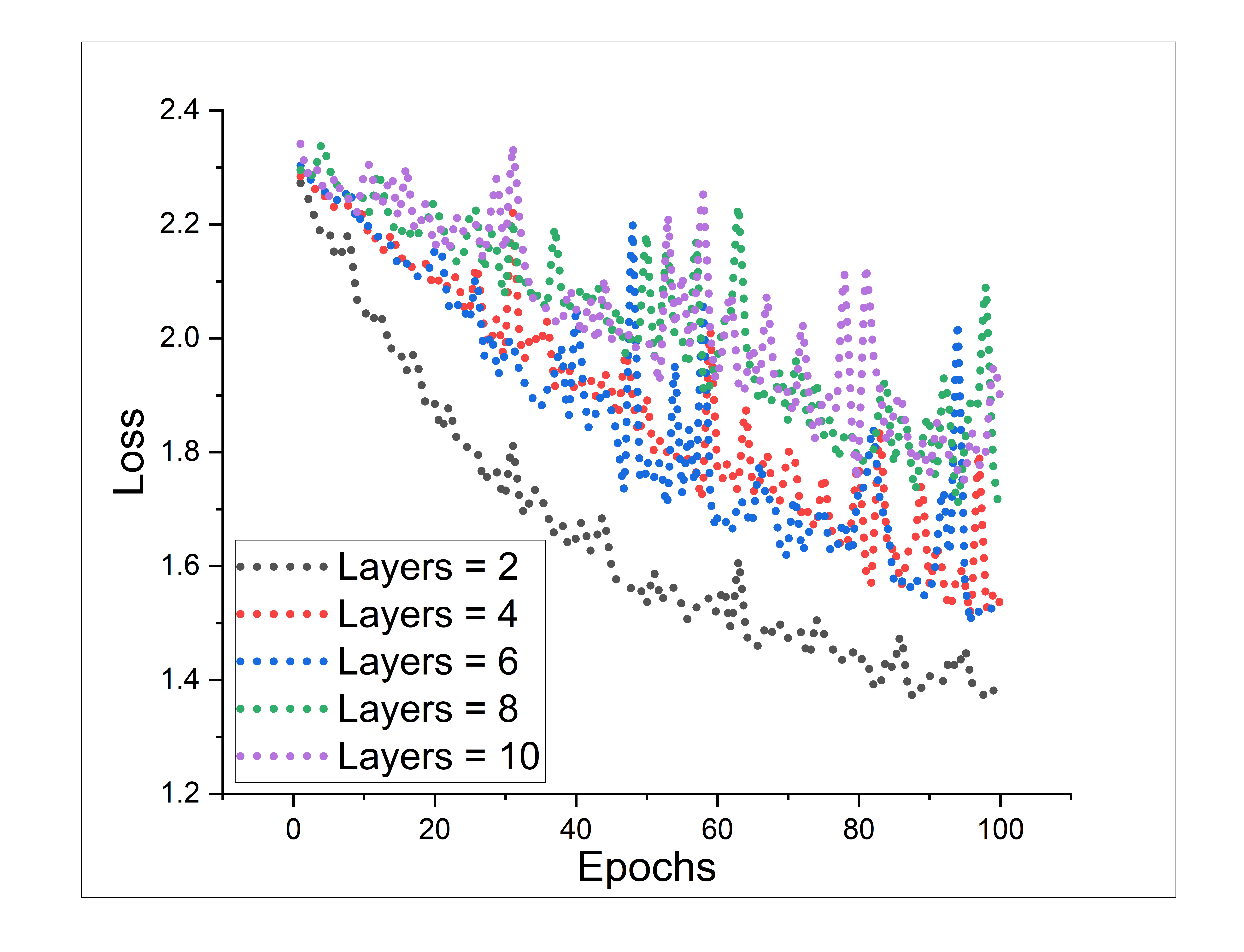}}

   \caption{Unentangled Ansatz: Loss trends for all $n$ and $m$ for amplitude encoding.}
    \label{fig:Loss_all_AmpE_NE}
\end{figure}

We now present the trainability vs. expressibility analysis for unentangled ansatz, while encoding the input data in qubit state vector. As discussed previously, for our dataset, the input dimensionality reduction for amplitude encoding is not required, and the PCA is therefore not applied. The experiments are the same as for entangled ansatz, i.e. changing the width and depth of quantum layers as shown in Figure \ref{fig:experiments}. Based on the obtained results, we observe a consistent decline in performance as $n$ increases, we therefore skip the experimentation for $n=14$ and all corresponding $m$. The accuracy and loss trends for unentangled ansatz are presented in Figure \ref{fig:AmpE_NE_all_acc} and \ref{fig:Loss_all_AmpE_NE}, respectively. 

We observe that, unlike the case of amplitude encoding in entangled ansatz, where for smaller width $n$ the allowed circuit depth $m$ is relatively bigger, in unentangled ansatz, i.e. when the entanglement is removed, the allowed circuit depth tends to become lower even for smaller $n$. For $n=6$, almost all corresponding depths $m$, yield somewhat same performance both in terms of accuracy and loss convergence as shown in Figure \ref{fig:AmpE_NE_all_acc} and \ref{fig:loss_ampE_NE_n=6} respectively. However, the smaller depth ($m=2$) is slightly better than other $m$, in contrast with entangled ansatz, where $m=8$ and $10$ turned out to be better for $n=6$ in terms of both the performance metrics (Figure \ref{fig:AmpE_WE_all_accuracies} and \ref{fig:AmpE_1a}). 
Similarly, for $n=8$ the relatively smaller depth quantum layers typically $m=4$ and $6$ are slightly better than other depths, with respect to both accuracy and convergence as shown in Figure \ref{fig:AmpE_NE_all_acc} and \ref{fig:loss_ampE_NE_n=8} respectively, which is again in contrast to entangled ansatz, where $n=8$ have better performance with maximum depth ($m=10$) and worst for smallest depth ($m=2$) as presented in Figures \ref{fig:AmpE_WE_all_accuracies} and \ref{fig:AmpE_2a}. 
When the $n$ is further increased to $10$ and $12$ the overall performance deteriorates (Figure \ref{fig:AmpE_NE_all_acc}), however, analogous to the results in entangled ansatz for bigger $n$, relatively smaller depths, typically $m=2$ and $4$ are relatively better than greater depth layers. 

Furthermore, as $n$ increases, the individual performance for all $m$ in both ansatz structures becomes more inconsistent, as shown in Figure \ref{fig:AmpE_NE_all_acc} and \ref{fig:AmpE_WE_all_accuracies}, however this inconsistency is more prominent in case of unentanglement ansatz in HQNNs for amplitude encoding. Also, in case of no entanglement in quantum layers, for bigger $n$ almost all circuit depths have a fluctuating journey to reach to a minimum value of cost function landscape, clearly exhibiting the existence of BP (unable to determine the cost minimizing direction). It is also worth mentioning here that for all $n$ and $m$ in unentangled ansatz, the HQNNs stuck in the local minima and fails to converge even after $100$ training iterations, and is insensitive to the regularization techniques (we used early stopping in this work to avoid overfitting), whereas in case of entangled ansatz, the model is more robust and and have relatively better and smoother journey to converge to minimum in a cost function landscape, exhibiting relatively lesser insensitivity to regularization.

\subsubsection{Trainability Vs. Expressibility of Unentangled Ansatz with Angle Encoding} \label{sec:results_TvsE_unentangled_ang}
\begin{figure}[H]
\centering
\includegraphics[scale= 0.6]{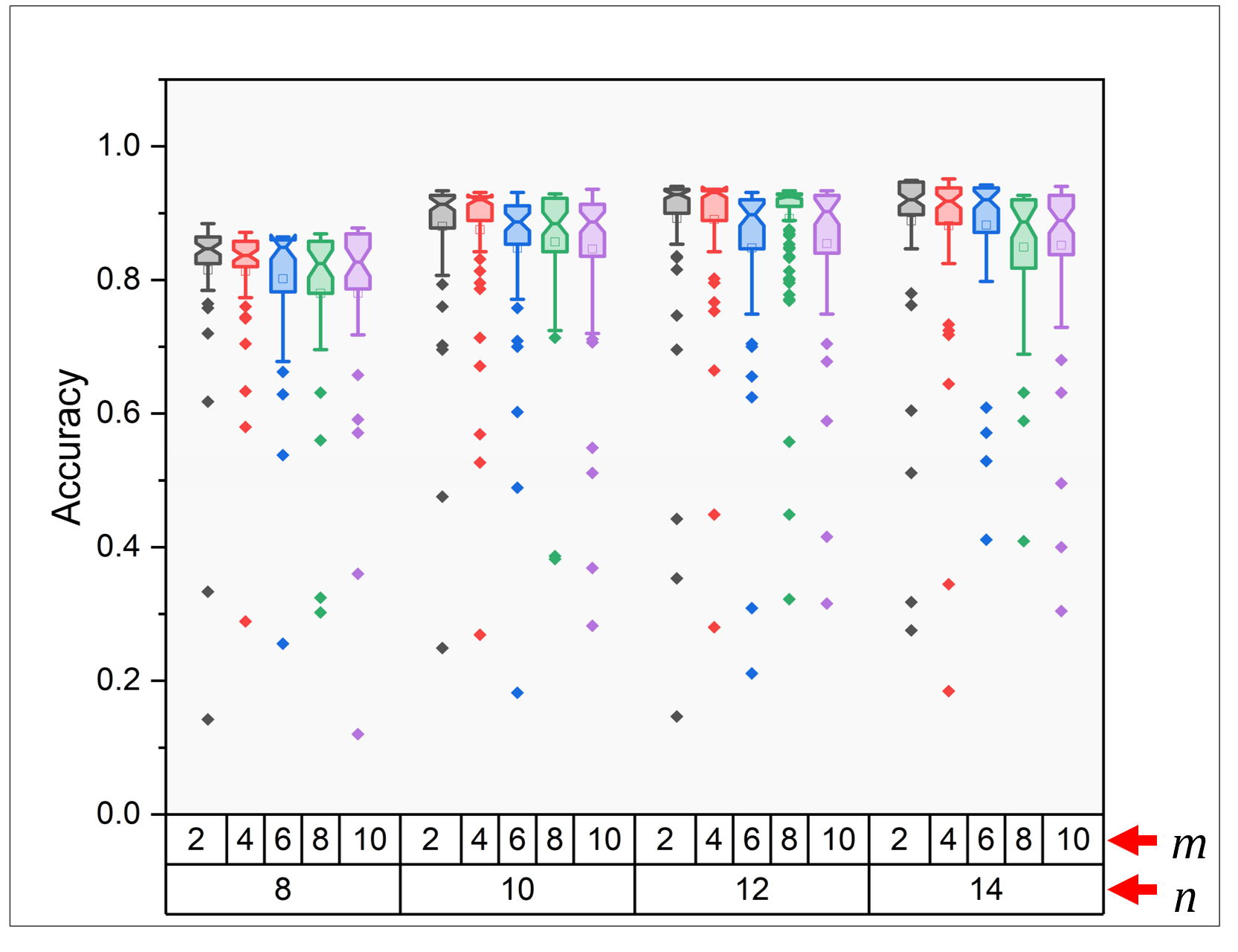}
	\caption{Unentangled Ansatz: Accuracy trends for all $n$ and $m$ for angle encoding. }
	\label{fig:AngE_NE_all_acc}
\end{figure}

\begin{figure}[H]
\centering

\subfloat[$n=6$]{\label{fig:loss_AngE_NE_n=8}
\includegraphics[width=.45\textwidth]{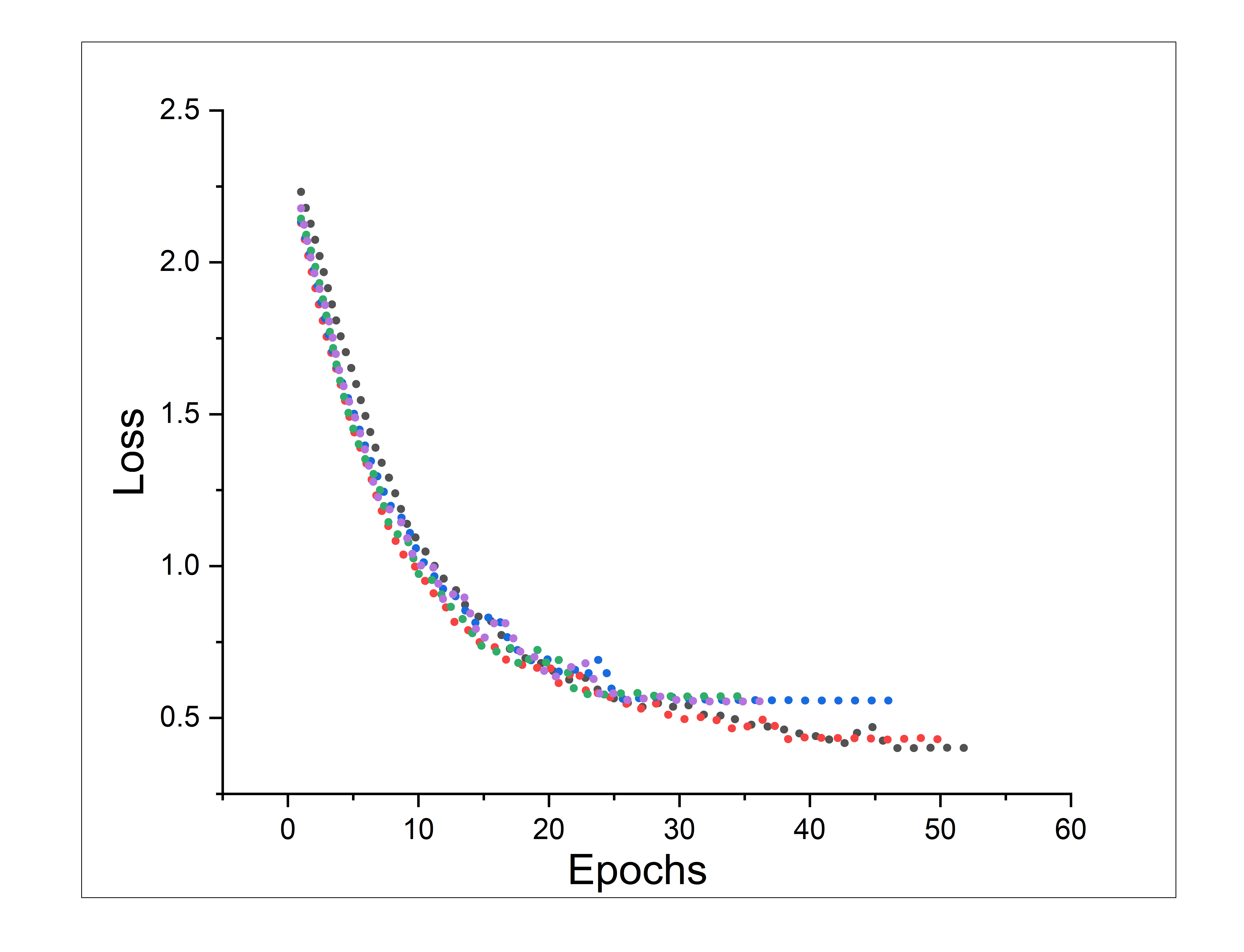}}\quad
\subfloat[$n=8$]{\label{fig:loss_AngE_NE_n=10}
\includegraphics[width=.45\textwidth]{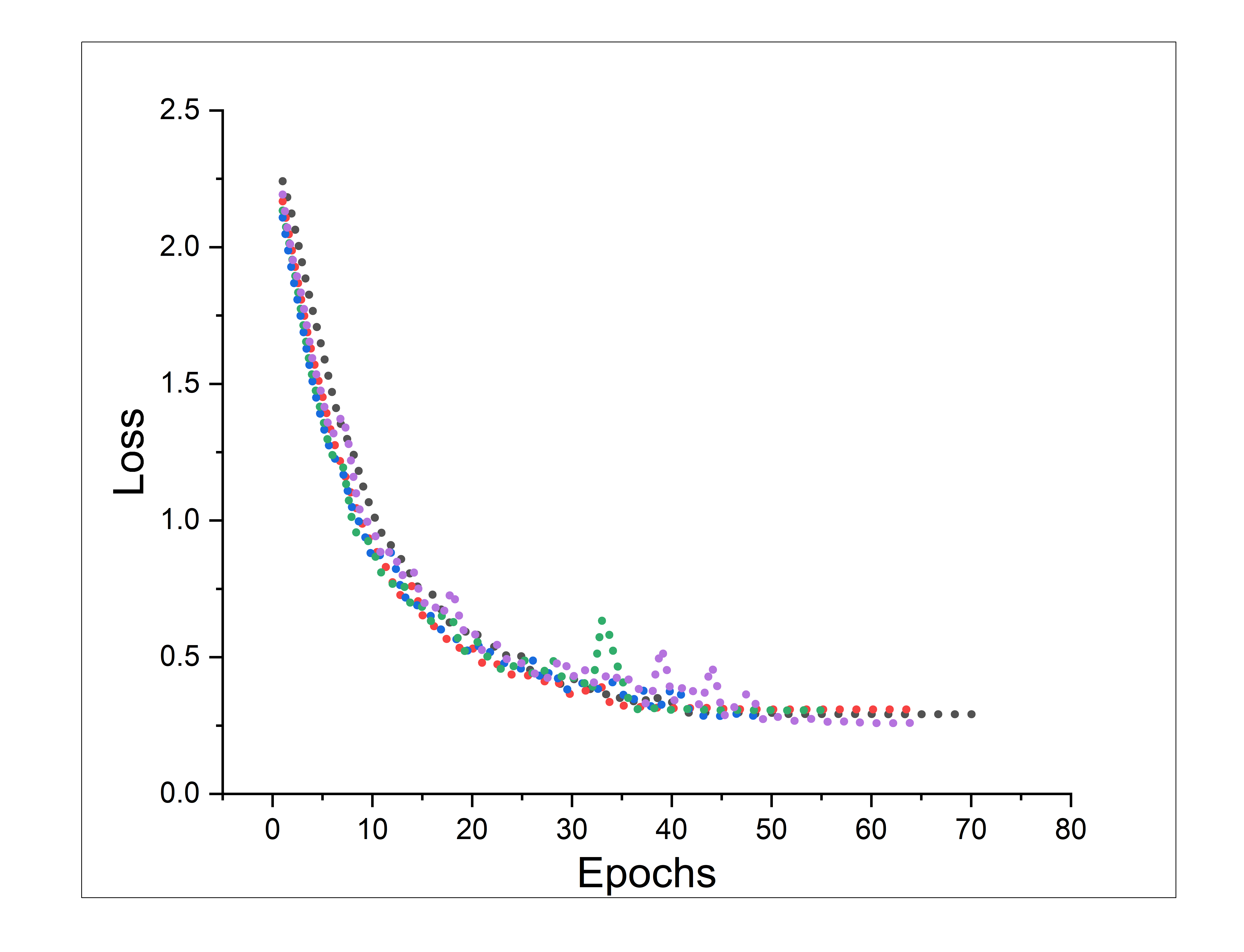}}

\medskip
\subfloat[$n=10$]{\label{fig:loss_AngE_NE_n=12}
\includegraphics[width=.45\textwidth]{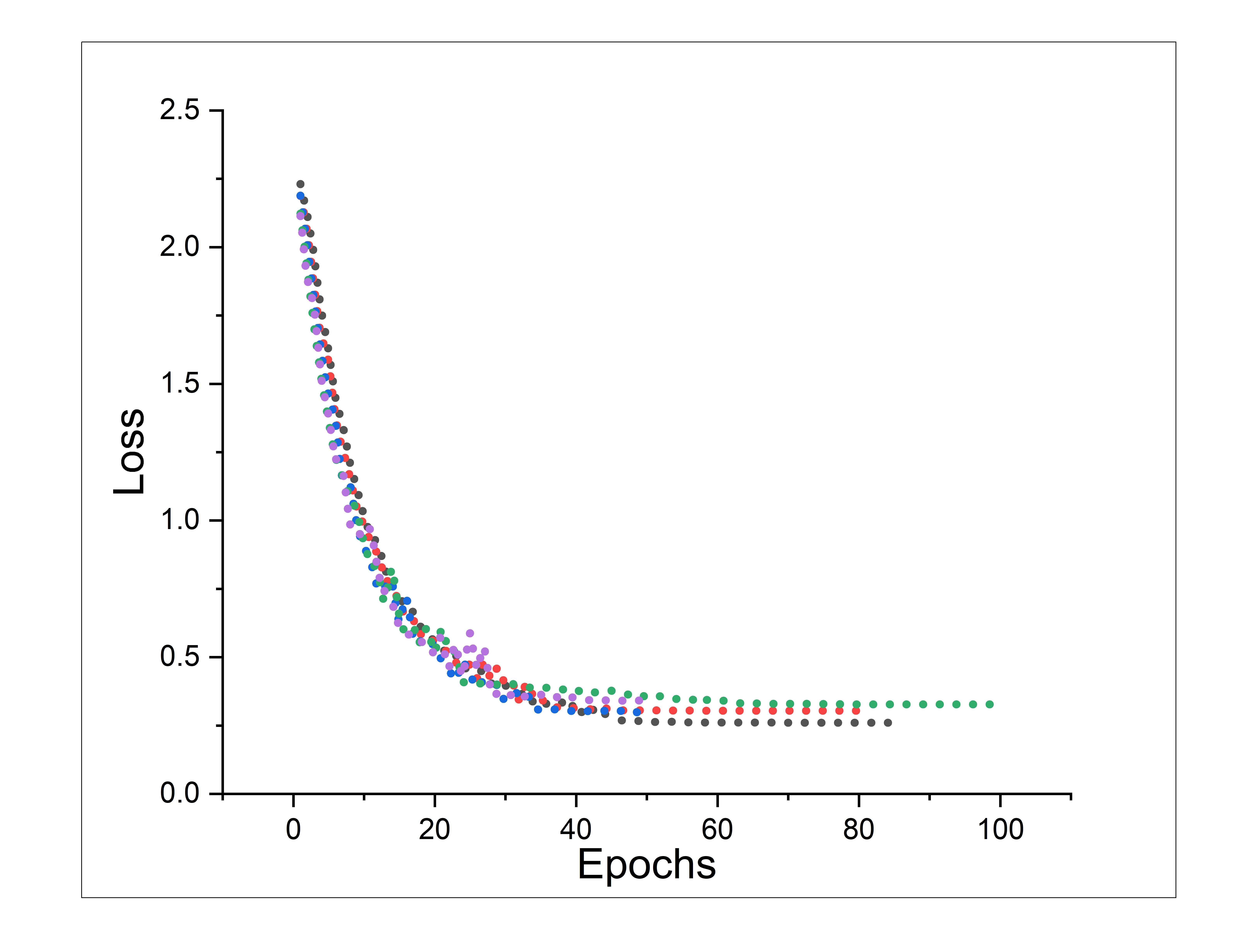}}\quad
\subfloat[$n=12$]{\label{fig:loss_AngE_NE_n=14}
\includegraphics[width=.45\textwidth]{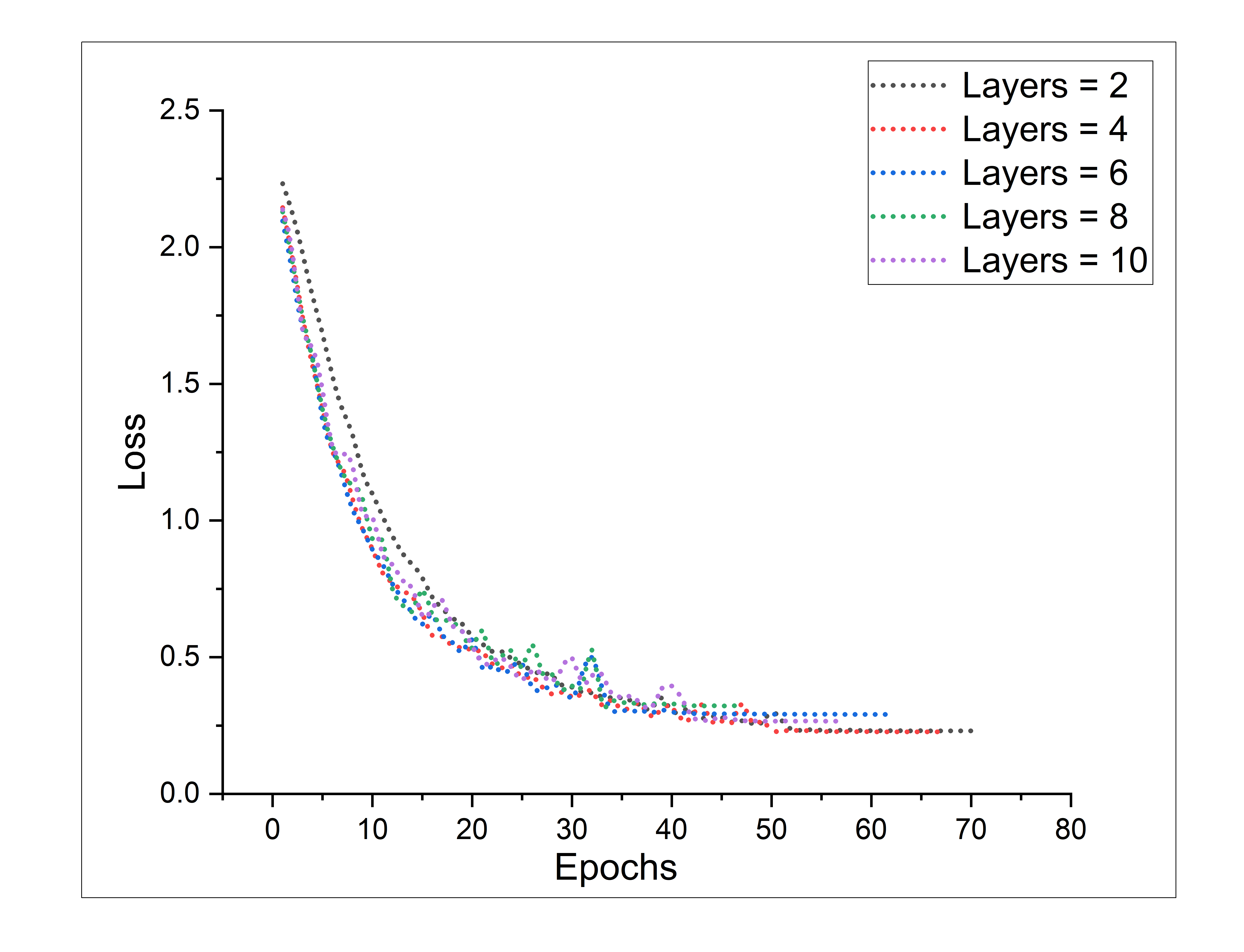}}

    \caption{Unentangled Ansatz: Loss trends for all $n$ and $m$ for angle encoding.}
    \label{fig:Loss_all_AngE_NE}
\end{figure}

We now present the trainability vs. expressibility analysis for unentangled ansatz when the data is being encoded in qubit rotation angles. Since the angle encoding requires the size of input data equal to that of number of qubits, PCA is applied to reduce input feature dimension. The experiments are performed from Figure \ref{fig:experiments}, starting with $n=8$.  The corresponding accuracy and loss convergence results for all the experiments are shown in Figure \ref{fig:AngE_NE_all_acc} and \ref{fig:Loss_all_AngE_NE} respectively.

We observe that the removal of entanglement in case of angle encoding affects the trainbility vs expressibility of HQNNs.
The loss convergence in all $n$ and $m$ (Figure \ref{fig:Loss_all_AngE_NE}), is not quite distinguishable and therefore we observe accuracy (Figure \ref{fig:AngE_NE_all_acc}) to analyze the effect of $n$ and $m$ on trainability.
Unlike angle encoding for entangled ansatz, where the expressibility tends to slightly reduce for bigger $n$ even with an increase in input features, for unentangled ansatz, the allowed circuit depth is greater, eventually leading to have more trainable parameters in hidden quantum layers, resulting in a better overall performance for relatively bigger $n$ and $m$. 
Moreover, for entangled ansatz smaller width quantum layers ($n=8$), have better accuracy with relatively wider quantum layers ($m=8$), as shown in Figure \ref{fig:AngE_WE_all_accuracies}, whereas for unentangled ansatz, the HQNN has a comparable performance for almost all the circuit depths \ref{fig:AngE_NE_all_acc}. 
For wider quantum layers ($n=14$) in unentangled ansatz, again all the depths have comparable performance. However, a smaller depths ($m=2$) deem more feasible, because not only it has a slightly better performance, it would also require less training time compared to deeper layers. 
In conventional ML, it is always conjectured that, for a reasonably complex model, increasing the input data yields better performance. 
In HQNNs, when the data is being encoded in rotation angles, we are bound to increase the input feature dimension, if the quantum layers width is required to be increased.

From the experimental results presented in this paper, we concur that for entangled ansatz, the HQNNs does not follow the speculated belief (in conventional ML) of better performance with more input data (Figure \ref{fig:AngE_WE_all_accuracies}), because of the so-called phenomenon of BP due to an increase in hidden quantum layers width. However, for unentangled ansatz, we observe that this very presumption (\emph{more data = better accuracy}), works and increasing the input feature dimension does infact lead to a better overall performance. Based on these observation, we can safely state that the removal of entanglement in case of angle encoding can potentially avoid (or delay) the BP. Furthermore, unlike the case of amplitude encoding for unentangled ansatz, where an increase in $n$ and almost all corresponding $m$, the model struggles to find the cost minimizing direction (Figure \ref{fig:Loss_all_AmpE_NE}), in case of angle encoding the journey to converge to minimum in cost function landscape is more smoother as shown in Figure \ref{fig:Loss_all_AngE_NE}. Lastly, although the learning process in unentangled ansatz for angle encoding is more reliable than amplitude encoding, it also exhibits some insensitivity to regularization technique specifically for bigger $n$ which can be seen in Figure \ref{fig:loss_AngE_NE_n=12}. However, the insensitivity to regularization is more prominent in case of amplitude encoding (Figure \ref{fig:Loss_all_AmpE_NE}).

Based on the trainability vs. expressibility analysis of both entangled and unentangled ansatzes, we concur that entanglement does affect the HQNNs training and eventually their performance, in correspondence with the underlying encoding strategy. This calls for a rather straight-forward comparison of ansatzes (with and without entanglement) to easily understand the role of entanglement in HQNNs. The comparison is presented in the following section. 


\subsection{Effect of Entanglement in HQNNs} \label{sec:ansatz_comparison}
In previous sections, the analysis of BP existence in HQNNs along with their trainability vs. expressibility is presented. However, that analysis does not explicitly highlight the importance of entanglement inclusion/removal in the underlying quantum layers. Entanglement is an important fundamental property of quantum mechanics and is a key to construct expressible PQCs in HQNNs. Consequently, it is vital to understand the role of entanglement in HQNNs for real-world applications.
In this section, we compare the results obtained for both ansatzes and analyze if entanglement plays any role in overall performance of HQNNs. We observe that the entanglement does affect the HQNN’s performance. However, whether its affect result in performance enhancement or degradation, is dependent on how the data is being encoded. 
Therefore, for understanding the role of entanglement in HQNNs, we compare both ansatz structures first with amplitude encoding 
and then with angle encoding.

\subsubsection{Effect of Entanglement in HQNNs - Amplitude Encoding}
\label{Entanglement_role_Amp}    
\begin{figure}[H]
\centering

\subfloat[$n=6$]{\label{fig:n=6_amp}
\includegraphics[width=.45\textwidth]{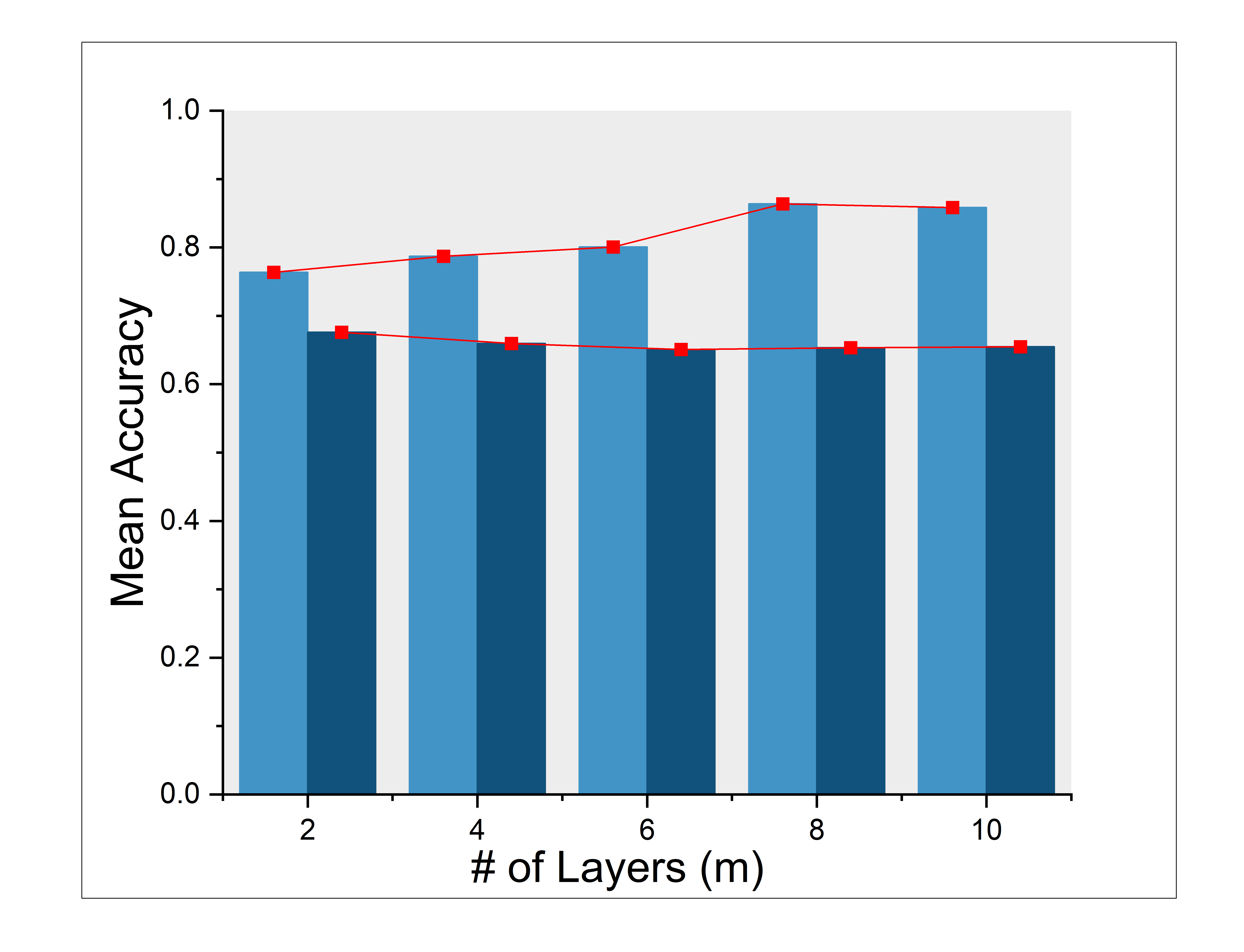}}\quad
\subfloat[$n=8$]{\label{fig:n=8_amp}
\includegraphics[width=.45\textwidth]{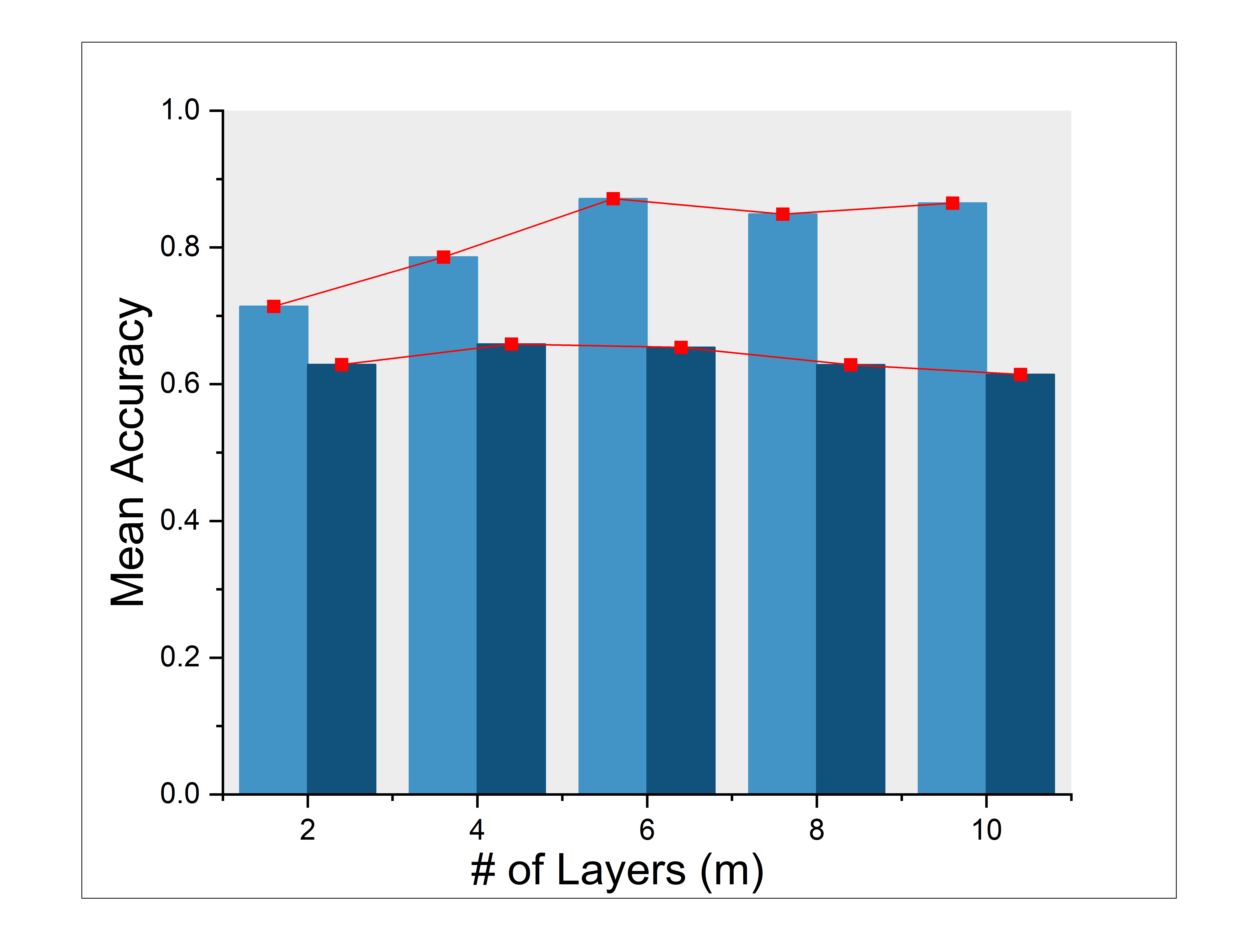}}

\medskip
\subfloat[$n=10$]{\label{fig:n=10_amp}
\includegraphics[width=.45\textwidth]{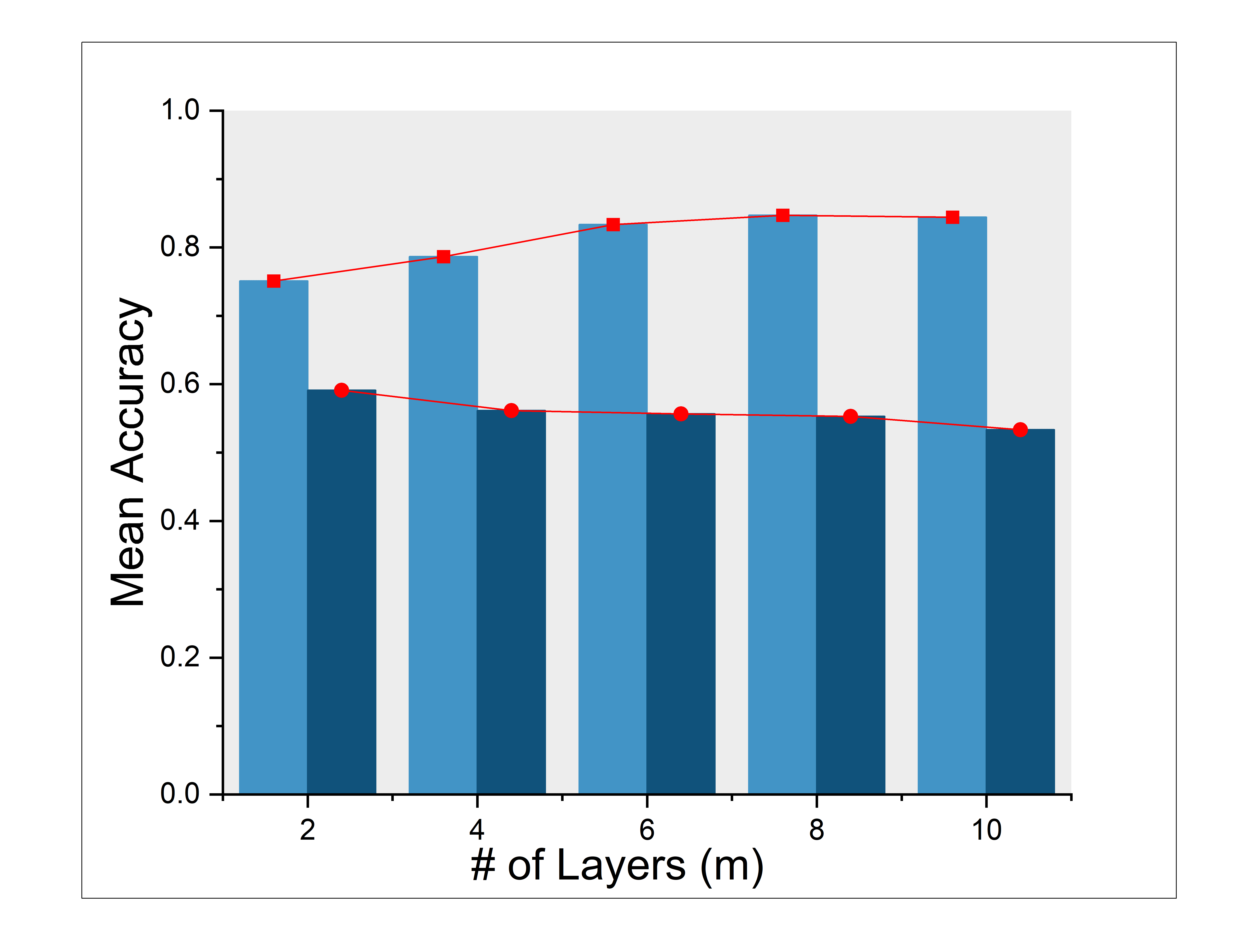}}\quad
\subfloat[$n=12$]{\label{fig:n=12_amp}
\includegraphics[width=.45\textwidth]{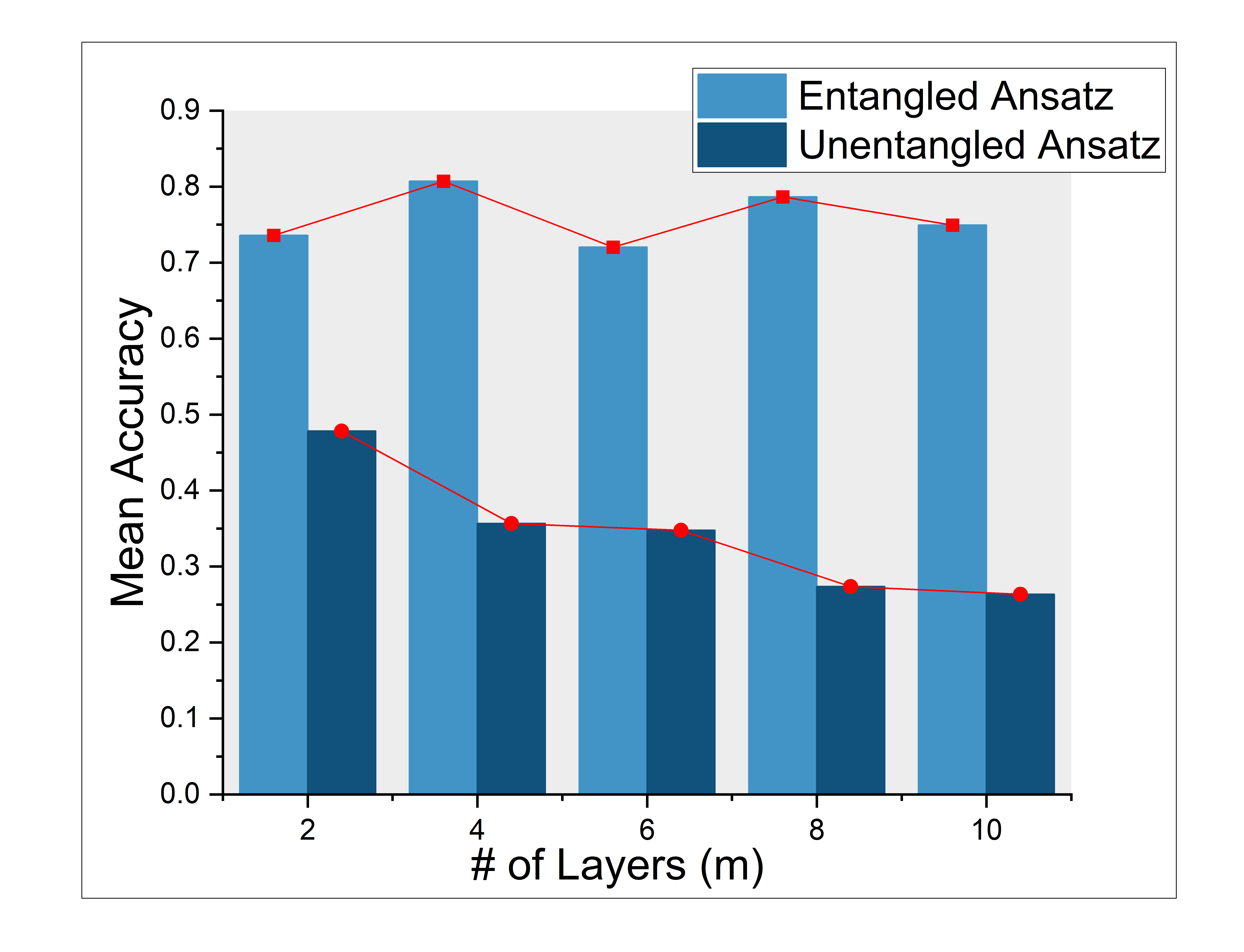}}

\caption{Performance comparison of both ansatz structures for amplitude encoding.} 
    \label{fig:Ansatz_comparison_amp_enc}
\end{figure}

When the classical data is being encoded in qubit state vector and then trained using PQC, the problem of vanishing gradients for unentangled ansatz quantum layers is quite prominent, as discussed in section \ref{sec:BP_results_unentangled_amp}, resulting in significant reduction in HQNN`s performance. A brief comparison of the performance for both ansatzes is shown in Figure \ref{fig:Ansatz_comparison_amp_enc}. We observe that for all $n$ and corresponding $m$, the performance of underlying HQNN with quantum layers for entangled ansatz is significantly better than unentangled ansatz. It can also be observed that the performance degrades as the number of qubits are increased. Based on the results shown and discussed, we concur that the inclusion of entanglement in quantum layers while constructing the HQNNs is not-at-all beneficial, when the data is encoded in qubit amplitudes. Consequently, it is recommended to use single-qubit parameterized unitaries only, when using amplitude embedding approach, for better performance and reduced training time.

\subsubsection{Effect of Entanglement in HQNNs - Angle Encoding}
\label{Entanglement_role_Ang}
\begin{figure}[H]
\centering

\subfloat[$n=8$]{\label{fig:n=8_ang}
\includegraphics[width=.45\textwidth]{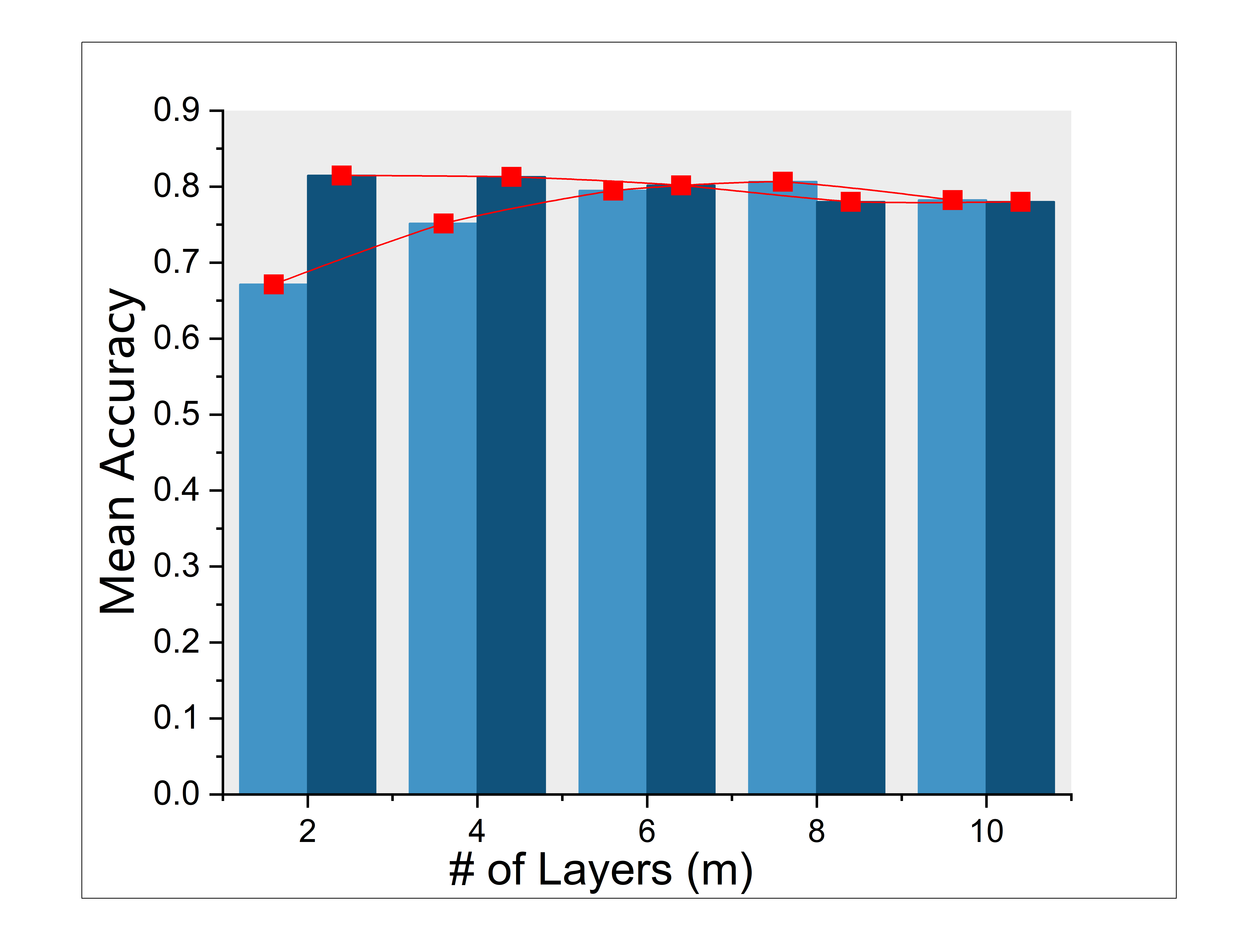}}\quad
\subfloat[$n=10$]{\label{fig:n=10_ang}
\includegraphics[width=.45\textwidth]{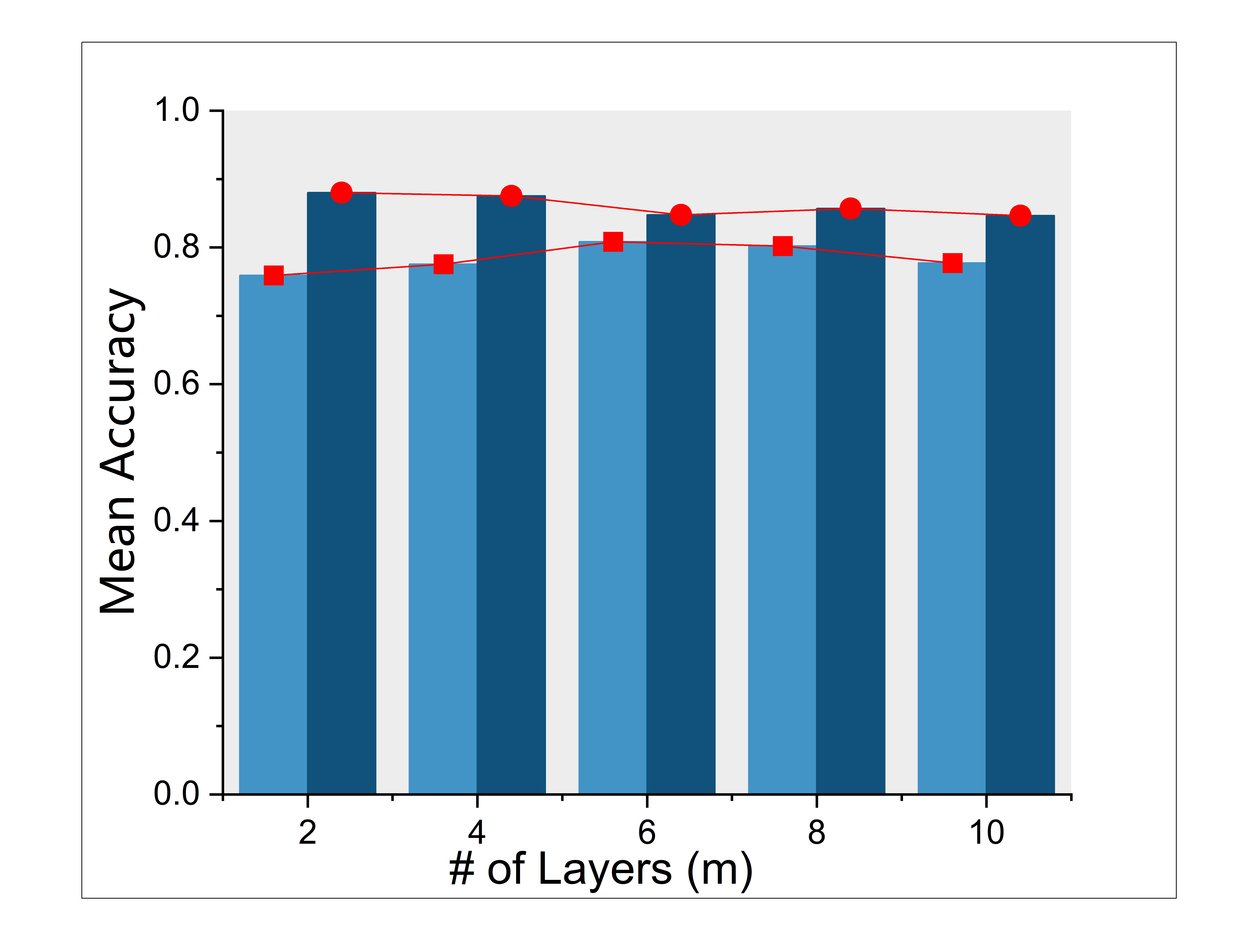}}

\medskip
\subfloat[$n=12$]{\label{fig:n=12_ang}
\includegraphics[width=.45\textwidth]{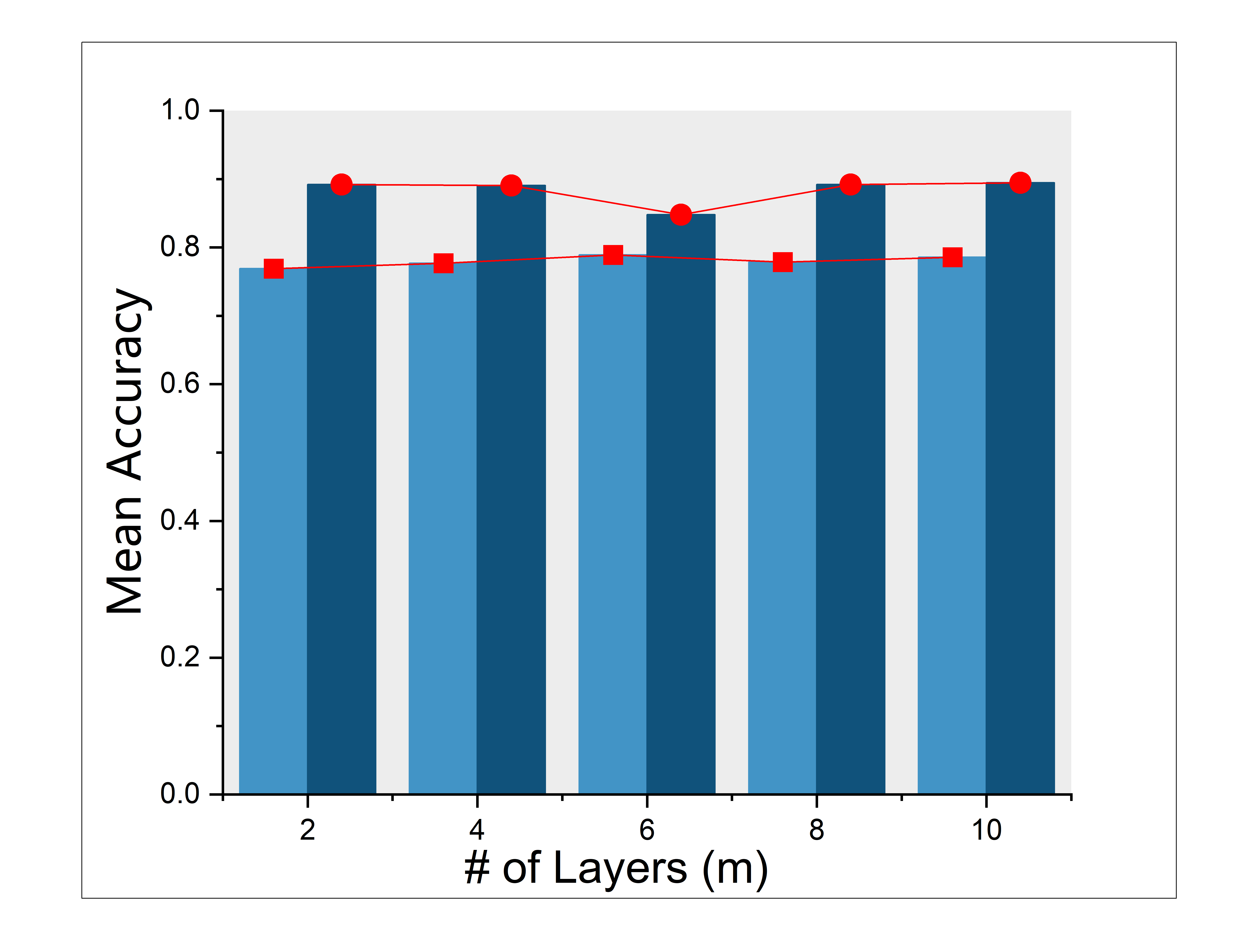}}\quad
\subfloat[$n=14$]{\label{fig:n=14_ang}
\includegraphics[width=.45\textwidth]{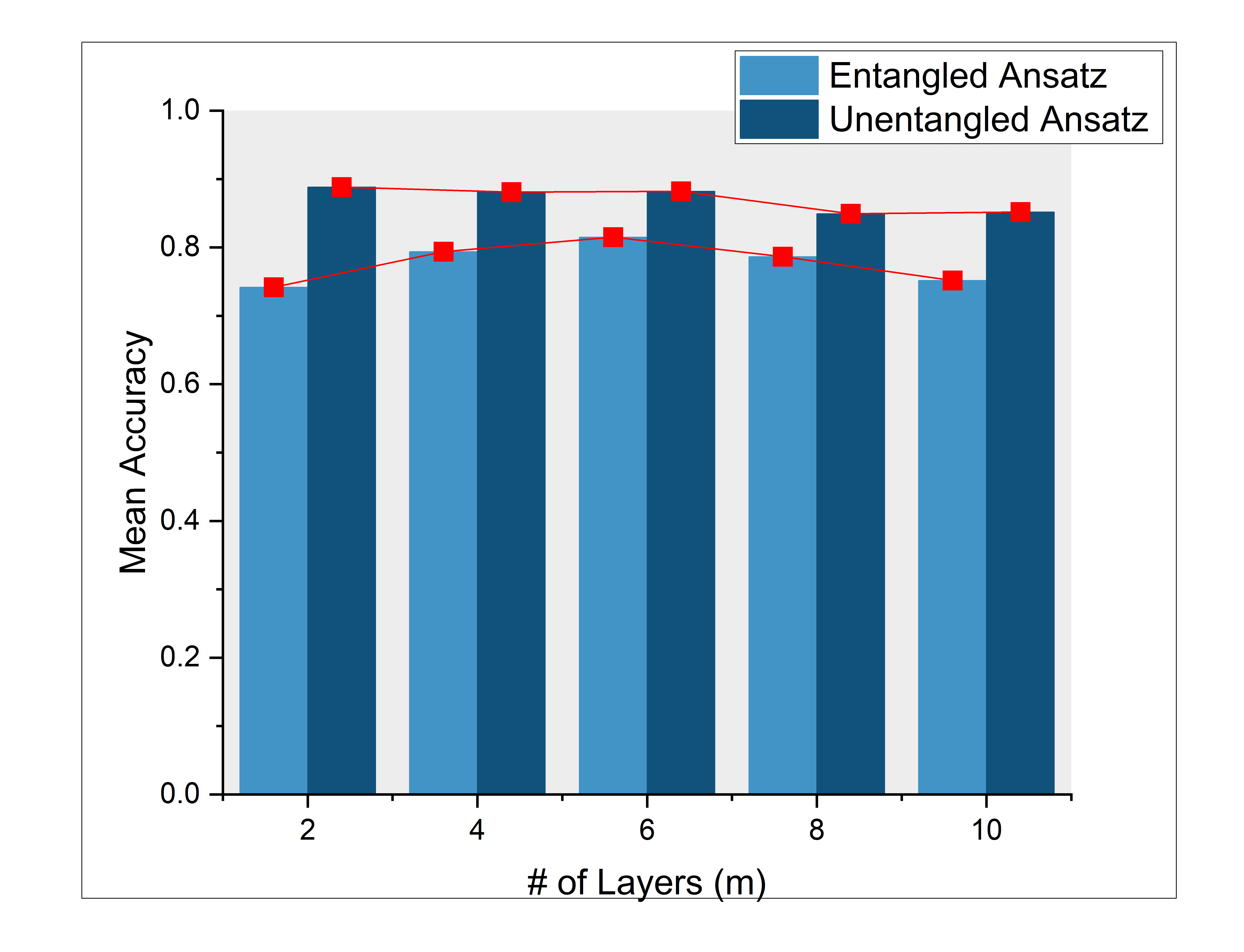}}

\caption{Performance comparison of both ansatz structures for angle encoding.} 
    \label{fig:Ansatz_comparison_angle_enc}
\end{figure}

Entanglement is into the play also, when the classical data is encoded into qubits rotation angles, which is then trained using parameterized quantum layers. However, unlike amplitude encoding, unentangled ansatz quantum layers results in performance enhancement in case of angle encoding. As shown before in Figure \ref{fig:AE_NE_1} and \ref{fig:AE_NE_2}, without any entanglement in quantum layers of underlying HQNNs, the allowed circuit depth is greater than that of quantum layers with entanglement \ref{fig:AngE_ansatz1a} and \ref{fig:AngE_ansatz1b}. The performance comparison of both ansatzes when the data is encoded qubit rotation angles, is shown in Figure \ref{fig:Ansatz_comparison_angle_enc}. Unlike amplitude encoding, here we observe that in general the model performs better when there is no entanglement included in quantum layers for all $n$ and $m$. The performance enhancement becomes more prominent when $n$ increases.  Based on the results, we concur that, removal of entanglement (unentangled ansatz) reasonably enhances the model’s performance when the data is encoded using angle encoding. Moreover, not only the allowed quantum layer(s) depth is greater than that of entangled ansatz quantum layers, but the training time is also reduced.

\subsection{Application-Oriented Evaluation of HQNNs for Classification Tasks}\label{sec:applications}
In previous sections, we discuss the results of our proposed framework based on accuracy and loss convergence. However, for classification tasks (target application in this paper) in ML, accuracy alone is not a sufficient metric to gauge model performance because it only shows the percentage of correct predictions out of the total predictions. Therefore a more diverse set of evaluation metrics are used for classification problems.

Although we consider a multi-class classification problem in this paper, explaining other evaluation metrics (precision, recall and F1-score) for multi-class classification is rather tricky. Therefore, for simplicity, we will demonstrate the need of more diverse evaluation metrics for a binary classification, which are directly applicable to multi-class classification also. 
In binary classification, there are two classes (positive and negative) for which the ML the model aims to predict the correct class. Conventionally, the accuracy in this case would be the sum of correctly predicted classes, regardless of what (correct) class was predicted.

When a positive sample is classified as negative, it is called False Negative (FN), and a negative sample predicted as positive is knows as False Positive (FP). When the positive and negative samples are correctly predicted to their respective classes, this is called True Positive (TP) and True Negative (TN), respectively. Classifying the performance based on these classes allows us to calculate other important metrics, namely: precision and recall. 
Precision tells us what proportion of predictions are truly positive whereas recall tells that what proportion of actual positives are correctly classified. Another important evaluation metric is F1-score, which is just the harmonic mean of precision and recall.  Mathematically, all these metrics can be calculated using the following equations.

\begin{align}
    \text{Recall} &= \frac{TP}{TP+FN}\\
    \text{Precision} &= \frac{TP}{TP+FP}\\
    \text{F1-Score} &= \frac{2\times \text{Recall} \times \text{Precision}}{\text{Recall} + \text{Precision}}
\end{align}

A high precision and recall scores are always desirable but in practice, classifiers are prone to errors and can result in different precision and recall scores. Therefore, a trade-of between these two scores may need to be made, and is highly application dependent. For instance, for a video recommendation system, a high precision would be more desirable to make sure that all potential videos are being recommended to the user. Similarly, a classifier to detect cancer in patients would need a high recall so that as many cancer patients as possible are correctly diagnosed. The F1-score typically is more useful for performance comparison between the classifiers. For instance, in case of two classifiers, where one classifier has a better precision while the other has a better recall, then the F1-score is an appropriate metric to pick the best classifier.   

We now briefly evaluate the HQNNs with respect to target (classification) application(s) for both the ansatz structures and data encodings used in this paper. 
Based on the results discussed in section \ref{sec:results_TvsE} and \ref{sec:ansatz_comparison}, we concur that entangled ansatz structures performs better with amplitude encoding, whereas unentangled ansatz yields better performance with angle encoding.
Therefore, we present the application-oriented evaluation of HQNNs only for the best performing ansatz structures with corresponding encodings (entangled ansatz with amplitude encoding
and unentangled ansatz for angle encoding).
The classifiers with both high precision and high recall (close to overall accuracy) are generally considered to be good classifiers. We observe that in HQNNs, the precision and recall scores follow the same trends as accuracy for all the respective experiments performed in this paper, making them more reliable for a wide range of applications.  

\subsubsection{Application-Oriented Evaluation - Amplitude Encoding} \label{applications_AmpE}

\begin{figure}[htp]
\centering
\subfloat[$n=6$]{\label{fig:allmetrics_AmpE_WE_n=6}
\includegraphics[width=.3\textwidth]{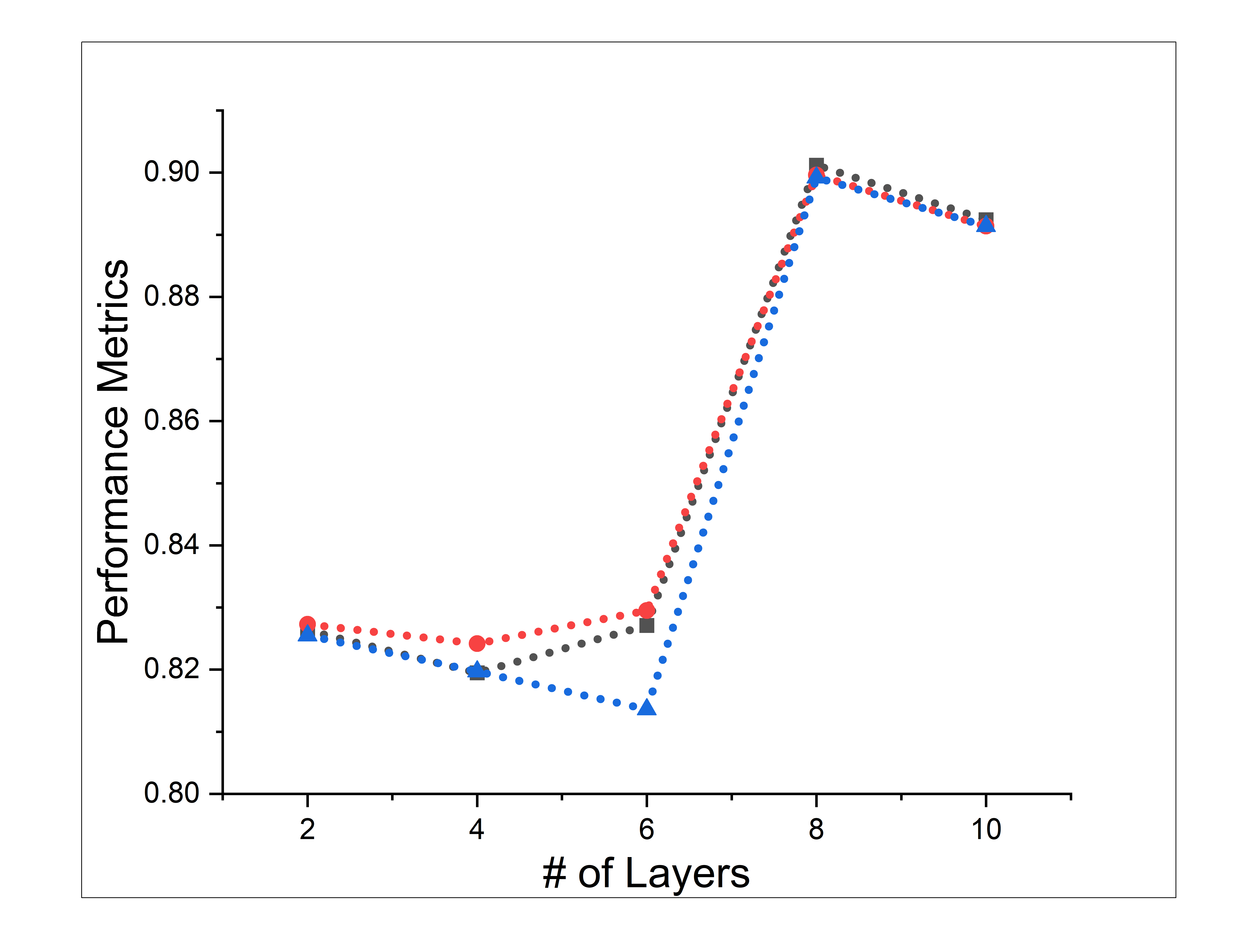}}\quad
\subfloat[$n=8$]{\label{fig:allmetrics_AmpE_WE_n=8}
    \includegraphics[width=.3\textwidth]{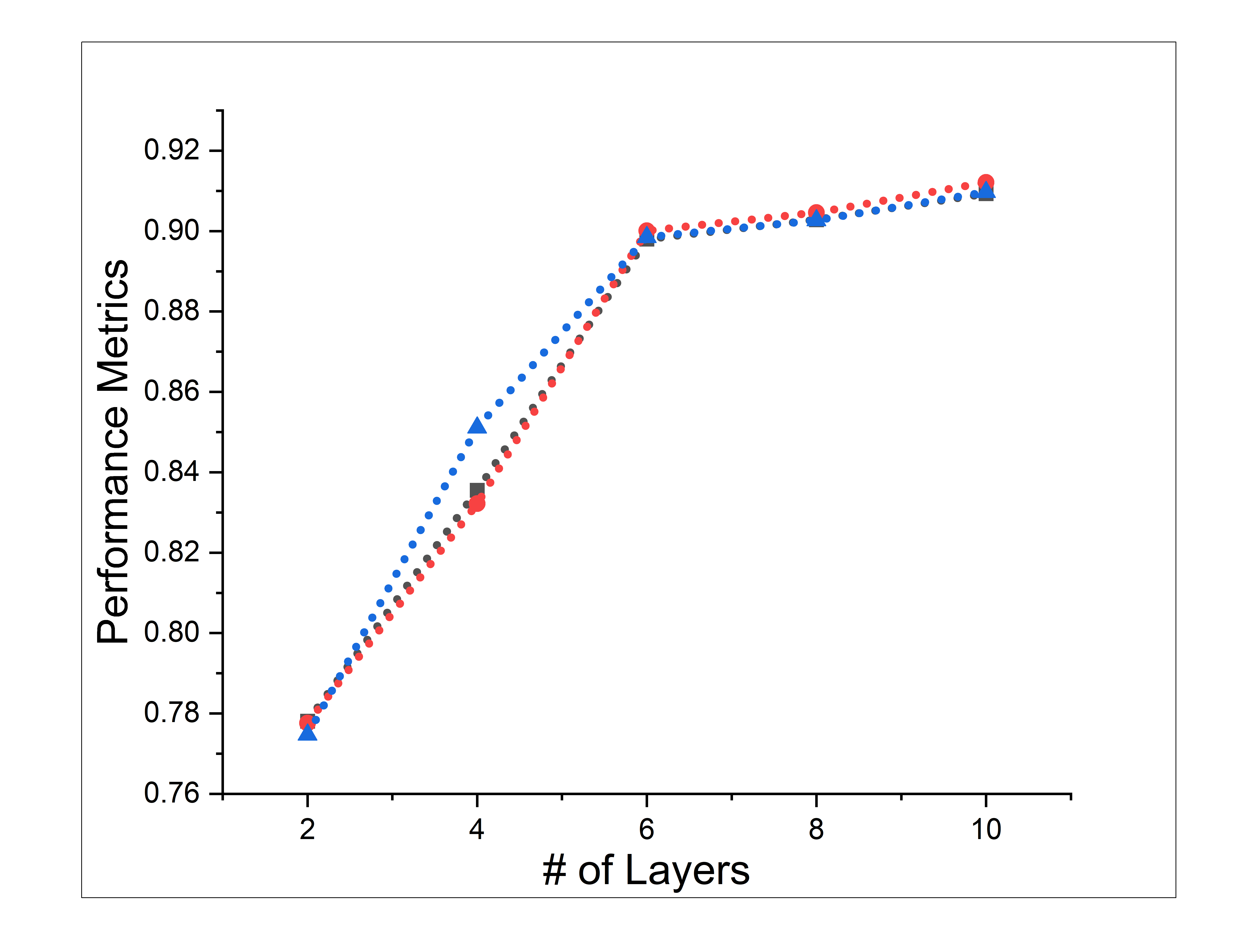}}\quad
\subfloat[$n=10$]{\label{fig:allmetrics_AmpE_WE_n=10}
\includegraphics[width=.3\textwidth]{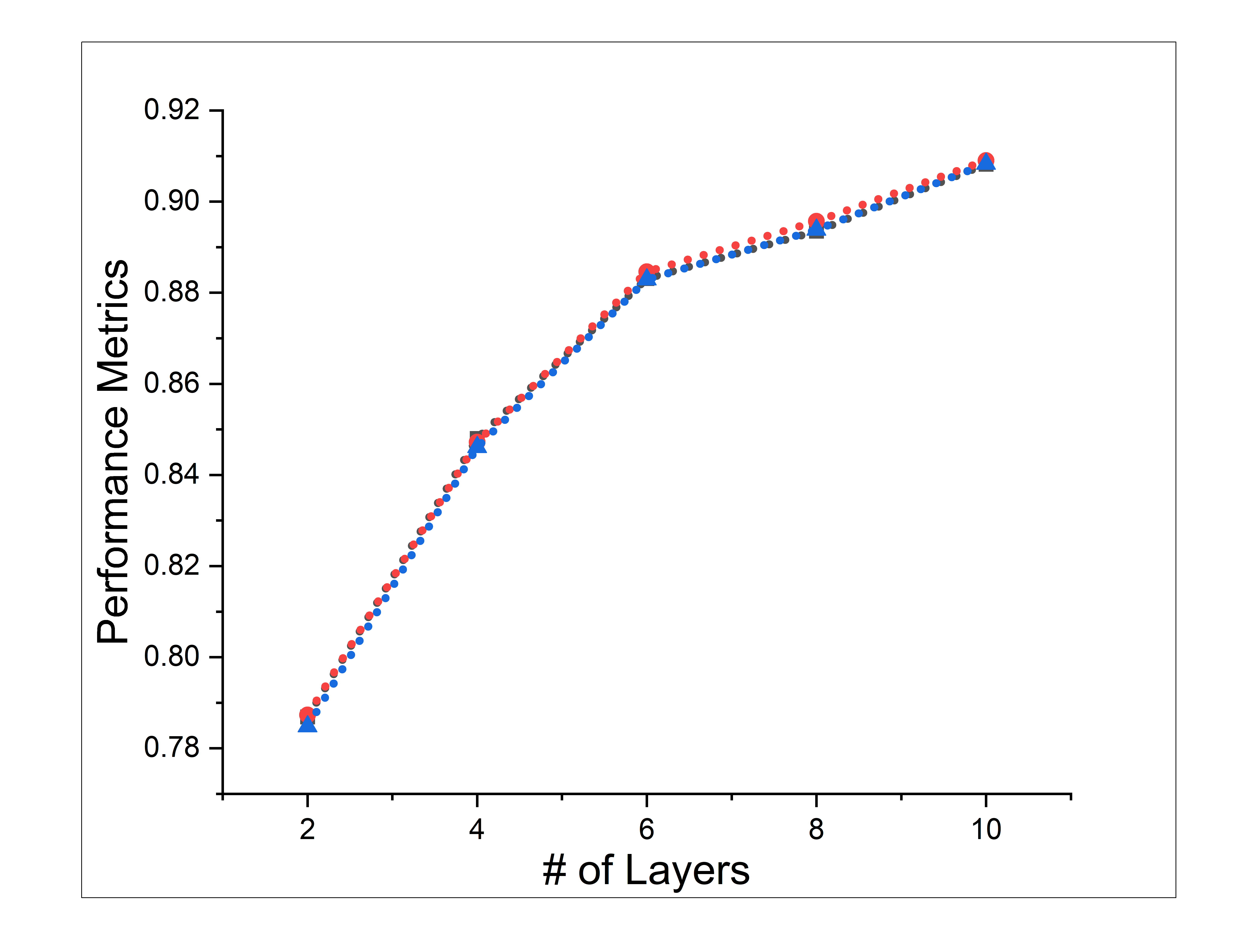}}

\medskip
\subfloat[$n=12$]{\label{fig:allmetrics_AmpE_WE_n=12}
\includegraphics[width=.3\textwidth]{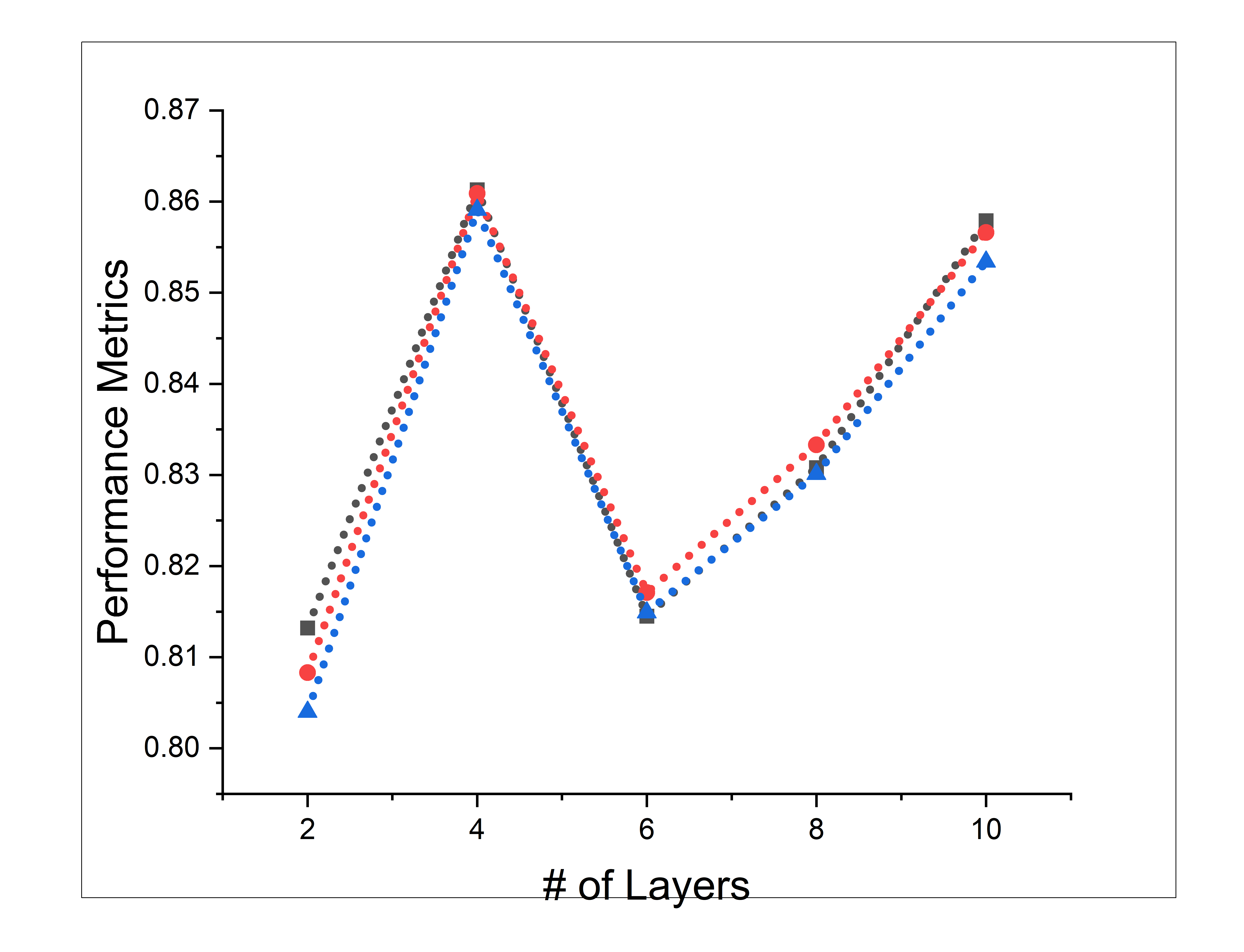}}\quad
\subfloat[$n=14$]{\label{fig:allmetrics_AmpE_WE_n=14}
\includegraphics[width=.3\textwidth]{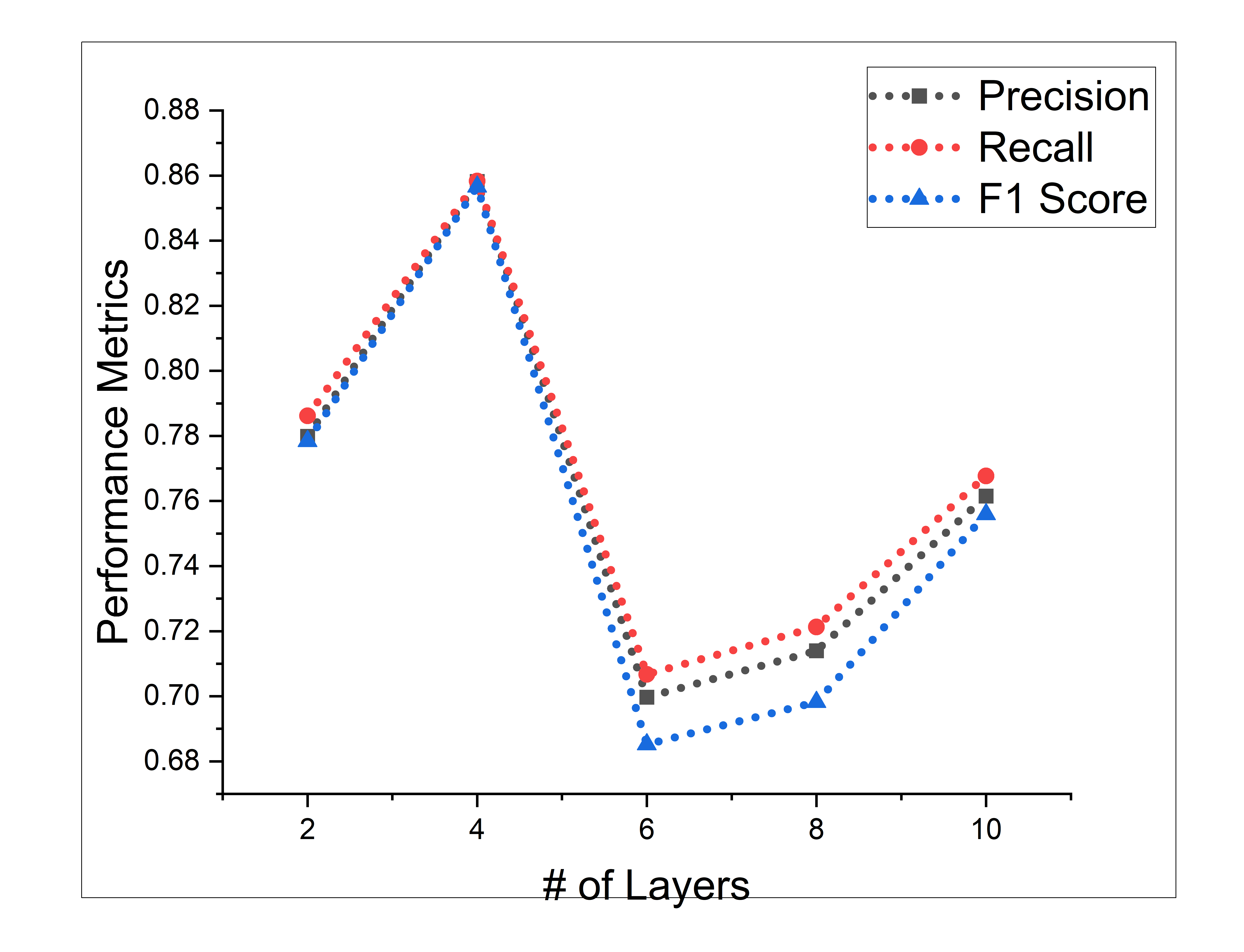}}

\caption{ Precision, Recall and F1-score for fixed $n$ and variable $m$ for amplitude encoding with entangled ansatz} 
\label{fig:all_metrics_AmpE_ansatz-1}
\end{figure}

We first present the application-oriented analysis for amplitude encoding and entangled ansatz. The precision, recall and F1-score are plotted as a function of $m$ for all $n$, as shown in Figure \ref{fig:all_metrics_AmpE_ansatz-1}.

Based on the results shown in Figure \ref{fig:all_metrics_AmpE_ansatz-1}, we observe that for all $n$ and $m$, the precision, recall and corresponding F1-score increase and decrease with almost the same ratio and can be equally applicable for both high precision and high recall applications. The general trends of precision, recall and F1-scores on entangled ansatz with amplitude encoding follows almost the same trend as that of the accuracy (Figure \ref{fig:AmpE_WE_all_accuracies}). This means that for smaller $n$, we require relatively bigger $m$ (and vice versa), to achiever a better recall and precision scores for the corresponding applications.

\subsubsection{Application-Oriented Evaluation - Angle Encoding} \label{applications_AngE}
\begin{figure}[H]
\centering
\subfloat[$n=8$]{\label{fig:allmetrics_AngE_NE_n=8}
\includegraphics[width=.3\textwidth]{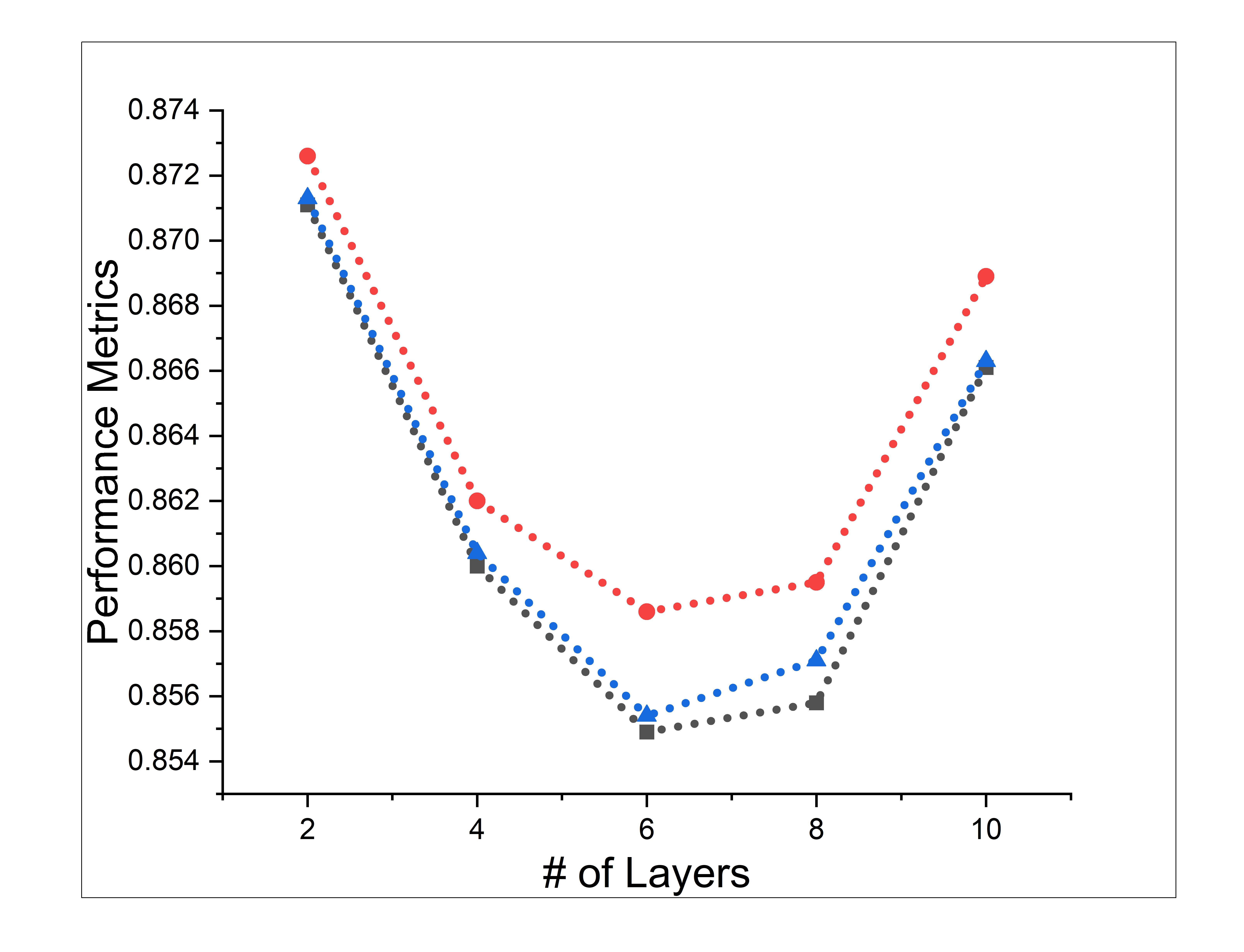}}\quad
\subfloat[$n=8$]{\label{fig:allmetrics_AngE_NE_n=10}
    \includegraphics[width=.3\textwidth]{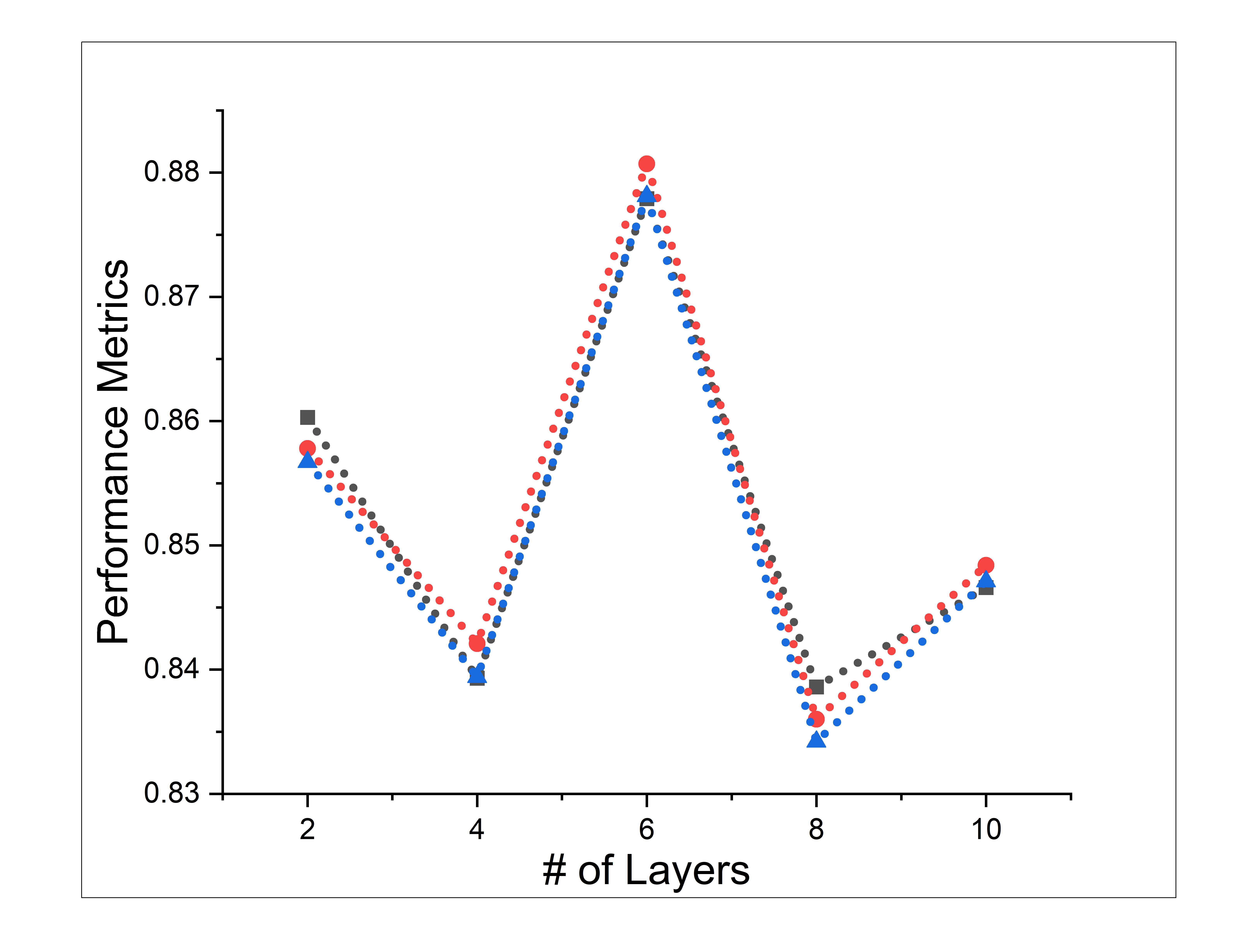}}\quad
\subfloat[$n=12$]{\label{fig:allmetrics_AngE_NE_n=12}
\includegraphics[width=.3\textwidth]{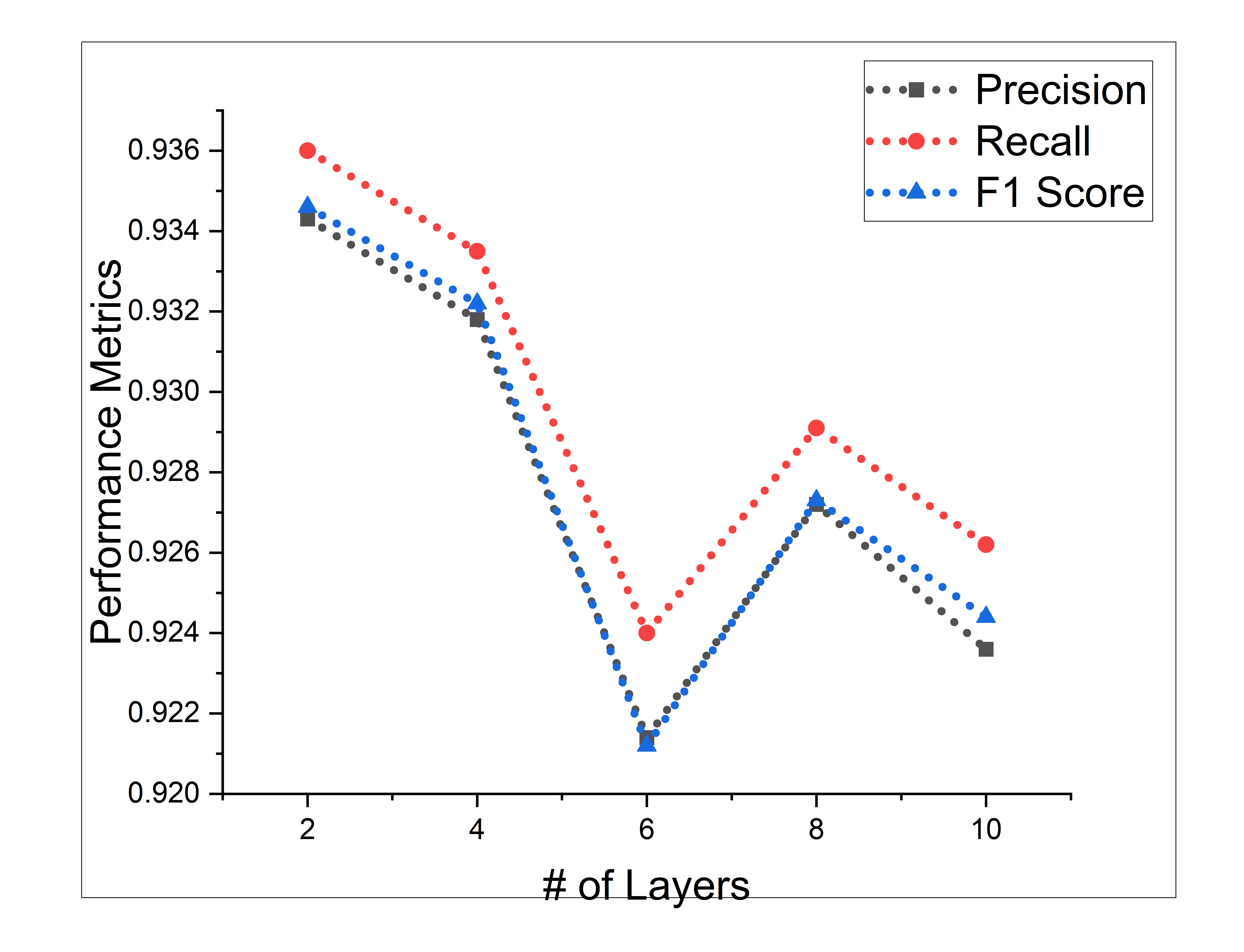}}\quad

    \caption{ Precision, Recall and F1-score for fixed $n$ and variable $m$ for angle encoding with unentangled ansatz}
    \label{fig:all_metrics_AngE_ansatz-2}
\end{figure}

Since the unentangled ansatz performs better with angle encoding, we only consider these results for application oriented analysis. The results of precision, recall and F1-score with fixed $n$ and variable $m$ are presented in Figure \ref{fig:all_metrics_AngE_ansatz-2}. All the performance metrics increase and decrease with same ratio, analogous to the case of amplitude encoding, and hence are equally suitable for either applications (High precision or recall). However, when encoding the data in qubit rotation angles yields slightly better recall than precision for all the training experiments, and is therefore more appropriate for high recall applications.

\section{Significance of Our Framework} \label{sec:significance}
The previous sections of this paper have presented and discussed a framework that focuses on the analysis of three major components of HQNNs. 
These are data encoding, ansatz expressibility, and entanglement inclusion/removal. 
In other words, there can be various possible scenarios coming from constraints on these three components of HQNN.
Therefore, the constraints on these components of HQNNs design, both individually and with respect to each other, are required to be considered. 
In this section, we highlight the significance of our proposed framework for various constraint scenarios by recommending specifications for different parameters of HQNNs design based on our results in section \ref{sec:results_TvsE}. Summaries of the recommendations for each constraint scenario can be found in Appendix \ref{appendix}.

\subsection{No Constraint}\label{sec:scenario_1}
In this scenario, we assume the designer has no constraint on any parameter and has the liberty
to choose any type of data encoding (amplitude or angle), an arbitrary ansatz width and depth (expressibility). Furthermore, the inclusion or removal of entanglement is also not a
constraint. 
In such a scenario, based on the results of our proposed framework, it is recommended to use angle encoding with no entanglement between qubits in the underlying quantum layers. 
Furthermore, based on both overall accuracy and loss convergence, the ansatz width of $n=10$ and depth of $m=4$ is recommended, 
as shown in Figure \ref{fig:AngE_NE_all_acc} and \ref{fig:loss_AngE_NE_n=10}. This is the best obtained result of our framework.


\subsection{Constraint on a Single Parameter} \label{sec:scenario_2}
In this scenario, we consider the possibility when there is a constraint on a single design parameter from data encoding, ansatz expressibility and entanglement in quantum layers. 

\subsubsection{Constraint on Data Encoding.} \label{sec:constraint_Enc} 
Here, we assume that there is a constraint on the data encoding scheme to be used. If it is required to encode the data in qubit amplitudes, then the entanglement is not recommended to be included in the underlying quantum layers. Moreover, the optimal width and depth of quantum layers is $n=6$ and $m=10$, as shown in Figure \ref{fig:AmpE_WE_all_accuracies} and \ref{fig:AmpE_1a} respectively. 
On the other hand, if it is constrained to use angle encoding then the same set of HQNNs can be selected as discussed in section \ref{sec:scenario_1}. 

\subsubsection{Constraint on Ansatz Expressibility.}

Ansatz expressibility has two factors: ansatz width and ansatz depth.
If the constraint is on ansatz width, then for smaller widths ($n=6, 8$), amplitude encoding with moderate depth ($m=6$) is more appropriate to use to improve the accuracy and convergence, as shown in Figure \ref{fig:AmpE_WE_all_accuracies} and \ref{fig:AmpE_2a}. In addition, entanglement is recommended to be included. 
On the other hand, for bigger widths ($n=10, 12 ,14$), angle encoding is a better choice. For the selection of other parameters in this scenario, the discussion in section \ref{sec:scenario_1} can be followed. 

If the constraint is on the depth of quantum layers, in the case of smaller depths ($m=2, 4$), angle encoding with unentangled ansatz works better. Moreover, the allowable width of quantum layers is greater, i.e., $n=10, 12, 14$. Although, it allows for wider quantum layers, using $n=10$ would be more appropriate due to the shorter training time. Similarly, for moderate depth to deeper quantum layers ($m=6,8,10$), both amplitude (with entanglement) and angle encoding (without entanglement) have comparable performance in terms of accuracy. However, with amplitude encoding, the allowable width of the quantum layers is relatively smaller ($n=6$) and angle encoding allows to have a wider quantum layer ($n=10, 12, 14$) and is hence less susceptible to BP, unlike amplitude encoding.


\subsubsection{Constraint on Entanglement.}
For the case when there is a constraint requiring entangling the qubits, amplitude encoding with ansatz width and depth of $n=8$ and $m=6$ can be used. 
On the other hand, when there is a constraint requiring that qubits should not be entangled, then angle encoding performs better, and the same specification as discussed in section \ref{sec:scenario_1} can be picked.

\subsection{Constraint on Two Parameters} \label{sec:scenario_3}
In this scenario, we consider the possibility when there is a constraint on any two of the parameters from data encoding, ansatz expressibility and entanglement inclusion/removal.

\subsubsection{Constraint on Data Encoding and ansatz expressibility.}
The ansatz expressibility has two factors: width and depth of quanutum layers. Therefore, we separately consider constraints on both these expressibility factors along with constraints on data encoding. 
We first consider constraints on the data encoding and ansatz width. In the case of both amplitude and angle encoding, for smaller widths ($n=8, 10$), the appropriate depth is greater ($m=8, 10$), whereas for bigger widths ($n=12, 14$) the appropriate depth reduces to $m=2, 4$. Similarly, in the case of constraint on ansatz depth and data encoding, again there is a trade-off between ansatz depth and width. Furthermore, in both cases (constraint on ansatz width or depth), it is recommended to entangle the qubits when encoding in qubit amplitudes, whereas for angle encoding, entangling qubits does not improve the performance. 

\subsubsection{Constraint on Data Encoding and Entanglement.}
We now consider a setting when there is a constraint on data encoding and entanglement inclusion/removal. If there is a constraint requiring the data to be encoded in qubit amplitudes, then irrespective of whether the qubits are entangled or not, the allowable width of quantum layers is relatively lower, i.e., ($n=6, 8$). The bigger widths lead to performance decline, indicating the presence of BP. However, in the case the qubits are entangled in underlying quantum layers, the allowable depth is bigger ($m=8,10$), whereas in case of no qubit entanglement, the allowable depth is reduced ($m=2, 4$).   

Similarly, When it is constrained to encode the data in qubit rotation angles, then irrespective of entanglement, the appropriate depths are bigger than amplitude encoding, ($n=10,12,14$), making it less prone to BP and eventually lead to better performance. However, in the case of entanglement inclusion, the allowable depth of quantum layers is greater ($m=8, 10$), whereas in the case of no entanglement, the allowable depth is reduced ($m=2, 4$).

\subsubsection{Constraint on expressibility and entanglement.}
We first consider the setting when there is a constraint on the width factor of expressibility while depth can be arbitrary. Moreover, the entanglement inclusion/removal is also constrained. 

When the qubits in underlying quantum layers are constrained to be entangled and the corresponding width of quantum layers is required to be smaller ($n=8, 10$), then amplitude encoding is better, with deeper quantum layers ($m=10$). However, if the required width of quantum layers is greater ($n= 12, 14$) along with entanglement inclusion, then angle encoding is relatively better and the appropriate depth of underlying quantum layers is smaller ($m=2$).
On the other hand, when the entanglement is constrained to be removed from quantum layers, then irrespective of width constraints, angle encoding is better than amplitude encoding and the appropriate depth is typically smaller ($m=2, 4$), for all corresponding widths used in this paper.

We now consider the setting when there is a constraint on depth factor of expressibility along with entanglement inclusion/removal. 
When the constraint is to have smaller depth ($m=2$ or $4$), and entanglement is constrained to be included also, then both the encodings (amplitude or angle) achieve similar performance and allows to have more wider quantum layers ($n=10$). However, when the constraint is to have more deeper quantum layers ($m=6, 8$ or $10$), along with entanglement inclusion, then amplitude encoding performs better than angle encoding. Moreover, the appropriate width of quantum layers is also greater ($n=8$).
When the entanglement is required to be removed, then irrespective of the depth, angle encoding is better than amplitude encoding and the appropriate width of underlying quantum layers is typically higher ($n=12$ or $14$).

\subsection{Constraint on All Three Parameters} \label{sec:scenario_4}

In this scenario, we consider the setting where there is a constraint on all the parameters (data encoding, ansatz expressibility and entanglement) of HQNN. As discussed earlier, the ansatz expressibility has two factors, i.e., width and depth of quantum layers. Therefore, we separately consider constraints on both these expressibility factors along with constraints on other data encoding and entanglement. 

\subsubsection{Constraint on data encoding, ansatz width and entanglement.}
When there is a constraint on ansatz width along with data encoding and entanglement inclusion/removal. 
When it is constrained to encode the data in qubit amplitudes, entanglement is constrained to be included, and width is required to be smaller (typically $n=6,8,10$), then to achieve a relatively better performance, the depth of underlying quantum layers should be greater (typically $m=10$). However, if the width is constrained to be bigger ($n=12, 14$), then the appropriate depth reduces to $m=4$ to achieve better performance. Similarly, when the entanglement is required to be removed, then irrespective of the constrained width, the appropriate depth is typically smaller ($m=2$).

On the other hand, when it is constrained to encode the data in qubit rotation angles, entanglement is required to be included and ansatz width is required to be smaller (typically $n=8,10$), then the appropriate ansatz depth is relatively greater ($m=8$). However, when the ansatz width is constrained to be bigger (typically $n=12,14$), then analogous to the case of amplitude encoding, the appropriate ansatz depth is reduced (typically $m=4$). Similarly, when the entanglement is required to removed and the data is required encoded in qubit angles, then irrespective of the constrained width, the smaller ansatz depth ($m=2$) is more appropriate to use.

\subsubsection{Constraint on data encoding, ansatz depth and entanglement.}
We now consider the setting when the ansatz depth is constrained along with data encoding and entanglement inclusion/removal. In such a setting the ansatz width can be arbitrary.
When the data is constrained to be included in amplitude encoding, entanglement is also required to be included and the ansatz depth is constrained to be smaller (typically $m=2,4$) then the appropriate ansatz width is slightly bigger ($n=10, 12$). However, when the ansatz depth is constrained to be moderate, i.e., $m=6$ the appropriate ansatz width reduces to $n=8$, which further reduces to $n=6$ for more deeper ansatz ($m=8,10$). Similarly, when it is constrained to remove entanglement, then irrespective of the depth constraint, the moderate ansatz width ($n=6$) is more appropriate to achieve relatively better performance. 

On the other hand, when the constraint is to encode the data in qubit angles, entanglement is required to be included and the ansatz depth is constrained to be smaller moderate (typically $m=2, 4,6$), the appropriate width is relatively bigger ($n=12,14$). However, if the ansatz depth is constrained to be bigger ($m=8,10$), the appropriate width reduces to $n=8$. 
Similarly, when the entanglement is constrained to be removed and data is constrained to be encoded in qubit angles, then irrespective of the depth constraint, the appropriate ansatz width is bigger, i.e, $n=12, 14$.  

\section{Conclusion} \label{conclusion}

Quantum machine learning (QML) has recently emerged as one of the potential applications of quantum computing, attempting to improve the classical machine learning by harnessing quantum mechanical phenomena. In QML, quantum neural networks (QNNs) are widely being explored because of the unparalleled success of their classical counterparts, namely neural networks (NNs). However, the practical applicability of QNNs is challenged by the phenomenon of barren plateaus (BP), where the gradients of parameters become exponentially small as the system size increases potentially making QNNs untrainable.
To this end, the primary components of QNNs, i.e., data encoding, ansatz expressibility, and entanglement between qubits have been identified as the potential sources of BP. All these components have been studied individually from the aspect of BP, however, these components exist simultaneously in a practical setting. Therefore, investigating their joint effect, with respect to each other is of significant importance for practical applications. 

In this paper, we propose a framework to empirically investigate the holistic effect of all the aforementioned components of QNNs for a practical application namely; multi-class classification. In a practical setting, because of the limitations of noisy intermediate-scale quantum devices, hybrid quantum neural networks (HQNNs) are widely being used to explore the potential quantum advantage in QNNs. Since the HQNNs completely replicate the general QNN architecture (with some classical input pre- and post-processing), the analysis of quantum parts of HQNN can be directly applicable to QNNs. 
The HQNNs we have used for our analysis, consist of the following sequence of operations; 1) input dimensionality reduction, 2) qubit initialization, 3) data encoding (classical to quantum feature mapping), 4) quantum ansatz (parameterized quantum circuit), 5) qubit measurements and 6) dense classical neuron layer to post-process the qubit measurement results and get the output.  

Our analysis focuses on the data encoding and quantum layers (their expressibility and entanglement inclusion removal), which are the main components of QNNs.  
For data encoding, we use two frequently used data encoding techniques, namely: amplitude and angle encoding. For ansatz expressibility, we change the width ($n$) and depth ($m$) of quantum layers. 
We train our HQNNs with underlying ansatz for $n$ and $m$. We consider two similar ansatz structures, entangled ansatz (which contains single-qubit parameterized unitaries and nearest neighbor entanglement), and unentangled ansatz (which contains single-qubit parameterized unitaries only). We first benchmark the mean accuracy of the training experiments and demonstrate the existence of BP in HQNNs. We observe that the BP in HQNNs does not follow a direct relation with the number of qubits but is dependent on the overall expressibility of quantum layers.   We then benchmark the overall accuracy and loss convergence of HQNNs and perform a comprehensive trainability vs. expressibility analysis. This analysis shows how the ansatz expressibility plays a role in the overall performance of HQNNs from the aspect of BP and how deep an ansatz can be for a given width before experiencing the BP, for each encoding. 

Furthermore, we observed that entanglement plays a role in the training landscapes of HQNNs and is dependent on the encoding type. When the data is encoded in qubit state vector, the entangled ansatz achieves better accuracy than the unentangled ansatz, demonstrating a positive impact of entanglement on the trainability of HQNNs in the case of amplitude encoding. On the contrary, when the data is encoded into qubit rotation angles, unentangled ansatz yields better accuracy than entangled ansatz, demonstrating a negative impact of entanglement on HQNNs trainability. 
We also briefly evaluate the HQNNs for classification applications considering other important evaluation metrics for classification problems, namely: precision, recall and F1-score. Finally, we illustrate the significance of our proposed framework by providing recommendations for different constraint scenarios (both alone and combined) on data encoding, ansatz expressibility and entanglement inclusion/removal in the underlying quantum layers.



\bibliographystyle{tfnlm}
\bibliography{interactnlmsample}

\begin{thebibliography}{10}
\providecommand{\url}[1]{\normalfont{#1}}
\providecommand{\urlprefix}{Available from: }

\bibitem{preskil:2018}
Preskill~J. Quantum {C}omputing in the {NISQ} era and beyond. {Quantum}. 2018
  Aug;\hspace{0pt}2:79.
  \urlprefix\url{https://doi.org/10.22331/q-2018-08-06-79}.

\bibitem{Arute:2019}
Arute~F, Arya~K, Babbush~R, et~al. Quantum supremacy using a programmable
  superconducting processor. Nature. 2019;\hspace{0pt}574(7779):505--510.

\bibitem{zhong:2020}
Zhong~HS, Wang~H, Deng~YH, et~al. Quantum computational advantage using
  photons. Science. 2020 Dec;\hspace{0pt}370(6523):1460–1463.
  \urlprefix\url{http://dx.doi.org/10.1126/science.abe8770}.

\bibitem{Wu:2021}
Wu~Y, Bao~WSe. Strong quantum computational advantage using a superconducting
  quantum processor. Phys Rev Lett. 2021 Oct;\hspace{0pt}127:180501.
  \urlprefix\url{https://link.aps.org/doi/10.1103/PhysRevLett.127.180501}.

\bibitem{Madsen:2022}
Madsen~LS, Laudenbach~Fe. Quantum computational advantage with a programmable
  photonic processor. Nature. 2022;\hspace{0pt}606(7912):75--81.
  \urlprefix\url{https://doi.org/10.1038/s41586-022-04725-x}.

\bibitem{Bergholm:2018}
Bergholm~V, Izaac~Je. Pennylane: Automatic differentiation of hybrid
  quantum-classical computations ; 2018.
  \urlprefix\url{https://arxiv.org/abs/1811.04968}.

\bibitem{Peruzzo:2014}
Peruzzo~A, McClean~J, Shadbolt~P, et~al. A variational eigenvalue solver on a
  photonic quantum processor. Nature Communications. 2014 jul;\hspace{0pt}5(1).
  \urlprefix\url{https://doi.org/10.1038\%2Fncomms5213}.

\bibitem{LaRose:2019}
LaRose~R, Tikku~A, O'Neel-Judy~{\'{E}}, et~al. Variational quantum state
  diagonalization. npj Quantum Information. 2019 jun;\hspace{0pt}5(1).
  \urlprefix\url{https://doi.org/10.1038\%2Fs41534-019-0167-6}.

\bibitem{anschuetz:2018}
Anschuetz~ER, Olson~JP, Aspuru-Guzik~A, et~al. Variational quantum factoring.
  arXiv. 2018;\hspace{0pt}\urlprefix\url{https://arxiv.org/abs/1808.08927}.

\bibitem{Farhi:2014}
Farhi~E, Goldstone~J, Gutmann~S. A quantum approximate optimization algorithm.
  arXiv. 2014;\hspace{0pt}\urlprefix\url{https://arxiv.org/abs/1411.4028}.

\bibitem{Klco:2018}
Klco~N, Dumitrescu~EF, McCaskey~AJ, et~al. Quantum-classical computation of
  schwinger model dynamics using quantum computers. Phys Rev A. 2018
  Sep;\hspace{0pt}98:032331.
  \urlprefix\url{https://link.aps.org/doi/10.1103/PhysRevA.98.032331}.

\bibitem{Klco:2019}
Klco~N, Savage~MJ. Digitization of scalar fields for quantum computing. Phys
  Rev A. 2019 May;\hspace{0pt}99:052335.
  \urlprefix\url{https://link.aps.org/doi/10.1103/PhysRevA.99.052335}.

\bibitem{Sharma:2020}
Sharma~K, Khatri~S, Cerezo~M, et~al. Noise resilience of variational quantum
  compiling. New Journal of Physics. 2020 apr;\hspace{0pt}22(4):043006.
  \urlprefix\url{https://doi.org/10.1088\%2F1367-2630\%2Fab784c}.

\bibitem{McClean:2018}
McClean~JR, Boixo~S, Smelyanskiy~VN, et~al. Barren plateaus in quantum neural
  network training landscapes. Nature Communications. 2018
  nov;\hspace{0pt}9(1).
  \urlprefix\url{https://doi.org/10.1038%2Fs41467-018-07090-4}.

\bibitem{Grant:2019}
Grant~E, Wossnig~L, Ostaszewski~M, et~al. An initialization strategy for
  addressing barren plateaus in parametrized quantum circuits. Quantum. 2019
  dec;\hspace{0pt}3:214.
  \urlprefix\url{https://doi.org/10.22331%2Fq-2019-12-09-214}.

\bibitem{Cerezo:2021}
Cerezo~M, Sone~A, Volkoff~T, et~al. Cost function dependent barren plateaus in
  shallow parametrized quantum circuits. Nature Communications. 2021
  mar;\hspace{0pt}12(1).
  \urlprefix\url{https://doi.org/10.1038%2Fs41467-021-21728-w}.

\bibitem{McClean:2016}
McClean~JR, Romero~J, Babbush~R, et~al. The theory of variational hybrid
  quantum-classical algorithms. New Journal of Physics. 2016
  Feb;\hspace{0pt}18(2):023023.
  \urlprefix\url{https://doi.org/10.1088/1367-2630/18/2/023023}.

\bibitem{Biamonte:2021}
Biamonte~J. Universal variational quantum computation. Physical Review A. 2021
  mar;\hspace{0pt}103(3).
  \urlprefix\url{https://doi.org/10.1103%2Fphysreva.103.l030401}.

\bibitem{goodfellow:2016}
Goodfellow~I, Bengio~Y, Courville~A. Deep learning. MIT press; 2016.

\bibitem{Kubler:2019}
Kübler~JM, Muandet~K, Schölkopf~B. Quantum mean embedding of probability
  distributions. Physical Review Research. 2019 dec;\hspace{0pt}1(3).
  \urlprefix\url{https://doi.org/10.1103%2Fphysrevresearch.1.033159}.

\bibitem{Suzuki:2020}
Suzuki~Y, Yano~H, Gao~Q, et~al. Analysis and synthesis of feature map for
  kernel-based quantum classifier. Quantum Machine Intelligence. 2020
  jun;\hspace{0pt}2(1).
  \urlprefix\url{https://doi.org/10.1007\%2Fs42484-020-00020-y}.

\bibitem{Date:2021}
Date~P, Arthur~D, Pusey-Nazzaro~L. {QUBO} formulations for training machine
  learning models. Scientific Reports. 2021 may;\hspace{0pt}11(1).
  \urlprefix\url{https://doi.org/10.1038%2Fs41598-021-89461-4}.

\bibitem{Date:2021a}
Date~P, Potok~T. Adiabatic quantum linear regression. Scientific Reports. 2021
  nov;\hspace{0pt}11(1).
  \urlprefix\url{https://doi.org/10.1038%2Fs41598-021-01445-6}.

\bibitem{Arthur:2020}
Arthur~D, Date~P. Balanced k-means clustering on an adiabatic quantum computer.
  arXiv. 2020;\hspace{0pt}\urlprefix\url{https://arxiv.org/abs/2008.04419}.

\bibitem{Cerezo:2021a}
Cerezo~M, Arrasmith~A, Babbush~R, et~al. Variational quantum algorithms. Nature
  Reviews Physics. 2021 aug;\hspace{0pt}3(9):625--644.
  \urlprefix\url{https://doi.org/10.1038%2Fs42254-021-00348-9}.

\bibitem{Farhi:2018}
Farhi~E, Neven~H. Classification with quantum neural networks on near term
  processors ; 2018. \urlprefix\url{https://arxiv.org/abs/1802.06002}.

\bibitem{Havlicek:2019}
Havl{\'{\i}}{\v{c}}ek~V, C{\'{o}}rcoles~AD, Temme~K, et~al. Supervised learning
  with quantum-enhanced feature spaces. Nature. 2019
  mar;\hspace{0pt}567(7747):209--212.
  \urlprefix\url{https://doi.org/10.1038%2Fs41586-019-0980-2}.

\bibitem{beer:2020}
Beer~K, Bondarenko~D, Farrelly~T, et~al. Training deep quantum neural networks.
  Nature communications. 2020;\hspace{0pt}11(1):1--6.

\bibitem{Cong:2019}
Cong~I, Choi~S, Lukin~MD. Quantum convolutional neural networks. Nature
  Physics. 2019 Aug;\hspace{0pt}15(12):1273–1278.
  \urlprefix\url{http://dx.doi.org/10.1038/s41567-019-0648-8}.

\bibitem{Du:2021a}
Du~Y, Hsieh~MH, Liu~T, et~al. Learnability of quantum neural networks. PRX
  Quantum. 2021 Nov;\hspace{0pt}2:040337.
  \urlprefix\url{https://link.aps.org/doi/10.1103/PRXQuantum.2.040337}.

\bibitem{Mitarai:2018}
Mitarai~K, Negoro~M, Kitagawa~M, et~al. Quantum circuit learning. Physical
  Review A. 2018 Sep;\hspace{0pt}98(3).
  \urlprefix\url{http://dx.doi.org/10.1103/PhysRevA.98.032309}.

\bibitem{kashif:2021}
Kashif~M, Al-Kuwari~S. Design space exploration of hybrid quantum--classical
  neural networks. Electronics. 2021;\hspace{0pt}10(23):2980.
  \urlprefix\url{https://www.mdpi.com/2079-9292/10/23/2980}.

\bibitem{Kashif:2022}
Kashif~M, Al-Kuwari~S. Demonstrating quantum advantage in hybrid quantum neural
  networks for model capacity. In: 2022 IEEE International Conference on
  Rebooting Computing (ICRC); 2022. p. 36--44.

\bibitem{Schuld:2020}
Schuld~M, Bocharov~A, Svore~KM, et~al. Circuit-centric quantum classifiers.
  Phys Rev A. 2020 Mar;\hspace{0pt}101:032308.
  \urlprefix\url{https://link.aps.org/doi/10.1103/PhysRevA.101.032308}.

\bibitem{Zhang:2020}
Zhang~K, Hsieh~MH, Liu~L, et~al. Toward trainability of quantum neural networks
  ; 2020. \urlprefix\url{https://arxiv.org/abs/2011.06258}.

\bibitem{Tacchino:2020}
Tacchino~F, Barkoutsos~P, Macchiavello~C, et~al. Quantum implementation of an
  artificial feed-forward neural network. Quantum Science and Technology. 2020
  oct;\hspace{0pt}5(4):044010.
  \urlprefix\url{https://doi.org/10.1088%2F2058-9565%2Fabb8e4}.

\bibitem{sharma:2020a}
Sharma~K, Cerezo~M, Cincio~L, et~al. Trainability of dissipative
  perceptron-based quantum neural networks ; 2020.
  \urlprefix\url{https://arxiv.org/abs/2005.12458}.

\bibitem{Killoran:2019}
Killoran~N, Bromley~TR, Arrazola~JM, et~al. Continuous-variable quantum neural
  networks. Phys Rev Research. 2019 Oct;\hspace{0pt}1:033063.
  \urlprefix\url{https://link.aps.org/doi/10.1103/PhysRevResearch.1.033063}.

\bibitem{Torrontegui:2019}
Torrontegui~E, Garc{\'{\i} }a-Ripoll~JJ. Unitary quantum perceptron as
  efficient universal approximator. {EPL} (Europhysics Letters). 2019
  mar;\hspace{0pt}125(3):30004.
  \urlprefix\url{https://doi.org/10.1209%2F0295-5075%2F125%2F30004}.

\bibitem{Wan:2017}
Wan~KH, Dahlsten~O, Kristj{\'{a}}nsson~H, et~al. Quantum generalisation of
  feedforward neural networks. npj Quantum Information. 2017
  sep;\hspace{0pt}3(1).
  \urlprefix\url{https://doi.org/10.1038%2Fs41534-017-0032-4}.

\bibitem{benedetti:2019}
Benedetti~M, Lloyd~E, Sack~S, et~al. Parameterized quantum circuits as machine
  learning models. Quantum Science and Technology. 2019
  nov;\hspace{0pt}4(4):043001.
  \urlprefix\url{https://doi.org/10.1088/2058-9565/ab4eb5}.

\bibitem{Hubregtsen:2020}
Hubregtsen~T, Pichlmeier~J, Stecher~P, et~al. Evaluation of parameterized
  quantum circuits: on the relation between classification accuracy,
  expressibility and entangling capability ; 2020.
  \urlprefix\url{https://arxiv.org/abs/2003.09887}.

\bibitem{Abbas:2021}
Abbas~A, Sutter~D, Zoufal~C, et~al. The power of quantum neural networks.
  Nature Computational Science. 2021 Jun;\hspace{0pt}1(6):403–409.
  \urlprefix\url{http://dx.doi.org/10.1038/s43588-021-00084-1}.

\bibitem{Rebentrost:2014}
Rebentrost~P, Mohseni~M, Lloyd~S. Quantum support vector machine for big data
  classification. Physical Review Letters. 2014 sep;\hspace{0pt}113(13).
  \urlprefix\url{https://doi.org/10.1103%2Fphysrevlett.113.130503}.

\bibitem{Biamonte:2017}
Biamonte~J, Wittek~P, Pancotti~N, et~al. Quantum machine learning. Nature. 2017
  sep;\hspace{0pt}549(7671):195--202.
  \urlprefix\url{https://doi.org/10.1038%2Fnature23474}.

\bibitem{Chia:2022}
Chia~NH, Gily{\'e}n~AP, Li~T, et~al. Sampling-based sublinear low-rank matrix
  arithmetic framework for dequantizing quantum machine learning. Journal of
  the ACM. 2022;\hspace{0pt}69(5):1--72.

\bibitem{Wang:2021}
Wang~S, Fontana~E, Cerezo~M, et~al. Noise-induced barren plateaus in
  variational quantum algorithms. Nature Communications. 2021
  nov;\hspace{0pt}12(1).
  \urlprefix\url{https://doi.org/10.1038%2Fs41467-021-27045-6}.

\bibitem{Wecker:2015}
Wecker~D, Hastings~MB, Troyer~M. Progress towards practical quantum variational
  algorithms. Phys Rev A. 2015 Oct;\hspace{0pt}92:042303.
  \urlprefix\url{https://link.aps.org/doi/10.1103/PhysRevA.92.042303}.

\bibitem{Bartlett:2007}
Bartlett~RJ, Musia\l{}~M. Coupled-cluster theory in quantum chemistry. Rev Mod
  Phys. 2007 Feb;\hspace{0pt}79:291--352.
  \urlprefix\url{https://link.aps.org/doi/10.1103/RevModPhys.79.291}.

\bibitem{lee:2018}
Lee~J, Huggins~WJ, Head-Gordon~M, et~al. Generalized unitary coupled cluster
  wave functions for quantum computation. Journal of chemical theory and
  computation. 2018;\hspace{0pt}15(1):311--324.

\bibitem{Cao:2019}
Cao~Y, Romero~J, Olson~JP, et~al. Quantum chemistry in the age of quantum
  computing. Chemical Reviews. 2019 aug;\hspace{0pt}119(19):10856--10915.
  \urlprefix\url{https://doi.org/10.1021%2Facs.chemrev.8b00803}.

\bibitem{Hadfield:2019}
Hadfield~S, Wang~Z, O’Gorman~B, et~al. From the quantum approximate
  optimization algorithm to a quantum alternating operator ansatz. Algorithms.
  2019;\hspace{0pt}12(2).
  \urlprefix\url{https://www.mdpi.com/1999-4893/12/2/34}.

\bibitem{Kandala:2017}
Kandala~A, Mezzacapo~A, Temme~K, et~al. Hardware-efficient variational quantum
  eigensolver for small molecules and quantum magnets. Nature. 2017
  sep;\hspace{0pt}549(7671):242--246.
  \urlprefix\url{https://doi.org/10.1038%2Fnature23879}.

\bibitem{Holmes:2022}
Holmes~Z, Sharma~K, Cerezo~M, et~al. Connecting ansatz expressibility to
  gradient magnitudes and barren plateaus. PRX Quantum. 2022
  Jan;\hspace{0pt}3:010313.
  \urlprefix\url{https://link.aps.org/doi/10.1103/PRXQuantum.3.010313}.

\bibitem{Marrero:2020}
Marrero~CO, Kieferová~M, Wiebe~N. Entanglement induced barren plateaus ; 2020.
  \urlprefix\url{https://arxiv.org/abs/2010.15968}.

\bibitem{Maciejewski:2021}
Maciejewski~FB, Baccari~F, Zimbor{\'{a}}s~Z, et~al. Modeling and mitigation of
  cross-talk effects in readout noise with applications to the quantum
  approximate optimization algorithm. Quantum. 2021 jun;\hspace{0pt}5:464.
  \urlprefix\url{https://doi.org/10.22331%2Fq-2021-06-01-464}.

\bibitem{Alam:2019}
Alam~M, Ash-Saki~A, Ghosh~S. Analysis of quantum approximate optimization
  algorithm under realistic noise in superconducting qubits ; 2019.
  \urlprefix\url{https://arxiv.org/abs/1907.09631}.

\bibitem{Xue:2019}
Xue~C, Chen~ZY, Wu~YC, et~al. Effects of quantum noise on quantum approximate
  optimization algorithm ; 2019.
  \urlprefix\url{https://arxiv.org/abs/1909.02196}.

\bibitem{Schuld;2021}
Schuld~M, Sweke~R, Meyer~JJ. Effect of data encoding on the expressive power of
  variational quantum-machine-learning models. Phys Rev A. 2021
  Mar;\hspace{0pt}103:032430.
  \urlprefix\url{https://link.aps.org/doi/10.1103/PhysRevA.103.032430}.

\bibitem{scikit-learn}
Pedregosa~F, Varoquaux~G, Gramfort~A, et~al. Scikit-learn: Machine learning in
  {P}ython. Journal of Machine Learning Research.
  2011;\hspace{0pt}12:2825--2830.

\bibitem{lecun:1998}
LeCun~Y. The mnist database of handwritten digits. http://yann lecun
  com/exdb/mnist/. 1998;\hspace{0pt}.

\bibitem{Du:2021}
Du~Y, Hsieh~MH, Liu~T, et~al. A grover-search based quantum learning scheme for
  classification. New Journal of Physics. 2021 feb;\hspace{0pt}23(2):023020.
  \urlprefix\url{https://doi.org/10.1088/1367-2630/abdefa}.

\bibitem{Arthur:2022}
Arthur~D, Date~P. A hybrid quantum-classical neural network architecture for
  binary classification ; 2022.
  \urlprefix\url{https://arxiv.org/abs/2201.01820}.

\bibitem{Schuld:2019}
Schuld~M, Killoran~N. Quantum machine learning in feature hilbert spaces. Phys
  Rev Lett. 2019 Feb;\hspace{0pt}122:040504.
  \urlprefix\url{https://link.aps.org/doi/10.1103/PhysRevLett.122.040504}.

\bibitem{schuld:2018}
Schuld~M, Petruccione~F. Supervised learning with quantum computers. Vol.~17.
  Springer; 2018.

\bibitem{Lloyd:2020}
Lloyd~S, Schuld~M, Ijaz~A, et~al. Quantum embeddings for machine learning ;
  2020. \urlprefix\url{https://arxiv.org/abs/2001.03622}.

\bibitem{LaRose:2020}
LaRose~R, Coyle~B. Robust data encodings for quantum classifiers. Physical
  Review A. 2020 sep;\hspace{0pt}102(3).
  \urlprefix\url{https://doi.org/10.1103\%2Fphysreva.102.032420}.

\bibitem{grant:2018}
Grant~E, Benedetti~M, Cao~S, et~al. Hierarchical quantum classifiers. npj
  Quantum Information. 2018;\hspace{0pt}4(1):1--8.

\bibitem{Cao:2020}
Cao~S, Wossnig~L, Vlastakis~B, et~al. Cost-function embedding and dataset
  encoding for machine learning with parametrized quantum circuits. Phys Rev A.
  2020 May;\hspace{0pt}101:052309.
  \urlprefix\url{https://link.aps.org/doi/10.1103/PhysRevA.101.052309}.

\bibitem{Schuld:2019a}
Schuld~M, Bergholm~V, Gogolin~C, et~al. Evaluating analytic gradients on
  quantum hardware. Physical Review A. 2019 mar;\hspace{0pt}99(3).
  \urlprefix\url{https://doi.org/10.1103%2Fphysreva.99.032331}.

\bibitem{Skolik:2021}
Skolik~A, McClean~JR, Mohseni~M, et~al. Layerwise learning for quantum neural
  networks. Quantum Machine Intelligence. 2021 jan;\hspace{0pt}3(1).
  \urlprefix\url{https://doi.org/10.1007%2Fs42484-020-00036-4}.

\bibitem{Volkoff:2021}
Volkoff~T, Coles~PJ. Large gradients via correlation in random parameterized
  quantum circuits. Quantum Science and Technology. 2021
  jan;\hspace{0pt}6(2):025008.
  \urlprefix\url{https://doi.org/10.1088%2F2058-9565%2Fabd891}.

\bibitem{Bengio:2006}
Bengio~Y, Lamblin~P, Popovici~D, et~al. Greedy layer-wise training of deep
  networks. Advances in neural information processing systems.
  2006;\hspace{0pt}19.

\bibitem{kashif:2023resqnets}
Kashif~M, Al-kuwari~S. Resqnets: A residual approach for mitigating barren
  plateaus in quantum neural networks ; 2023.
  \urlprefix\url{https://arxiv.org/abs/2305.03527}.

\bibitem{Arrasmith:2021}
Arrasmith~A, Cerezo~M, Czarnik~P, et~al. Effect of barren plateaus on
  gradient-free optimization. Quantum. 2021 oct;\hspace{0pt}5:558.
  \urlprefix\url{https://doi.org/10.22331%2Fq-2021-10-05-558}.

\bibitem{Cerezo:2021aa}
Cerezo~M, Sone~A, Volkoff~T, et~al. Cost function dependent barren plateaus in
  shallow parametrized quantum circuits. Nature Communications. 2021
  mar;\hspace{0pt}12(1).
  \urlprefix\url{https://doi.org/10.1038\%2Fs41467-021-21728-w}.

\bibitem{Kashif:2023}
Kashif~M, Al-Kuwari~S. The impact of cost function globality and locality in
  hybrid quantum neural networks on nisq devices. Machine Learning: Science and
  Technology. 2023 jan;\hspace{0pt}4(1):015004.
  \urlprefix\url{https://dx.doi.org/10.1088/2632-2153/acb12f}.

\bibitem{sklearn}
Pedregosa~F, Varoquaux~G, Gramfort~A, et~al. Scikit-learn: Machine learning in
  {P}ython. Journal of Machine Learning Research.
  2011;\hspace{0pt}12:2825--2830.

\bibitem{Kingma:2014}
Kingma~DP, Ba~J. Adam: A method for stochastic optimization ; 2014.
  \urlprefix\url{https://arxiv.org/abs/1412.6980}.

\end{thebibliography}


\begin{thebibliography}{99}

\bibitem{Jen05}
Jenkins~PF. Making sense of the chest x-ray: a hands-on guide. New York (NY):
  Oxford University Press; 2005.

\bibitem{Sch02}
Schott~J, Priest~J. Leading antenatal classes: a practical guide. 2nd ed.
  Boston (MA): Books for Midwives; 2002.

\bibitem{Wen95}
Wenger~NK, Sivarajan~Froelicher~E, Smith~LK, et~al. Cardiac rehabilitation.
  Rockville (MD): Agency for Health Care Policy and Research (US); 1995.

\bibitem{Sha78}
Shakelford~RT. Surgery of the alimentary tract. Philadelphia (PA): W.B.
  Saunders; 1978. Chapter 2, Esophagoscopy; p. 29--40.

\bibitem{AG98}
Ambudkar~SV, Gottesman~MM, editors. {ABC} transporters: biomedical, cellular,
  and molecular aspects. San Diego (CA): Academic Press; 1998. (Methods in
  enzymology; vol. 292).

\bibitem{Smi75}
Smith~CE. The significance of mosquito longevity and blood-feeding behaviour in
  the dynamics of arbovirus infections. Med Biol. 1975;53:288--294.

\bibitem{Men05}
Meneton~P, Jeunemaitre~X, de~Wardener~HE, et~al. Links between dietary salt
  intake, renal salt handling, blood pressure, and cardiovascular diseases.
  Physiol Rev. 2005 Apr;85:679--715.

\bibitem{DCK03}
Dostorovsky~JO, Carr~DB, Koltzenburg~M, editors. Proceedings of the 10th World
  Congress on Pain; 2002 Aug~17--22; San Diego, CA. Seattle: IASP Press; c2003.

\bibitem{Hor98}
Horrobin~DF, Lampinskas~P. The commercial development of food plants used as
  medicines. In: Prendergast~HD, Etkin~NL, Harris~DR, et~al., editors. Plants
  for food and medicine. Proceedings of the Joint Conference of the Society for
  Economic Botany and the International Society for Ethnopharmacology; 1996
  Jul~1--6; London. Kew (UK): Royal Botanic Gardens; 1998. p. 75--81.

\bibitem{Ant03}
Antani~S, Long~LR, Thoma~GR, et~al. Anatomical shape representation in spine
  x-ray images. Paper presented at: VIIP 2003. Proceedings of the 3rd IASTED
  International Conference on Visualization, Imaging and Image Processing;
  2003 Sep~8--10; Benalmadena, Spain.

\bibitem{Zha05}
Zhao~C. Development of nanoelectrospray and application to protein research and
  drug discovery [dissertation]. Buffalo (NY): State University of New York at
  Buffalo; 2005.

\bibitem{Rog05}
Roguskie~JM. The role of \emph{Pseudomonas aeruginosa} 1244 pilin glycan in
  virulence [master's thesis]. [Pittsburgh (PA)]: Duquesne University; 2005.

\bibitem{SRW05}
Savage~E, Ramsay~M, White~J, et~al. Mumps outbreaks across England and Wales
  in 2004: Observational study. BMJ. 2005;330(7500):1119--1120 [cited 2005
  May~31]; Available from: http://bmj.bmjjournals.com/cgi/reprint/330/7500/1119.

\bibitem{Gau05}
Gaul~G. When geography influences treament options. Washington Post (Maryland
  Ed.). 2005 Jul~24;Sect.~A:12 (col.~1).

\bibitem{BGC04}
Berrino~F, Gatta~G, Crosignani~P. [Case-control evaluation of screening
  efficacy]. Epidemiol Prev. 2004 Nov--Dec;28:354--359. Italian.

\bibitem{PI51}
Piaget~J, Inhelder~B. La gen{\`e}se de l'id{\'e}e de hasard chez l'enfant
  [The origin of the idea of chance in the child]. Paris: Presses
  Universitaires de France; 1951.

\end{thebibliography}


\appendix
\
\section{Design guidelines for all constraint scenarios} \label{appendix}

Here we provide detailed recommendations about the design parameters of an HQNN based on the results of our framework.
We consider all possible constraints while designing the HQNNs typically the quantum layers. 
Since our proposed framework focuses on the analysis of data encoding, ansatz expressibility, and entanglement, the constraints on these components of HQNNs design are considered both individually and with respect to each other. It is important to note that when multiple depths for a single width (or vice versa) are recommended, they achieve almost the same accuracy, i.e., ($\pm1\%$). In such cases, we comment on the convergence rate to select the optimal width and depth. If all the recommended widths and depths converge on the same rate, then the smallest one would be more appropriate to use because of the relatively faster training. 


\subsection{No Constraint} \label{app:no_constraint}
    
\begin{table}[H]
        \centering
        
        \begin{tabular}{|>\centering p{3.2cm}|>\centering p{2cm}|>\centering p{2cm}|>\centering p{2.2cm}|>\centering p{3.5cm}|}
            \hline 
            Recommended Data Encoding & Optimal Width (n) & Optimal Depth (m) & Entanglement Inclusion & Remarks
            \tabularnewline
            \hline
            \multirow{3}{*}{Angle Encoding} &10 & 2,4 & No & 
            \tabularnewline 
             \cline{2-4}
              &12 &2,4,8 &No &$m=4$ converges faster
              \tabularnewline
             \cline{2-4}
            &14 &2,4,6 &No &
            \tabularnewline
            \hline
              
        \end{tabular}
        \caption{Recommendations when there is no constraint on any design parameter}
        \label{tab:no_constraint}
    \end{table}

  
\subsection{Constraint on a Single parameter} \label{app:constraint_one_param}

\begin{table}[H]
        \centering
        
        \begin{tabular}{|>\centering p{3.2cm}|>\centering p{2cm}|>\centering p{2cm}|>\centering p{2cm}|>\centering p{3.8cm}|}
            \hline \multirow{2}{*}{Constraint} & \multicolumn{3}{c|}{Recommendations} & \multirow{2}{*}{Remarks} \tabularnewline
            \cline{2-4}
             & Optimal Width (n) & Optimal Depth (m) & Entanglement Inclusion &\tabularnewline
             \hline
             
             \multirow{2}{*}{Amplitude Encoding} & $6$ & $8,10$ & Yes & $m=10$ converges faster\tabularnewline
             \cline{2-5}
             & $8$ & $6, 8,10$ & Yes & $m=6$ converges faster \tabularnewline
             \hline
             
            Angle Encoding &\multicolumn{4}{c|}{\bfseries Refer to Table \ref{tab:no_constraint}} \tabularnewline
           
            \hline

        \end{tabular}
        \caption{Recommendations when there is constraint only on type of data encoding}
        \label{tab:Constraint_data_enc}
    \end{table}


\begin{table}[H]
        \centering
        
        \begin{tabular}{|>\centering p{3.2cm}|>\centering p{2cm}|>\centering p{2cm}|>\centering p{2cm}|>\centering p{3.8cm}|}
            \hline 
            Constraint &\multicolumn{3}{c|}{Recommendations} &\multirow{2}{*}{Remarks} \tabularnewline
            
        \cline{1-4}
        
        Ansatz Width & Optimal Depth (m) & Data Encoding &Entanglement Inclusion & \tabularnewline
             
         \hline  
            
        8 &6,8,10 &Amplitude &Yes &$m=6$ converges faster
        \tabularnewline
        \hline
         10 &2,4 &Angle &No &
         \tabularnewline
         \cline{1-4}
        12 &2,4,8 &Angle &No & $m=4$ converges faster
         \tabularnewline
         \cline{1-4}
        14 &2,4,6 &Angle &No & 
         \tabularnewline
        \hline

        \end{tabular}
        \caption{Recommendations when there is constraint only on ansatz width}
        \label{tab:Constraint_ansatz_width}
    \end{table}


\begin{table}[H]
        \centering
        
        \begin{tabular}{|>\centering p{3.2cm}|>\centering p{2cm}|>\centering p{2cm}|>\centering p{2cm}|>\centering p{3.8cm}|}
            \hline 
          Constraint &\multicolumn{3}{c|}{Recommendations} &\multirow{2}{*}{Remarks} \tabularnewline
            
        \cline{1-4}
        
       Ansatz Depth & Optimal Width (n) & Data Encoding &Entanglement Inclusion & \tabularnewline
        \hline  
        
        2 &10,12,14 &Angle &No &Similar convergence for \tabularnewline
         \cline{1-4}
         
         4 &10,12,14 &Angle &No & all recommended widths
         \tabularnewline
          \hline
          
          \multirow{2}{*}{6} &8 &Amplitude &Yes & - 
          \tabularnewline
          \cline{2-5}
           &14 &Angle &No &-
           \tabularnewline
          \hline
          \multirow{2}{*}{8} &6 &Amplitude &Yes & - 
          \tabularnewline
          \cline{2-5}
           &12 &Angle &No &-
           \tabularnewline
          
          \hline
          \multirow{2}{*}{10} &6,8 &Amplitude &Yes & $n=6$ converges faster 
          \tabularnewline
          \cline{2-5}
           &12 &Angle &No &$n=12$ converges faster
           \tabularnewline
          
          \hline
          
        \end{tabular}
        \caption{Recommendations when there is constraint only on ansatz depth}
        \label{tab:Constraint_ansatz_depth}
    \end{table}


\begin{table}[H]
        \centering
        
        \begin{tabular}{|>\centering p{3.2cm}|>\centering p{2cm}|>\centering p{2cm}|>\centering p{2cm}|>\centering p{3.8cm}|}
            \hline 
          \multirow{2}{*}{Constraint} &\multicolumn{3}{c|}{Recommendations} &\multirow{2}{*}{Remarks} \tabularnewline
          \cline{2-4}
          
          &Data Encoding &Optimal Width (n) &Optimal Depth (m) & 
          \tabularnewline
          \hline
          
          Entanglement  &\multirow{2}{*}{Amplitude} &6 &8,10 &$m=10$ converges faster
          \tabularnewline
          \cline{3-5}
           Required& &8 &6,10 &$m=6$ converges faster
           \tabularnewline
            \hline
            
         Entanglement  &\multirow{3}{*}{Angle} &10 &2,4 &
          \tabularnewline
          \cline{3-4}
           \multirow{2}{*}{Not Required} & &12 &2,4,8 &$n=12$ converges faster
           \tabularnewline
        \cline{3-4}
         & &14 &2,4,6 &
         \tabularnewline

          \hline
          
        \end{tabular}
        \caption{Recommendations when there is constraint only on entanglement}
        \label{tab:Constraint_entanglement}
    \end{table}


\subsection{Constraint on a Two parameters} \label{app_two_params_constraint}

 \begin{table}[H]
        \centering
        
        \begin{tabular}{|>\centering p{3.2cm}|>\centering p{2cm}|>\centering p{2cm}|>\centering p{2cm}|>\centering p{3.8cm}|}
            \hline 
          \multicolumn{2}{|c|}{Constraint} &\multicolumn{2}{c|}{Recommendations}
         &\multirow{2}{*}{Remarks}
          \tabularnewline
         \cline{1-4}
         
          Data Encoding & Ansatz Width (n) & Optimal Depth (m) &Entanglement Inclusion &
          \tabularnewline
          
         \hline \multirow{4}{*}{Amplitude} &8 &6,10 &Yes &$m=6$ converges faster \tabularnewline
          \cline{2-5}
           &10 &6,8,10 &Yes &$m=8$ converges faster
          \tabularnewline
          \cline{2-5}
          
          &12 &4 &Yes &-
          \tabularnewline
          \cline{2-5}
          
          &14 &2 &Yes &-
          \tabularnewline
          \cline{2-5}
          
          \hline

          \multirow{4}{*}{Angle} &8 &6,8 &No &$m=8$ converges faster \tabularnewline
          \cline{2-5}
           &10 &6,8 &No &$m=6$ converges faster
          \tabularnewline
          \cline{2-5}
          
          &12 &4,6 &No &$m=6$ converges faster
          \tabularnewline
          \cline{2-5}
          
          &14 &4 &No &-
          \tabularnewline
          \cline{2-5}
          
          \hline

        \end{tabular}
        \caption{Recommendations when there is constraint data encoding and ansatz width}
        \label{tab:Constraint_enc_width}
    \end{table}


 \begin{table}[H]
        \centering
        
        \begin{tabular}{|>\centering p{3.2cm}|>\centering p{2cm}|>\centering p{2cm}|>\centering p{2cm}|>\centering p{3.8cm}|}
            \hline 
          \multicolumn{2}{|c|}{Constraint} &\multicolumn{2}{c|}{Recommendations}
         &\multirow{2}{*}{Remarks}
          \tabularnewline
         \cline{1-4}
         
          Data Encoding & Ansatz Depth (m) & Optimal Width (n) &Entanglement Inclusion &
          \tabularnewline
          
         \hline \multirow{4}{*}{Amplitude} &2 &6,10 &Yes &$n=6$ converges faster \tabularnewline
          \cline{2-5}
           &4 &12 &Yes &-
          \tabularnewline
          \cline{2-5}
          
          &6 &8 &Yes &-
          \tabularnewline
          \cline{2-5}
          
          &8 &6,8,10 &Yes &$n=6$ converges faster
          \tabularnewline
          \cline{2-5}
          
          &10 &6,8 &Yes &$n=6$ converges faster
          \tabularnewline
          \hline

         \hline \multirow{4}{*}{Angle} &2 &10,12,14 &No &\multirow{2}{*}{$n=6$ converges faster} \tabularnewline
          \cline{2-4}
           &4 &10,12,14 &No &
          \tabularnewline
          \cline{2-5}
          
          &6 &14 &No &-
          \tabularnewline
          \cline{2-5}
          
          &8 &12 &No &-
          \tabularnewline
          \cline{2-5}
          
          &10 &12,14 &No &$n=14$ converges faster
          \tabularnewline
          \hline 
           
        \end{tabular}
        \caption{Recommendations when there is constraint data encoding and ansatz depth}
        \label{tab:Constraint_enc_depth}
    \end{table} 


 \begin{table}[H]
        \centering
        
        \begin{tabular}{|>\centering p{3.2cm}|>\centering p{2cm}|>\centering p{1.7cm}|>\centering p{1.8cm}|>\centering p{4.5cm}|}
            \hline 
          \multicolumn{2}{|c|}{Constraint} &\multicolumn{2}{c|}{Recommendations}
         &\multirow{2}{*}{Remarks}
          \tabularnewline
         \cline{1-4}
         
          Data Encoding & Entanglement Inclusion & Optimal Width (n) &Optimal Depth (m) &
          \tabularnewline
          \hline

        \multirow{2}{*}{Amplitude} &  \multirow{2}{*}{Yes} &6 &8,10 &$m=10$ converges faster \tabularnewline
        \cline{3-5}
        & &8 &6,10 &$m=6$ converges faster \tabularnewline
        \hline
        \multirow{2}{*}{Amplitude} &  \multirow{2}{*}{No} &6 &2 &- \tabularnewline
        \cline{3-5}
        & &8 &4,6 &similar convergence rate for  \tabularnewline 
         &&&&both recommended depths \tabularnewline 
          
         \hline 
         
        \multirow{3}{*}{Angle} &  \multirow{3}{*}{Yes} &8 &8 &- \tabularnewline
        \cline{3-5}
        & &12 &10 &-\tabularnewline 
         \cline{3-5}
        & &14 &6 &-\tabularnewline \hline
         \multirow{3}{*}{Angle} &  \multirow{3}{*}{No} &10 &2,4 &\multirow{3}{*}{$n=12$ converges faster} \tabularnewline
        \cline{3-4}
        & &12 &2,4,8 &-\tabularnewline 
         \cline{3-4}
        & &14 &2,4,6 & \tabularnewline 
          
        \hline
        \end{tabular}
        \caption{Recommendations when there is constraint data encoding and entanglement}
        \label{tab:Constraint_enc_entanglement}
    \end{table} 


 \begin{table}[H]
        \centering
        
        \begin{tabular}{|>\centering p{3.2cm}|>\centering p{2cm}|>\centering p{2cm}|>\centering p{2cm}|>\centering p{3.8cm}|}
            \hline 
          \multicolumn{2}{|c|}{Constraint} &\multicolumn{2}{c|}{Recommendations}
         &\multirow{2}{*}{Remarks}
          \tabularnewline
         \cline{1-4}
         
         Entanglement Inclusion   & Ansatz Width (n) &Optimal Depth (m) & Data Encoding &
          \tabularnewline
          \hline
         Yes &8 &6 &Amplitude &- \tabularnewline
         \hline
         Yes &10 &8,10 &Amplitude &$m=10$ converges faster \tabularnewline
         \hline
         Yes &12 &4 &Amplitude &- \tabularnewline
         \hline
        Yes &14 &4,6 &Angle &$m=6$ converges faster \tabularnewline
         \hline
        
        No &8 &2,4,6 &Angle &Similar convergence \tabularnewline
         \cline{1-4}
         No &10 &2,4 &Angle &rate for all the reco- \tabularnewline
         \cline{1-4}
         No &12 &2,4,8 &Angle &mmended widths but \tabularnewline
         \cline{1-4}
         No &14 &2,4,6 &Angle &$n=12$ converges slightly faster \tabularnewline
         \hline
        
        \end{tabular}
        \caption{Recommendations when there is constraint entanglement and ansatz width}
        \label{tab:Constraint_entanglement_width}
    \end{table} 
    

 \begin{table}[H]
        \centering
        
        \begin{tabular}{|>\centering p{3.2cm}|>\centering p{2cm}|>\centering p{2cm}|>\centering p{2cm}|>\centering p{3.8cm}|}
            \hline 
          \multicolumn{2}{|c|}{Constraint} &\multicolumn{2}{c|}{Recommendations}
         &\multirow{2}{*}{Remarks}
          \tabularnewline
         \cline{1-4}
         
         Entanglement Inclusion   & Ansatz Depth (m) &Optimal Width (n) & Data Encoding &
          \tabularnewline
          \hline
        
        \multirow{2}{*}{Yes} &\multirow{2}{*}{2} &10 &Amplitude &- \tabularnewline
        \cline{3-5}
          &  & 10,12 &Angle &$n=10$ converges faster \tabularnewline
          \hline
        
         \multirow{2}{*}{Yes} &\multirow{2}{*}{4} &12 &Amplitude &- \tabularnewline
        \cline{3-5}
          &  &14 &Angle &- \tabularnewline
          \hline
        
        Yes &6 &8 &Amplitude &- \tabularnewline
        \hline
        
         Yes &8 &8,10 &Amplitude &$n=8$ converges faster \tabularnewline
        \hline
        
         Yes &10 &8 &Amplitude &- \tabularnewline
        \hline
        
        \multirow{2}{*}{No} &\multirow{2}{*}{2}  &\multirow{2}{*}{10,12,14}  &\multirow{2}{*}{Angle}  &\multirow{2}{*}{Almost same convergence }
        \tabularnewline
        &&&&\tabularnewline
        &&&&rate for all widths but \tabularnewline
        \cline{1-4}
        No &4 &12,14 &Angle &$n=12$ converges slightly faster   \tabularnewline
        
        \hline
       
        \end{tabular}
        \caption{Recommendations when there is constraint on entanglement and ansatz depth}
        \label{tab:Constraint_entanglement_depth}
    \end{table} 
    

\subsection{Constraint on a All three parameters} \label{app_constraint_three_params}
 \begin{table}[H]
        \centering
        
        \begin{tabular}{|>\centering p{2.8cm}|>\centering p{2cm}|>\centering p{2cm}|>\centering p{2.7cm}|>\centering p{3.8cm}|}
            \hline 
          \multicolumn{3}{|c|}{Constraint} &Recommendations
         &\multirow{2}{*}{Remarks}
          \tabularnewline
         \cline{1-4}
         
         Data Encoding  & Ansatz Depth (m) & Entanglement Inclusion &Optimal Width (n) & 
          \tabularnewline
          \hline
        
        \multirow{10}{*}{Amplitude} &2 &Yes &6,10 &$n=10$ converges faster \tabularnewline
        \cline{2-5}
         &4 &Yes &12 &- \tabularnewline
        \cline{2-5}
          &6 &Yes &8 &- \tabularnewline
         \cline{2-5}
          &8 &Yes &6 &- \tabularnewline
          \cline{2-5}
        &10 &Yes &6,8 &$n=6$ converges faster \tabularnewline
        \cline{2-5}

        &2 &No &6 &- \tabularnewline
        \cline{2-5}
        &4 &No &6,8 &\multirow{2}{*}{$n=6$ converges faster} \tabularnewline
        \cline{2-4}
        &6 &No &6,8 & \tabularnewline
        \cline{2-5}
        &8 &No &6 &- \tabularnewline
        \cline{2-5}
         &10 &No &6 &- \tabularnewline
        \hline

        \multirow{10}{*}{Angle} &2 &Yes &10,12 &$n=10$ converges faster \tabularnewline
        \cline{2-5}
         &4 &Yes &14 &- \tabularnewline
        \cline{2-5}
          &6 &Yes &10,14 & $n=10$ converges faster \tabularnewline
         \cline{2-5}
          &8 &Yes &8,10 &$n=8$ converges faster \tabularnewline
          \cline{2-5}
        &10 &Yes &8,12 &$n=12$ converges faster \tabularnewline
        \cline{2-5}
       
       &2 &No &10,12,14 &  \tabularnewline
        \cline{2-4}
        
        &4 &No &12,14 & All width have almost   \tabularnewline
        \cline{2-4}
        
        &6 &No &14 & same convergence rates  \tabularnewline
        \cline{2-4}
        
        &8 &No &12 & but $n=12$ converges slightly faster \tabularnewline
        \cline{2-4}
        
        &10 &No &10,12,14 & \tabularnewline
        \cline{2-4}
        
        \hline

        \end{tabular}
        \caption{Recommendations when there is constraint on data encoding, ansatz depth and entanglement}
        \label{tab:Constraint_enc_depth_entanglement}
    \end{table}


 \begin{table}[H]
        \centering
        
        \begin{tabular}{|>\centering p{2.8cm}|>\centering p{2cm}|>\centering p{2cm}|>\centering p{2.7cm}|>\centering p{3.8cm}|}
            \hline 
          \multicolumn{3}{|c|}{Constraint} &Recommendations
         &\multirow{2}{*}{Remarks}
          \tabularnewline
         \cline{1-4}
         
         Data Encoding  & Ansatz Width (n) & Entanglement Inclusion &Optimal Depth (m) & 
          \tabularnewline
          \hline
        
        \multirow{9}{*}{Amplitude} &6 &Yes &8,10 &$m=10$ converges faster \tabularnewline
        \cline{2-5}
         &8 &Yes &6,10 &$m=6$ converges faster \tabularnewline
        \cline{2-5}
          &10 &Yes &8,10 &$m=10$ converges faster \tabularnewline
         \cline{2-5}
          &12 &Yes &4 &- \tabularnewline
          \cline{2-5}
        &14 &Yes &4 &- \tabularnewline
        \cline{2-5}

        &6 &No &2 &- \tabularnewline
        \cline{2-5}
        &8 &No &4,6 &$m=4$ converges faster \tabularnewline
        \cline{2-5}
        &10 &No &2 &- \tabularnewline
        \cline{2-5}
        &12 &No &2 &-  \tabularnewline
        \hline

        \multirow{10}{*}{Angle} &8 &Yes &6,8 &$m=8$ converges faster \tabularnewline
        \cline{2-5}
         &10 &Yes &6,8 &$m=6$ converges faster \tabularnewline
        \cline{2-5}
          &12 &Yes &2,4,6,8,10 & $m=6$ converges faster \tabularnewline
         \cline{2-5}
          &14 &Yes &4,6,8 &$m=8$ converges faster \tabularnewline
          \cline{2-5}
        
       &8 &No &2,4,6 &  \tabularnewline
        \cline{2-4}
        
        &10 &No &2,4 & All width have almost   \tabularnewline
        \cline{2-4}
        
        &12 &No &2,4,8 & same convergence rates  \tabularnewline
        \cline{2-4}
        
        &14 &No &2,4,6 & but $m=2$ will be better due to shorter training time \tabularnewline
        \cline{2-4}

        \hline

        \end{tabular}
        \caption{Recommendations when there is constraint on data encoding, ansatz width and entanglement}
        \label{tab:Constraint_enc_width_entanglement}
    \end{table}

\end{document}